\documentclass[12pt, preprint]{emulateapj}

\slugcomment{submitted to The Astrophysical Journal}

\shorttitle{Coronal Evolution of the Sun in Time}
\shortauthors{Telleschi et al.}

\begin{document}

\title{Coronal Evolution of the Sun in Time: High-Resolution X-Ray Spectroscopy of Solar Analogs with 
        Different Ages}

\author{Alessandra Telleschi, Manuel G\"udel and Kevin Briggs}
\affil{Paul Scherrer Institut, W\"urenlingen and Villigen, CH-5232 Villigen PSI, Switzerland}
\email{atellesc@astro.phys.ethz.ch, guedel@astro.phys.ethz.ch, briggs@astro.phys.ethz.ch }

\author{Marc Audard}
\affil{Columbia Astrophysics Laboratory, Mail code 5247, 550 West 120$^{th}$ Street, New York, NY 10027}
\email{audard@astro.columbia.edu}

\author{Jan-Uwe Ness}
\affil{Department of Physics (Theoretical Physics), University of Oxford, 1 Keble Road, Oxford OX1 3NP, England, UK}
\email{ness@thphys.ox.ac.uk}

\and

\author{Stephen L. Skinner}
\affil{Center for Astrophysics and Space Astronomy, University of Colorado, Boulder, CO 80309-0389}
\email{skinners@casa.colorado.edu}

\begin{abstract}
We investigate the long-term evolution of X-ray coronae of solar analogs 
based on high-resolution X-ray spectroscopy and photometry with {\it XMM-Newton}.
Six nearby main-sequence G stars with ages between $\approx 0.1$~Gyr and $\approx 1.6$~Gyr
and rotation periods between $\approx 1$~d and 12.4~d have been observed. 
We use the X-ray spectra to derive coronal element abundances of C, N, O, Ne, Mg, Si, S, and Fe
and the coronal emission measure distribution (EMD). We find that the abundances
change from an inverse-First Ionization Potential (FIP) distribution in stars with
ages around 0.1~Gyr to a solar-type FIP distribution  in stars at ages of 0.3~Gyr and beyond.
This transformation is coincident with a steep decline of non-thermal 
radio emission. The results are in qualitative agreement with a simple 
model in which the stream of electrons in magnetic fields suppresses 
diffusion of low-FIP ions from the chromosphere into the corona.
The coronal emission measure distributions show shapes characterized by  power-laws on
each side of the EMD peak. The latter shifts from temperatures of about 10~MK in
the most rapidly rotating, young stars to temperatures around 4~MK in the oldest
target considered here. The power-law index on the cooler side of the EMD exceeds
expected slopes for static loops, with typical values being 1.5--3. We interpret this
slope with a model in which the coronal emission is due to a superposition of stochastically
occurring flares, with an occurrence rate that is distributed in radiated energy $E$ as a 
power-law, $dN/dE \propto E^{-\alpha}$, as previously
found for solar and stellar flares. We obtain the relevant power-law index $\alpha$ from the slope 
of the high-temperature tail of the EMD. 
Our EMDs indicate $\alpha \approx 2.2-2.8$, in excellent agreement with 
values previously derived from light curves of magnetically active stars. 
Modulation with time scales reminiscent of flares is found in the light 
curves of all our targets. Several strong flares are also observed. 
We use our $\alpha$ values to simulate light curves and compare
them with the observed light curves. We thus derive the range of flare energies 
required to explain the light-curve modulation. More active stars require a larger range of flare energies 
than less active stars within the framework of this simplistic model.
In an overall scenario, we propose
that flaring activity plays a larger role in more active stars. 
In this model, the higher flare rate is responsible both for
the higher average coronal temperature and the high coronal 
X-ray luminosity, two parameters that are
indeed found to be correlated. 
\end{abstract}


\section{Introduction}

The solar magnetic field has steadily declined during the Sun's evolution on the main sequence.
Studies of stellar clusters and individual field stars with approximately known ages have shown
that the principal parameter determining magnetic activity on a star is its spin rate which, together
with convection, controls the operation of the internal magnetic dynamo. As a star spins down
due to angular momentum loss via a magnetized stellar wind, its dynamo action weakens, thus
continuously reducing magnetic activity expressed in the outer stellar atmosphere. The spin-down
history of a solar-like star has been studied in detail by using
open-cluster samples (e.g., \citealt{bouvier90, soderblom93}) and accompanying 
theoretical calculations (e.g., \citealt{pinsonneault89, macgregor91}). For a given main-sequence stellar mass
in the spectral domain of G and K stars,  the rotation period is a rather well-defined function of the 
star's age if the latter exceeds a few 100~Myr, regardless of the initial Zero-Age Main-Sequence
(ZAMS) rotation rate. For near-ZAMS stars, in contrast, the rotation rate depends on the pre-main sequence
evolution and may range between rather modest values (rotation periods $P$ of several days) and rates of 
the so-called ultra-fast rotators (periods of $\la 1$~d), regardless of the precise stellar age. This 
spread in rotation rate is well  illustrated by stellar cluster samples, such as the Pleiades, in
which G and K stars are still close to the ZAMS \citep{soderblom93, stauffer94}.

Magnetic activity expresses itself in the outer stellar atmospheres through various phenomena
such as dark magnetic spots, bright chromospheric plages, chromospheric emission lines, and
coronal radio and X-ray emissions. The coronal emissions display the largest range of variation
in response to the surface magnetic activity level. Whereas starspots may cover a few percent of  
the photosphere of the most active solar analogs, thus producing a photometric wave with a modulation 
depth of a few percent at best, the range of X-ray luminosity $L_X$ in a solar analog spans  at least
three orders of magnitude between spun-down inactive examples such as the Sun or $\beta$ Hyi (with $L_X$
between a few times $10^{26}$~erg~s$^{-1}$ and a few times $10^{27}$~erg~s$^{-1}$) and rapid rotators
at the ``saturation level'' ($L_X \approx 10^{-3}L_{\rm bol} \approx [2-4] \times 10^{30}$~erg~s$^{-1}$
for a solar-mass star; \citealt{maggio87, guedel97a}). Even the magnetic cycle of the Sun induces an X-ray 
luminosity variation over one order of magnitude \citep{micela03}. Such systematic cyclic or semi-cyclic variations in turn 
limit the accuracy with which we can attribute ``characteristic'' X-ray luminosity levels to stars of a given 
age if other solar analogs are subject to similar magnetic cycles in X-rays, as seems to be the case \citep{dorren94,
dorren95}.

Short-wavelength ultraviolet and X-ray emissions of a star like the Sun not only serve as a valuable 
diagnostic to study stellar spin-down and the operation of the internal dynamo, they are also pivotal
for the evolution of the outer atmospheres of planets, in particular the photochemistry in the
early atmospheres of Earth-like planets \citep{canuto82,ribas05}. Motivated by the interest in understanding the 
Sun's and our solar system's past, we have been studying the ``Sun in Time'' from the early evolutionary stages
on the ZAMS to the terminal stage on the main sequence, at ages of 5-10~Gyr. This study 
encompasses various wavelength regimes, including radio \citep{guedel01b},
optical and ultraviolet \citep{dorren94, guinan03}, and X-ray wavelengths \citep{dorren95, guedel97a, guedel98b}.
The latter two studies comprise a detailed description of various aspects of the X-ray emission of a
solar analog during its main-sequence evolution, based on low-resolution X-ray spectroscopic data
from the {\it ROSAT} and {\it ASCA} satellites. Among the principal findings of that
work was a clear trend of decreasing coronal electron temperatures as a solar analog ages, following
the decrease in overall X-ray luminosity from ages of $\approx 0.1$~Gyr all the way to
ages of nearly 10~Gyr. The authors speculated that the decreasing heating efficiency
is due to a decreasing level of coronal flaring owing to a smaller filling factor in
older stars. In this picture, the flare rate is responsible both for bringing dense material
into the corona and heating it to high temperatures, analogous to the behavior of individual flares 
observed on the Sun.  A large flare rate in younger stars with a higher magnetic filling factor
could thus produce X-ray coronae that are more luminous and at the same time are hotter. A larger 
filling factor is, however, not necessarily needed to make the corona more luminous or hotter, as
we will discuss in this paper.

The new generation of X-ray observatories, {\it XMM-Newton} and {\it Chandra}, offers entirely
new access to stellar coronal physics by providing high-resolution X-ray spectroscopy with
considerable sensitivity. Apart from more detailed studies of the thermal stratification,
they also allow us to derive the coronal abundances of individual elements and to measure
electron densities. Both parameters are important for our understanding of the 
physical processes in stellar coronae. It is now well established that coronae of 
magnetically active stars show various anomalies in their composition, in particular
an overall depletion of metals (e.g., \citealt{drake94, white94}) and a relative
underabundance  of elements with a low ($< 10$~eV) First Ionization Potential (FIP) compared to
elements with a high ($> 10$~eV) FIP (e.g., \citealt{brinkman01, guedel01a, drake01}).
In the inactive Sun, in contrast, element abundances are arranged according to the so-called FIP-effect,
with low-FIP elements being overabundant in the corona relative to their photospheric values
and relative to the higher-FIP elements \citep{meyer85a, meyer85b, feldman92}. Similar trends
have been noted in inactive stellar coronae \citep{laming96, drake97, guedel02}. Because
the elemental composition and the element fractionation ultimately derive from the photospheric
gas and the physical processes heating it and transporting it to coronal heights, they may
be important tracers for the coronal heating mechanism. For example, it has been suggested
that the Ne enrichment seen in active stars may be related to increased levels of flaring \citep{brinkman01}.

This paper presents a study of six solar analogs at young and intermediate ages observed
with the instruments onboard {\it XMM-Newton}. The stars range from near-ZAMS ages (0.1 Gyr) to ages between
approximately 1--2~Gyr when the rotation period has increased to about 12 days. Together with
the Sun, they cover almost the entire path of coronal main-sequence evolution. 
The present paper is in many ways a follow-up and continuation of the work presented by
\citet{guedel97a}.

Our work emphasizes the similarity of stellar mass, surface gravities
and internal structure, i.e., we confine this study to the main sequence and treat rotation
(or equivalently, age) as our principal free parameter. In a complementary
study, \citet{scelsi05} investigate three G-type stars, all at very high activity levels
but at largely differing evolutionary stages and with different internal structure and
surface gravities, including the young main-sequence 
star  EK Dra, a pre-main sequence (weak-lined T Tau) star, 
and the evolved Hertzsprung-gap giant 31 Com. 
They find very similar emission measure distributions in the latter
two active stars, regardless of the differences in their fundamental
parameters.

The outline of our presentation is as follows. In Sect.~\ref{targets} and~\ref{obs}  we discuss
our targets, the observations, and the principal procedures of the data reduction, respectively.
Special attention has been paid to the analysis and interpretation of the spectroscopic
data in order to recognize the principal potential and the limitations of the spectral
inversion, i.e., the derivation of emission measure distributions and the element abundances.
In Sect.~\ref{spec} we describe the spectra of the six targets.
We discuss our methods in Sect.~\ref{analysis}. Sect.~\ref{results} presents our results, while
we discuss various features and models in Sect.~\ref{discussion}. Finally, Sect.~\ref{conclusions} contains
our conclusions.

\section{Targets}\label{targets}

\subsection{General Properties}

{\it XMM-Newton} data of six  young and intermediate-age solar analog stars have been analyzed. 
The stars are all of early-to-mid spectral type G on the main sequence. Their ages range from approximately 0.1 Gyr for 47 Cas B and
EK Dra to $\approx 1.6$~Gyr for $\beta$ Com. These ages have been determined using various proxies
such as rotation periods (for the older targets), or memberships in moving groups of 
known ages (for the younger targets). We have selected these targets because they have been studied
in much detail before (e.g., \citealt{dorren94}), have well-measured
fundamental parameters, and are well-behaved representatives of their
age class. A detailed summary and discussion is given
in \citet{dorren94} and \citet{guedel97a}. The distances quoted are derived from the
{\it Hipparcos} parallaxes \citep{perryman97}, and the rotation periods have been measured photometrically.
The general properties of the stars are listed in Table~\ref{stars}, where they are also compared with solar values. 
The X-ray luminosities $L_X$ given there refer to measurements from {\it ROSAT} photometry in the 0.1--2.4~keV
band \citep{guedel97a, guedel98a}, and to the spectral modeling discussed in the present paper, also for the 0.1--2.4~keV band,
and additionally for the 0.1--10~keV band. For the X-ray luminosity of the Sun, we use a representative value of  
$\log L_X=27.3$ as in \citet{guedel97a}. This value is in agreement with the $L_X$ of $\alpha$ Cen determined by 
\citet{raassen03}, $L_X=1.6\times10^{27}$~erg~s$^{-1}$. A steady decline of $L_X$ with increasing 
age and increasing rotation period $P$ is evident.

The most active target, 47 Cas B, is the fainter component in the 
47 Cas binary system. It has not been characterized 
optically as it has been individually detected only by radio 
methods \citep{guedel98b} and indirectly from {\it Hipparcos} 
measurements as a companion to the optically bright F0~V star 
47 Cas = HR~581 = HD~12230. The radio position is clearly 
offset from the position of the F0 star. It is, furthermore, 
very unlikely that early F stars display luminous and spectrally  
hard X-ray emission \citep{panzera99}. On the other hand, all 
characteristics of the X-ray detection fit well to an early G-type
star with an age similar to that of the Pleiades. In particular, 
the X-ray luminosity  corresponds to the saturation level of an early 
G star ($L_X \approx 10^{-3}L_{\rm bol}$, \citealt{vilhu84}), thus 
excluding a corona of a later-type star as its origin.

A saturated corona requires, for a G star, a rotation period smaller than 
that of EK Dra. A periodic signal with a period of $\approx 1$ 
day was reported from the {\it ROSAT}  All-Sky Survey observations, and 
was attributed to stellar rotation \citep{guedel95}.
This period also coincides  with the rotation period of the fastest
early-G type rotators in the Pleiades \citep{soderblom93}. 
We refer the reader to the detailed discussion in \citet{guedel95} 
and \citet{guedel98b}. 
We tried to quantify the maximum contribution
of the F0 star to the X-ray luminosity of the 47 Cas system. 
{\it ROSAT} studies of the Pleiades
\citep{stauffer94, micela96, micela99}  found only few stars 
with spectral type A7-F3 that have $\log L_X > 29.4 $ (i.e., 
more than 10\% of the $L_X$ of 47 Cas), and {\it all of them} have
known late-type companions, which are likely to be the
sources of the high X-ray luminosity \citep{mermilliod92}.
In addition, the typical X-ray spectrum of F-type stars in the
Pleiades is softer than that of G-type stars \citep{gagne95}, 
so we expect the F primary to provide even less flux from hot plasma.
In summary, we thus expect the F0 star to contribute 
less than 10\% to the total detected X-ray flux of the 47 Cas system.
We therefore add this target to our list of G-type stars, 
being the only likely solar analog that is accessible to high-resolution X-ray spectroscopy
and that is more active than EK Dra, in fact reaching the saturation level.

\subsection{Photospheric Composition} \label{sec:phcomp}

When measuring coronal abundances, $A$, of elements, a principal
problem is to what standard set of abundances the results should
refer to. Often, stellar coronal abundances are cited with respect
to the solar photospheric composition. But because the stellar
coronal plasma ultimately originates in the respective {\it stellar}
photosphere, coronal abundances may reflect the composition
of the latter, and abundance ``anomalies'' may simply
be due to a non-solar composition of the stellar photospheric
material. Fortunately, the photospheric abundances of
most of our objects have  been measured.
The reported photospheric abundances of our sample of
stars (except 47 Cas) are given in Table~\ref{tab:photab}. 
The entries refer to individual measurements or to catalogs 
compiled from previous studies.

As one sees from this literature survey, almost all Fe abundances reported
for our targets are compatible with solar photospheric values. Moreover,
several measurements for other elements exist, including high-FIP elements
such as C, N, and O, and again most of the reports are compatible with
solar abundances. We find a trend for super-solar abundances in $\beta$ Com,
by perhaps 10--20\%. Nevertheless, the photospheric-abundance
tabulations for our targets are incomplete and the given measurements scatter,
making it impossible to reliably express our coronal abundances relative
to the respective photospheric values. However, the above summary makes
it clear that our targets (except for 47 Cas) must show a near-solar
composition, and we therefore adopt the solar photospheric abundances
as the reference composition throughout the paper. As for 47 Cas, to our knowledge no photospheric abundances
of the brighter F star in the system have been reported. However,
this system is young and is a member of the Local Association \citep{guedel98b},
for which we can reasonably assume near-solar abundances, similar to EK Dra,
which is also a member of the Local Association.

\section{Observations and Data Reduction}\label{obs} 

\subsection{Observations} 

Our target stars were observed with the Reflection Grating Spectrometers
(RGS, \citealt{herder01}) and the European Photon Imaging Cameras (EPIC, \citealt{strueder01, turner01}) 
onboard {\it XMM-Newton} \citep{jansen01}. The RGS are suited 
for high-resolution spectroscopy in the wavelength range between 6--38~\AA, with a resolution of 
$ \Delta \lambda \approx 60-76~$m\AA, hence a resolving power of $\lambda/\Delta \lambda \approx 100-500$.
The EPICs observe between $\approx$ 0.15--15~keV, providing a moderate 
energy resolution of approximately $ E/ \Delta E = 20-50$.

 We provide a detailed log of the observations in Table~\ref{log}.

\subsection{Data Reduction}\label{reduction}

The data were reduced using the Science Analysis System (SAS) version 5.4.1. We applied the standard 
processing performed by the RGS metatask \texttt{rgsproc} and the EPIC MOS task \texttt{emchain}.  The PN data (used only for the light curves)
were reduced using the PN task \texttt{epchain}.

For the RGS data reduction, we extracted the first-order spectra from a spatial cut including 95\% of 
the cross-dispersion Point Spread Function (PSF) (\texttt{xpsfincl=95} in \texttt{rgsproc}) and an energy cut including  95\% 
of the pulse-height distribution (\texttt{pdistincl=95}). The background spectra were extracted above and 
below the source spectra, by excluding 97\% of the source cross-dispersion PSF.

For all stars except  $\pi^1$ UMa, $\chi^1$ Ori, and $\beta$ Com the MOS data were taken in the small-window 
mode. In this mode, only the central $100 \times 100$ pixels of the middle CCD are operational, whereas the outer CCDs 
work in full mode. The small-window mode allows for shorter integration times for the source and thus avoids the brighter sources 
from becoming  piled up.  The disadvantage of this mode is, however, that the central window is too small to reliably
extract a background region. We therefore selected a source-free region in one of the outer CCDs. 
The target source itself was extracted from a circle as large as possible in the small window (with a radius
of approximately 50$^{\prime\prime}$).

For $\pi^1$ UMa,  MOS1 was in full window mode whereas  MOS2 was in small-window mode. 
The exposure time was 5000~s longer in MOS1. Pile-up 
was studied for both MOS cameras with the SAS \texttt{epatplot} task. This task makes use of the 
relative ratio of single- and double-pixel events, that deviate in case of pile-up. Since no evidence for 
pile-up was found for this star, we decided to use MOS1.

For $\chi^1$ Ori, MOS1 data were taken in timing mode and MOS2 data in full-window 
mode. We did not use data taken in timing mode because reliable background subtraction cannot be performed. We found that pile-up was not negligible in the MOS2 data and therefore used an annular 
extraction region for the source, with inner and outer radius of 80 and 1200 detector pixels (4 and 60 arcsec), respectively, 
in order to remove the piled-up central part of the PSF.

For $\beta$ Com both MOS1 and MOS2 observed in full window mode. We found that pile-up did not affect this source.

In those cases where the background area was selected in one of the outer CCDs, the background count rates could have been
underestimated with respect to the source extraction region due to vignetting. 
We studied background behavior effects using the \texttt{eboxdetect} SAS task. This task detects all sources 
in the field of view; these were subsequently cut out from the image. We then quantified the remaining background 
count rate per unit area as a function of the distance from the center of the field of view, by analyzing 
the average rates in annuli at different radii. We found a decrease of at most 10\% from the image center
to the central portions of the outer CCDs, which is 
negligible when compared to calibration uncertainties, particularly in the light of the high
count rates of our sources. The only possible but small influence could be on the data at the highest energies
where the source flux drops below the background flux. We therefore decided not to consider spectral 
data at energies where the background count rate exceeded the source count rate.

We used the \texttt{rgsrmfgen} task for RGS and the \texttt{rmfgen} and \texttt{arfgen} tasks for MOS to generate
response and ancillary files appropriate for the specified extraction regions. 
To avoid bins with very few counts, we grouped the spectral channels: the MOS data were grouped to a 
minimum of 25 counts per bin, whereas the RGS data were grouped to a minimum of  10 counts per bin. 
The very weak continuum level in the RGS spectra of $\beta$ Com made this grouping scheme not well suited; instead, we
regularly rebinned the entire RGS spectra by a factor of five.

Finally, we studied the agreement between the observed wavelengths of bright emission lines with their
tabulated rest wavelengths (stellar radial velocities being negligible); in cases where small systematic offsets due to residual calibration uncertainties 
of order 10~m\AA\ were found, we slightly updated the assumed boresight coordinates accordingly and repeated the data reduction.

\section{Description of the Spectra}\label{spec}

\begin{figure}
\begin{center}
\includegraphics[width=0.73\linewidth]{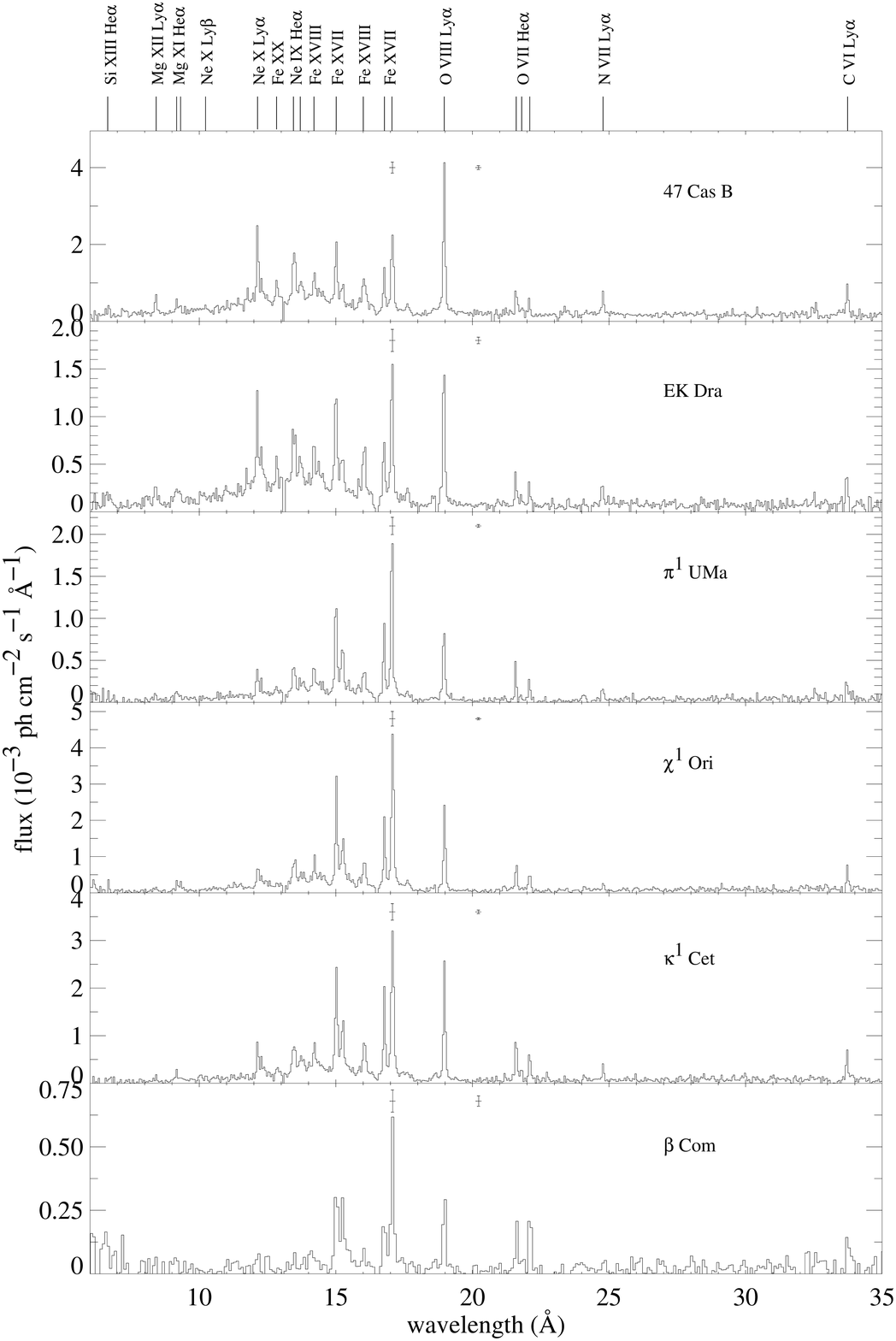}
\end{center}
\caption{Fluxed, coadded RGS 1 \& 2 spectra of the six solar analogs, ordered from high (top) to low (bottom) activity. Examples of error bars at the wavelength of Fe\,{\sc xvii} and at $\lambda$=20 \AA, a nearly line-free region, are overplotted. }\label{rgs_spectra}
\end{figure}

Before quantitatively evaluating the observed spectra, we discuss some general features.
Fig.~\ref{rgs_spectra} reveals four significantly different types of spectra: i) the most
active star, 47 Cas B, shows typical features of a very hot plasma, namely a well-developed
bremsstrahlung continuum, lines of Mg\,{\sc xi} and {\sc xii}, and comparatively
high flux ratios  
of O\,{\sc viii}~$\lambda 18.97$/O\,{\sc vii}~$\lambda 21.60$,
of Ne\,{\sc x}~$\lambda 12.13$/Ne\,{\sc ix}~$\lambda 13.45$, and
of Fe\,{\sc xviii}~$\lambda 14.20$/Fe\,{\sc xvii}~$\lambda 15.01$.
ii) The spectrum of EK Dra is significantly cooler, which is 
in particular evident from the smaller O\,{\sc viii}/O\,{\sc vii}
flux ratio, a more modest continuum level compared to the line strengths, and a marked dominance of
the Fe\,{\sc xvii} lines.
iii) The third group consists of the intermediately active, intermediately old 
stars $\pi^1$ UMa, $\chi^1$ Ori, and  $\kappa^1$ Cet which all show very similar 
spectra in which the Fe\,{\sc xvii} lines are much stronger than those of Fe\,{\sc xviii},
and the O\,{\sc viii}/O\,{\sc vii} flux ratio is further reduced.
Note that the maximum formation temperature of Fe\,{\sc xvii} is only $\approx 5$~MK.
The continuum has become very weak in these spectra.
iv) Finally, the spectrum of the least active target, $\beta$ Com, is largely dominated by lines of Fe\,{\sc xvii}, and the O\,{\sc viii}/O\,{\sc vii}
flux ratio approaches unity.

\begin{figure}
\begin{center}
\includegraphics[width=0.95\linewidth]{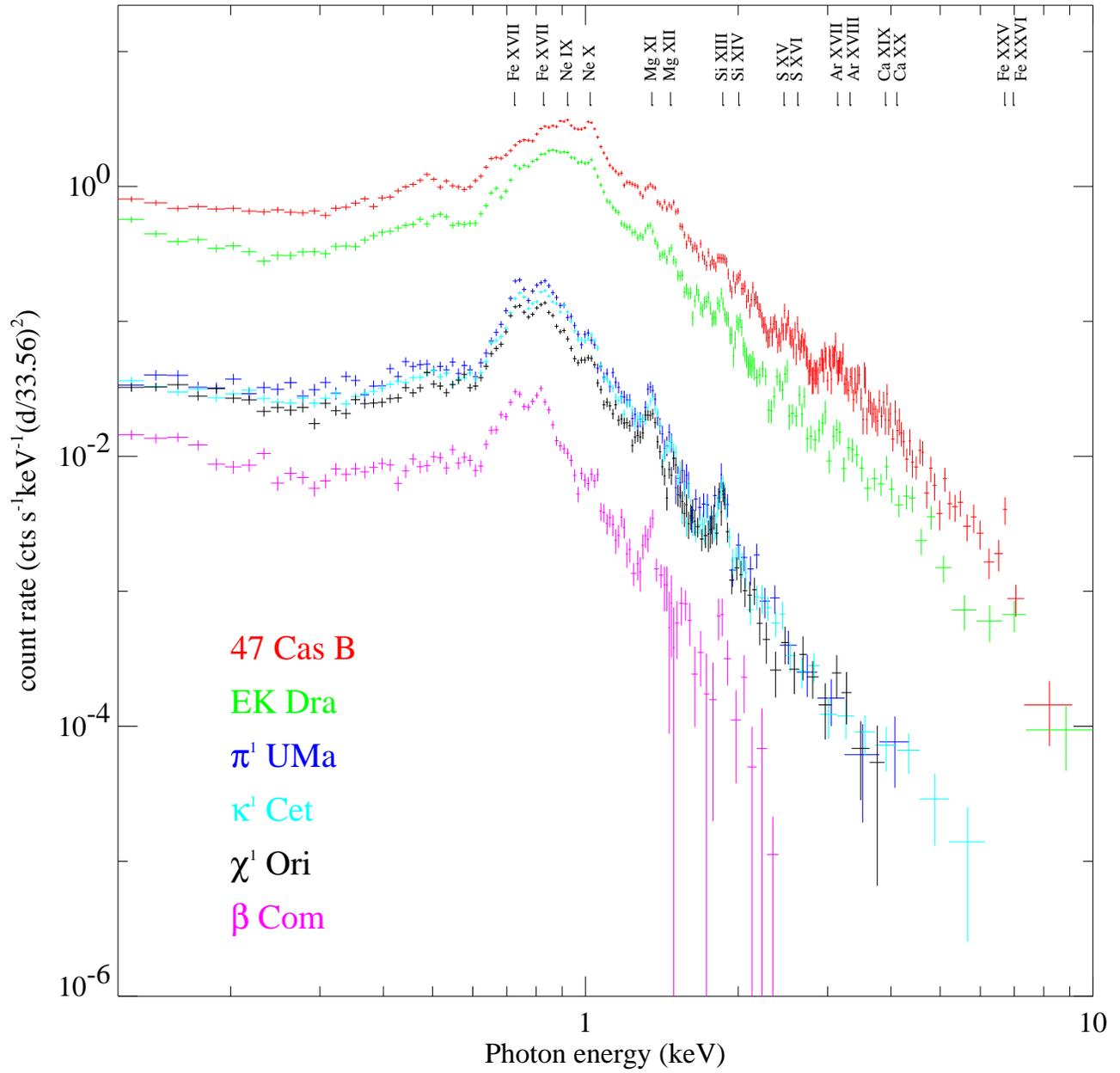}
\end{center}     
\caption{MOS spectra of the six solar analogs, normalized to a distance of 33.56 pc, the distance of 47 Cas B. The overall luminosity of the corona decreases from top to bottom as indicated by the labels.}\label{mos_spectra}
\end{figure}

Turning to the EPIC MOS spectra in  Fig.~\ref{mos_spectra}, further indicators support this picture. 
In this figure, we have renormalized the MOS spectra so as to represent the stars at a common distance
of 33.56~pc, identical to the distance of 47 Cas B.  
A marked decrease of the overall emission level is seen as the  stellar
age proceeds from the sample of Pleiades-age stars (47 Cas B, EK Dra) to the older sample.
The most active stars show shallower continuum spectra between $\approx 2$ and 10~keV than the 
less active targets, indicative of the higher overall plasma temperatures of the former.
The dominant Fe\,{\sc xvii} lines (at 0.826 keV and 0.727 keV) are well developed in the spectra of the less active  stars.

\section{Data Analysis}\label{analysis}

Because the novel aspect of our X-ray data is the high spectral resolution of the RGS, allowing us
to access individual emission lines from different elements, we used the spectral data from both 
RGS instruments but used EPIC data only in so far as they contribute additional information
not accessible by RGS, such as spectral data at wavelengths shorter than 6~\AA. To keep maximum weight
in our data analysis on the RGS data, we restricted the more sensitive EPIC spectral data to one of
the three cameras. We chose MOS1 or MOS2, because the MOS spectral resolution is superior to that
of the pn camera, and the S/N at higher energies balances well with the S/N provided by
RGS at lower energies. Also, the cross-calibration between MOS and RGS is better understood (they use the same mirrors).
Only for the light curves  did we make use of the data of other
EPIC cameras as well.

For the analysis of each target, we used the exposure time covered simultaneously by all three detectors. 
Each light curve except that of  $\beta$ Com  contained one well-developed flare. In order to avoid 
systematic bias by any of these flares (i.e., increasing the average $L_X$ and possibly
also increasing the characteristic coronal temperatures), we eliminated the flare intervals from
consideration. We will, however, briefly present and discuss the individual light curves separately
in Sect.~\ref{lc}. The exposure times remaining for our spectral analysis are listed in Table~\ref{log} (last column).

\subsection{Spectral Inversion}

An observed coronal X-ray spectrum is the superposition 
of the spectra emitted by various coronal features with 
different temperatures, volumes, densities, and possibly
different composition in terms of chemical elements. The inversion
of an observed spectrum to obtain the underlying physical 
parameters is therefore a highly degenerate problem, with 
numerous solutions describing essentially the same spectrum. 

For our analysis we consider a coronal model with the following, 
observationally tested assumptions of the physical parameters.
First, the plasma is assumed to be in collisional ionization 
equilibrium, a model that appears to be sufficiently
good as long as no very rapid change in the heating rate of the plasma
is taking place \citep{mewe99}. 
We also assume that the coronal plasma is effectively optically 
thin and that lines with high oscillator strengths are not
subject to resonance scattering, an assumption that has been shown 
to be justified for the stars in our sample \citep{ness03}.
Furthermore, the density-dependence of the populations
of metastable levels is neglected, i.e., the spectrum is computed
in the low-density limit. This approximation is supported by the flux ratios 
in the He-like triplet of O\,{\sc vii} in our observations, and 
it appears to be a reasonable assumption for most
coronal plasmas \citep{ness04, testa04}.

At this point, then, the observed spectrum
is essentially a function of the distribution of optically
thin coronal features in volume, temperature, and elemental composition;
conventionally, the thermal distribution is described by
the differential emission measure distribution (DEM),
\begin{equation}\label{dem}
Q(T)  = {n_en_HdV\over d{\rm ln}T}
\end{equation}
where $n_e$ and $n_H$ are the electron and hydrogen number densities, respectively, 
and $dV$ is a differential volume element at 
temperature $T$. 
The DEM determines the line flux $F_j$ of any given emission line $j$ through
\begin{equation}\label{lineflux}
F_j = {1\over 4\pi d^2}   \int A\varphi_j(T) Q(T)d{\rm ln}T.
\end{equation}
Here, $\varphi_j$ is the line power per unit emission measure (``emissivity'' henceforth), and
$A$ is the abundance of the element emitting the relevant line, with
respect to some standard tabulation (such as ``solar photospheric abundances'')
used for the computation of $\varphi_j$. 

Non-solar abundances in stellar coronae
are now well established (e.g., \citealt{brinkman01, drake01, audard03a}), and
we recall that the corona of the Sun itself shows considerable deviations from
the photospheric mixture. However, it is also known that the solar coronal 
abundances vary greatly  from feature to feature, some showing a marked
FIP bias, with others showing photospheric composition \citep{feldman92, laming95}.
Our spectra are - like most existing stellar X-ray spectra - insufficient to 
characterize abundances at various separate temperatures; therefore, we will assume 
$T$-independent abundances to at least recognize overall trends in our stellar sample.

The spectral inversion problem is mathematically
ill-posed. Statistical scatter due to photon statistics in line fluxes, even if amounting
to only a few percent, may introduce considerable scatter in the reconstructed
emission measure distribution \citep{craig76}. We expect a strong amplification
of such effects from the uncertainties in the atomic physics parameters (which may amount
up to a few tens of percent for the emissivities)  and from calibration uncertainties (up to several
percent in certain spectral regions). 

Therefore, the spectral inversion can essentially
be performed in a meaningful way only by constraining the problem suitably, for example
by subjecting the reconstruction to smoothness conditions for the resulting, discrete Emission 
Measure Distribution (EMD), or by iterating the problem using a pre-defined convergence 
algorithm that reconstructs a preferred, physically meaningful solution (see, e.g., \citealt{kaastra96a} and 
\citealt{guedel97b} for a discussion of various algorithms with different bias). 

We have chosen here to perform the inversion using two widely different
algorithms that we briefly summarize as follows:
\begin{itemize}

\item Method 1 (M1). We fitted the spectrum using synthetic template spectra calculated for
      a set of physical parameters; the parameters were varied  until the 
      fit was optimized. This method has conventionally been used for low-resolution
      spectra but also for high-resolution grating spectra, in particular as 
      implemented in the SPEX \citep{kaastra96b} and the
      XSPEC \citep{arnaud96} software packages. We, however, modified  the conventional
      approach by including almost exclusively segments of the spectrum that are dominated by
      bright lines for which comparatively robust atomic physical parameters should
      be available, together with some nearly line free regions. 
      
\item Method 2 (M2). Here, we worked with a list of discrete line fluxes that we extracted
      from the observed spectrum. If the formation temperatures of these lines
      cover the range of relevant coronal electron temperatures, then the EMD
      can be reconstructed by successive approximation, essentially inverting
      a system of equations like Eq.~(\ref{lineflux}). 
      
\end{itemize}
We emphasize the complementarity of our two approaches: method 1 uses all tabulated emission lines and their
blends within the selected spectral segments, while we confine method 2 to a minimum
number of lines, namely the brightest lines available, required to derive meaningful
EMDs and the most essential abundances. 

Both methods are subject to separate biases. Some of the many lines considered in method 1
may be poorly fitted, owing to discrepancy in the tabulated emissivities.
If the latter scatter around their true values, the resulting fit may show reduced systematic
errors compared to a fit based on one single line per ionization stage.
The individual lines in  method 2 may be better described, but their small number 
makes the inversion process rather sensitive to any systematic uncertainty in any
of the line emissivities, and some a priori estimate for blend contribution must be considered. 
Lines not tabulated in the line-emissivity lists affect both methods, either by altering
the measured line flux in method 2 through unrecognized blending, or by adding 
a continuum-like base level if many weak lines contribute in either of the analysis methods.

A comparison of method 2 with a method using polynomial EMDs to calculate synthetic spectra has previously been presented by \citet{audard04}
for the extremely active and hot corona of the FK Com-type star YY Men.  These authors found
that both the reconstructed EMDs and the derived abundances closely agreed
for the two methods. Here, we study and compare the results   
for stars across a large range of activity and coronal temperatures. 

To further assess potential sources of systematic error, we applied two different databases to each of
the two methods. The spectral models are based on the assumption of coronal
ionization equilibrium (CIE). We used the atomic parameters
from the MEKAL database in SPEX vers.2.0 (\citealt{mewe85};  \citealt{kaastra96b})  and from the 
Astrophysical Plasma  Emission Code vers.1.3.1 (APEC, \citealt{smith01}) in XSPEC \citep{arnaud96}.
The MEKAL emissivities are computed using the ionization balance of \citet{arnaud92} for Fe and 
\citet{arnaud85} for the other elements. The APED emissivities (emissivities used in the APEC code) 
are computed using the ionization balance of \citet{mazzotta98}.

\subsection{Method 1: Synthetic Spectra} 
 Here, we used an approach essentially identical to the one described in \citet{audard03a},
 apart from slightly different wavelength ranges (see below). 
 Because numerous emission lines are poorly  described by the spectral line lists in use, we confined  
 our analysis to a restricted number of spectral regions that contain bright lines for which the 
 atomic physics is believed to be relatively well known, and some nearly line-free regions 
 of the spectrum where the continuum dominates. The wavelength ranges
 of the selected regions are summarized in Table~\ref{region}. As we decided to put more weight 
 on the data from the RGS with its higher 
 spectral resolution, we used EPIC MOS data only in the wavelength range shortward of 9.35~\AA, 
 where the RGS effective area is small and its calibration is less accurate. We thus cut the RGS data 
 shortward of 8.3 \AA, so that the data from the RGS and the EPIC instruments overlapped around the Mg\,{\sc xi} 
 and Mg\,{\sc xii} lines. 
 As mentioned in Section 4, we discarded the high-energy part of the EPIC spectrum where the background flux
 exceeded the source flux (see Table~\ref{region}).

The physical model was defined as a combination of 10 optically thin, thermal CIE models, and a photoelectric absorption component. The photoelectric 
absorption was frozen by defining a fixed interstellar hydrogen column density.
The latter ranged between negligible values ($< 10^{18}$~cm$^{-2} $) for the closer stars and  $7 
\times 10^{18}$~cm$^{-2}$ for 47 Cas B. These values are consistent with hydrogen column densities given by \citet{guedel97b} 
and \citet{audard99} and with values of the interstellar hydrogen density given 
by \citet{paresce84}. However, even the largest values used here were too small to significantly
alter our model spectra even at the longest wavelengths in the RGS.

\begin{figure}
\begin{center}
\includegraphics[width=1.0\linewidth]{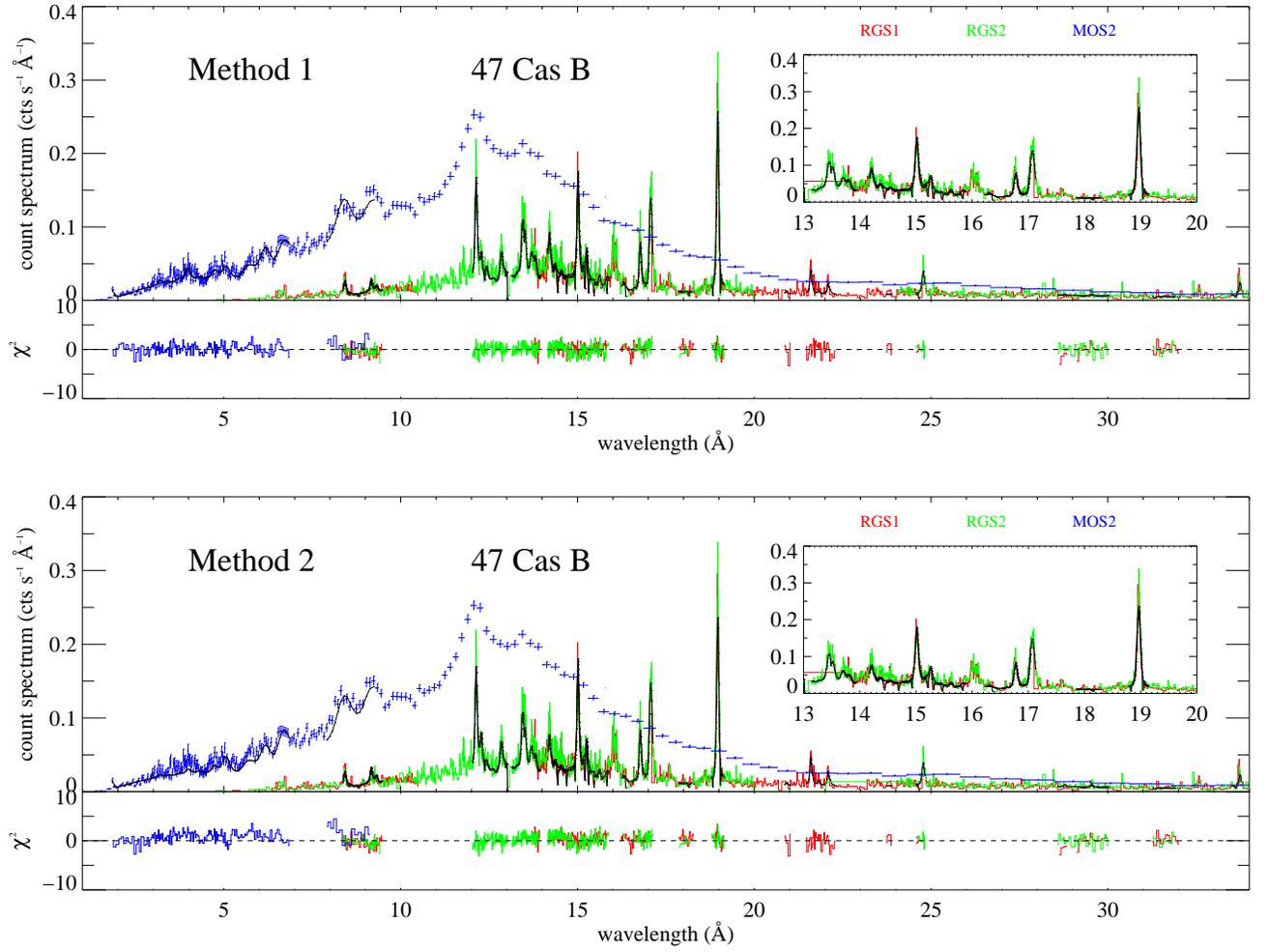}
\end{center}
\caption{{\bf Upper panel:} data and fitted spectrum for 47 Cas B using method 1 with APEC;
         {\bf lower panel:} same for method 2. The synthesized spectra from the best-fit 
         parameters are shown in black for the ranges that were used in method 1. Error bars  
         are displayed only for  the MOS data in order to avoid confusion. The insets show the
         important spectral portion containing the Fe L shell lines. The lower plots in each
         figure show the contributions of each bin to the $\chi^2$ value. Note that for method 2, we show the identical
         layout for illustration and comparison purposes although the iteration procedure did not make use of 
         a binned spectrum.}
\label{specfit}
\end{figure}

Each of the ten CIE temperatures was confined to within one of ten adjacent temperature intervals 
of equal width (0.22~dex) in $\log T$; to optimize the multi-temperature fit, given the relatively 
broad bins, we treat the temperature values as free parameters within the bounds of the respective 
bin intervals, as well as the associated emission measures. The temperature intervals covered the range from
$\log T$=6 to $\log T$=8.2 ($T$ is given in K). The abundances 
of C, N, O, Ne, Mg, Ar, Si, and S were also used as free parameters. Only the abundances
found with the 10-temperature fit were used in a second step for the DEM reconstruction. The Chebychev polynomial DEM 
code in SPEX, with polynomials of order 6 and 8, was used to describe
the DEM that best fitted the given spectrum. An example of a best fit is shown 
in Fig.~\ref{specfit}, which also illustrates the spectral ranges used for this procedure. 
 
\subsection{Method 2: Inversion of Line-Flux Lists}

\subsubsection{EMD Reconstruction}\label{sec:emdrec}

\begin{figure}[!h]
\begin{center}
\includegraphics[width=0.95\linewidth]{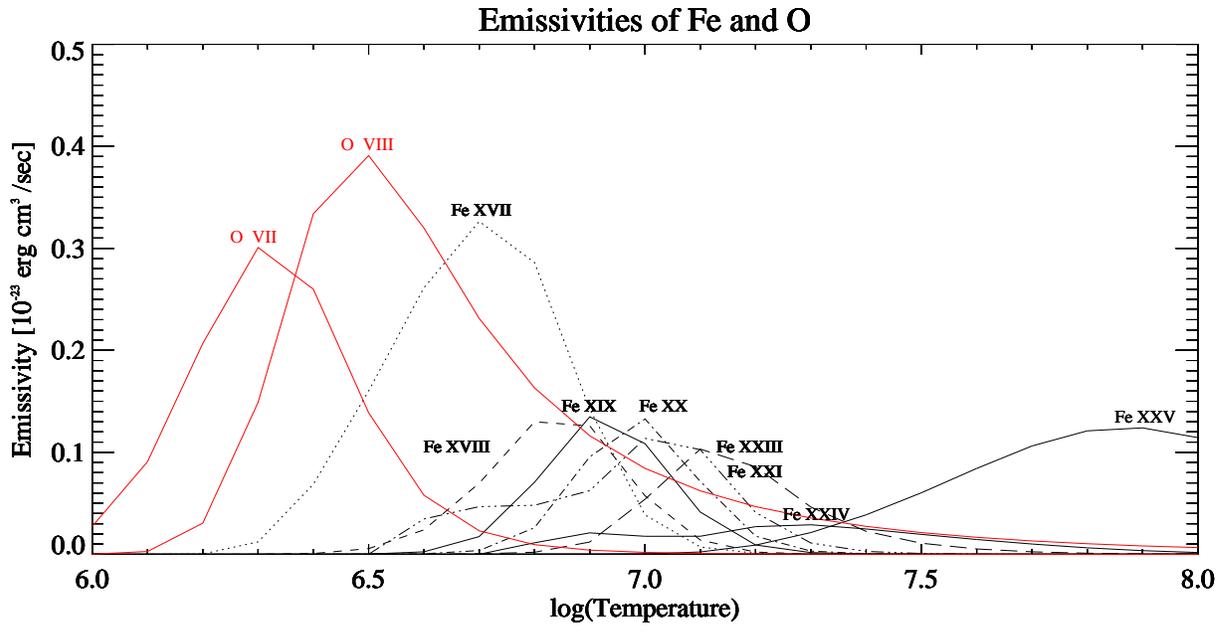}
\end{center}
\caption{Emissivities for Fe and O line blends from the MEKAL database, plotted on a logarithmic grid, assuming solar abundances \citep{anders89}.}
\label{fig:emiss}
\end{figure}

We reconstructed the discrete EMD starting from fluxes of individual lines. 
In order to obtain an EMD independently of the element abundances, we 
selected a few well-defined, bright Fe lines from the ionization stages 
of Fe\,{\sc xvii} to Fe\,{\sc xxv} (as far as measurable, 
see Table~\ref{tab:lines}). The Fe\,{\sc xvii}-{\sc xxiv} lines were extracted 
from the RGS spectrum, whereas the unresolved Fe\,{\sc xxv} line triplet was extracted 
as a single blend system from the MOS spectrum.
 The  emissivities  of the Fe\,{\sc xvii}-{\sc xxiv} 
lines cover the temperature range between 3~MK and 20~MK rather well, 
and for the more active stars (47 Cas B and EK Dra), Fe\,{\sc xxv} constrains 
the hottest temperatures, up to about 100 MK (Figure~\ref{fig:emiss}).
In order to 
obtain information on the cooler part of the EMD in which no  
Fe lines are detected, we used the flux ratio between the OVIII $\lambda$18.97 and the  OVII $\lambda$21.60 resonance lines,  
which itself is independent of abundances. The lines used for each star are listed in Table~\ref{tab:lines} and Table~\ref{tab:lines_xspec}, 
together with their measured luminosities.

To avoid cross-talk with the abundance determination, we selected 
only lines of Fe that are not strongly blended with lines of {\it other} 
elements (with the exception of Fe\,{\sc xix}, see below; the 
Fe\,{\sc xvii} lines at 17\AA\ were not used because they 
are partly obscured by one of the RGS CCD gaps). Nevertheless,
our line features usually still contain a number of weaker blends 
from various Fe ions. We thus considered all Fe blending lines close to
the principal line as being part of our Fe feature and we therefore 
computed new $T$-dependent emissivity curves for each such Fe {\it blend} system.
The line feature, composed of all Fe lines in the chosen wavelength range, 
was thus subsequently treated like a single emission line. 
In the RGS spectrum, all tabulated blending Fe lines within
$\pm 0.06$~\AA\ (one FWHM on each side) of the wavelength  
of the principal line were considered. In the MOS spectrum, 
the range was chosen to be $\pm 0.15$~\AA. 
 
In a first step, we extracted the fluxes of the relevant Fe lines and of the O\,{\sc viii}~$\lambda 18.97$
and O\,{\sc vii}~$\lambda 21.60$ resonance lines from the spectrum. This can be problematic, 
because the lines have broad wings due to the RGS PSF, and the determination of the underlying continuum is difficult.  Moreover, 
in some cases, blends with lines from different elements  may still be present, in 
particular in the case of the brighter lines of  Fe\,{\sc xix} that are blended with Ne\,{\sc ix} lines.
While the effect of these blends cannot be assessed a priori without knowledge of the thermal structure
and the abundances, we can obtain useful approximations as follows. Because most ions are represented
by several different emission lines, and the continuum can be interpolated from values obtained
in nearly line-free regions of the spectrum, a good approximation of the contaminating blends and of the baseline continuum level in fact comes from the fit we derived in method 1. We emphasize that we have not used that model to derive parameters, but only to obtain an approximate description of
the baseline spectral flux distribution on which the line fluxes of interest are superimposed. The latter being the fluxes of 
Fe lines, we set the Fe abundances in the best-fit model from method 1 to zero, retaining only the emission lines and the 
continuum from all other elements. We  note that neglecting Fe does slightly change the continuum as well, but
the influence is on the order of a few percent in the temperature and wavelength
range of interest if solar abundances are assumed. Anticipating sub-solar abundances in active stars
with a strong continuum as reported in the previous literature, the contribution of Fe to the continuum will be even smaller.
The errors thus introduced into the line-flux extraction are negligible. An equivalent procedure was then
also  applied to extract the O\,{\sc viii} and O\,{\sc vii} line fluxes, although in these cases, blending 
was not significant.

Before extracting individual lines, we need to make sure that the baseline continuum
level has been correctly fitted individually in each spectrum. We checked the relevant fits in a 
nearly line-free region of the RGS spectrum, namely immediately shortward and longward of the 
O\,{\sc viii} $\lambda 18.97$ resonance line by multiplying the EM by a suitable factor. 
We found that an optimization of the continuum level in these wavelength regions required a change 
in flux of only a few percent. It is not
clear whether this is due to slight cross-calibration problems or to some bias in the
fit. Such bias may be introduced by the fact that the MOS spectral portion, determining 
mostly the hotter part of the DEM, has a relatively high S/N ratio which may lead to some over- or
underestimation of the hot EM. Because the latter produces continuum at all wavelengths, 
a slight offset could also affect the soft part of the RGS spectrum. Given the small 
magnitude of the effect, we are unable to ascribe such an offset either to (possibly wavelength-dependent)
cross-calibration problems or to fit bias as described above. Our final results indicate
that any such effect is of minor consequence and does not need to be considered for the overall fit in method 1. For the extraction of line fluxes in method 2, however,
we need to optimize the continuum level so that the line flux can be properly defined as
an excess above this baseline level.

In a similar manner, we extracted the Fe\,{\sc xxv} $\lambda 1.85$ line by first optimizing the nearby
continuum level starting from the overall best-fit provided by method 1. Again, the continuum renormalization
required for this procedure was at most a few percent in the spectral region of interest.
 
To actually extract the line fluxes, we added $\delta$ function lines to the model, located at the theoretical wavelengths of the principal lines in the blends under
consideration, and convolved them with the instrument response. The  line fluxes were then obtained by fitting
the amplitudes of the $\delta$ functions to the observed lines and the adjacent  narrow wavelength ranges. 
If the wings of two Fe lines overlapped significantly, we fitted two $\delta$ functions simultaneously. Note that the fluxes
thus derived may be different for APEC and MEKAL because of slightly different atomic physics, fit parameters, and blend contributions.

The obtained Fe, O\,{\sc viii} and O\,{\sc vii} line fluxes were converted to luminosities
as presented in Table~\ref{tab:lines} and \ref{tab:lines_xspec} for MEKAL and APEC emissivities,
respectively, based on {\it Hipparcos} distances as quoted in Table~\ref{stars}. They were then used to reconstruct the EMD, as follows:

A first, smooth estimate of the EMD was derived from the emissivities at the maximum line formation temperature $T_m$ of each Fe line. 
The emissivities are based on solar abundances as given by \citet{anders89}. 
Our EMD was defined on a grid with a grid point separation of  $\Delta \log T$=0.1~dex 
in the range between $\log T = 5.5-8$ for SPEX ($T$ in K) (APEC: $\log T = 5.7-8$) 
for the more active stars 47 Cas B and EK Dra, and in the range between 
$\log T = 5.5-7.1$ (APEC: $\log T = 5.7-7.1$) for the less active stars 
$\chi^1$ Ori, $\kappa^1$ Cet and $\beta$ Com. For $\pi^1$ UMa, the EMD was 
defined in the range between $\log T = 5.5-7.3$ (APEC: $\log T = 5.7-7.3$). 
(The different low-temperature limits reflect the different availability of 
emissivities in the different codes, but we note that the range 
of $\log T = 5.5 -  5.7$ is irrelevant for any line we use for the EMD
reconstruction, and it is also not important for the emissivities
of any lines of other elements analyzed here.) 
As a starting condition, the EMD was extrapolated to temperatures cooler than $T_m$ 
of Fe\,{\sc xvii} ($\log T = 6.7$) by using a slope of 2, and to temperatures higher than 
$T_m$ of the hottest Fe ion in use (Fe\,{\sc xxv},  $\log T = 7.9$ for 47 Cas 
and EK Dra,  Fe\,{\sc xix}, $\log T = 6.9$ for $\beta$ Com, and  Fe\,{\sc xxi}, $\log T = 7.0$ 
for the  other stars) by using a slope of --2. These extrapolations were performed to the
limits of the respective temperature ranges  as defined above.
These starting conditions were suggested by the slopes
of the initial EMDs defined by the EM values at the different $T_m$.
Also,
we subsequently found that the low-$T$ slope converges to values around 2 even if
the starting slope was largely different (see Sect.~\ref{sec:flares}).
For each line, the flux $F_C$ predicted from the initial EMD was calculated according to the following equation:
\begin{equation}
\label{predflux}
F_C(x) = \sum_i {\rm EM}(i) \cdot  \varphi(x,i)
\end{equation}  
where the sum is over the temperature bins $i$; EM$(i)$ is the emission measure in the $i$th temperature bin,
and $\varphi (x,i)$ is the emissivity of the given line $x$ at this grid point. The calculated fluxes were then 
compared with the measured fluxes, $F(x)$. The EM in each bin was iteratively corrected 
using the algorithm described by \citet{withbroe75}:
\begin{equation}
\label{withbroe}
{\rm EM}^{n+1}(i) = {\rm EM}^{n}(i) \cdot \frac{\sum_{x} \frac{F(x)}{F_C(x)} \cdot \varphi(x,i)+\frac{R_C}{R} \cdot 
               \varphi(\mathrm{O\,VII},i)}{\sum_{x} \varphi(x,i)+\varphi(\mathrm{O\,VII},i)}
\end{equation}  
where $i$ is the index of the temperature bins  and $n$ is the iteration number. 
The last term in the numerator and  denominator was added to also obtain convergence of the line-flux ratio 
between O\,{\sc viii} and O\,{\sc vii}, where we have used the measured ($R$) and the predicted ($R_C$) flux ratio. 
The EMD was then iterated until we reached a pre-set convergence criterion. 

In the presence of considerable systematic uncertainties in the modeling, a reasonable convergence criterion should
be set, although its statistical meaning may be marginal. A straightforward goal is to achieve convergence in
such a way that on average the squared deviation between model and fit for a line flux is equal to
the variance of the same line flux. This leads to a reduced-$\chi^2$-like expression,
\begin{equation}
\label{for:chi}
\psi^2 = \frac{1}{N_x+1}\left(\sum_x\left[ \frac{(F[x]-F_C[x])^2}{(\sigma[x])^2}\right] +\frac{(R-R_C)^2}{(\sigma_R)^2}\right)
\end{equation}  
where the last term relates to the O\,{\sc viii}/O\,{\sc vii} flux ratio. Here, $N_x$ is the number of lines used,
and $\sigma(x)$ is the  error assigned to the measured line flux, which comprises the error from the finite photon 
statistics and an assumed systematic uncertainty from the atomic physics (see Sect.~\ref{sec:err}). Similarly, $\sigma_R$ is the
derived error in the O\,{\sc viii}/O\,{\sc vii} flux ratio.  We iterated until this expression reached
a value of unity, or if this did not occur, until it no longer significantly decreased. We would like to mention that 
the fit parameters, i.e., the EMs per bin, are not independent of each other owing to the broad emissivity
curves of each line; given the large systematic uncertainties in the line fluxes, we feel that a more detailed 
convergence criterion is not warranted. Our main goal is to stop the iteration at a reasonable level
to avoid over-interpretation of superficial features in the EMD that may arise from iterating too deeply - see our discussion in Sect.~\ref{sec:emd}.

At this point, then, we have found an EMD from fluxes of the single element Fe,  under the assumption 
of solar metallicity as used for the tabulated emissivities, and from a ratio of O line-fluxes.
Therefore, there  still remains a normalization factor for the EMD and the absolute level of the Fe abundance to be determined (see Sect.~\ref{ab}). We will use the observed continuum level to fix the EMD normalization and,
at the same time, the absolute Fe abundance, as explained in the following.

\subsubsection{Abundances}\label{ab}

To determine the abundances of the elements, we extracted all lines of interest (see Tables~\ref{tab:lines} and \ref{tab:lines_xspec}) in a similar manner as we extracted the Fe and O lines before. Here, however we used the EMD constructed from our  Fe and O line fluxes in order to describe the continuum, instead of the 10-$T$ model. This EMD is known only up to a normalization constant depending on the absolute Fe abundance, which we derived as follows: we constructed a set of spectra from the calculated EMD with different Fe abundances such that the product of the Fe abundance and
the EM$(T)$ is constant for any $T$, still using the approximate abundances of the other elements from method 1 to estimate the contributions from blends and to obtain a more accurate description of the continuum. 
From the spectrum that best fitted the nearly line-free regions  long- and shortward of the O\,{\sc viii} Ly$\alpha$ line at 18.97 \AA\,, we obtained the absolute Fe abundance. We then extracted the line fluxes of all interesting elements using $\delta$-line models, as described before. Note that the line fluxes of Mg, Si, S, and, if available, Fe\,{\sc xxv}, were extracted from the MOS spectra while all other lines were extracted from the RGS spectra. As done for the RGS spectra before, 
we adjusted  the model continuum  to the observation in the line-free 
regions of the MOS instrument at high energies, before the line-flux measurement. 

The predicted line fluxes of these elements (see Table~\ref{tab:lines} 
and \ref{tab:lines_xspec}) were then calculated from the EMD, 
using their catalogued emissivities, which were again based on
 solar photospheric abundances. The ratios between 
the predicted and the measured fluxes provided the abundances 
relative to Fe, $A/A(\rm Fe)$, with respect to the corresponding 
solar ratios. In some cases, we measured the fluxes of more 
than one line for a single element. In these cases, we calculated the abundance for each 
line and  computed averages, using $1/\sigma_{A}^2$ as weights, where $\sigma_{A}$
is the error in the abundance (see Sect.~\ref{sec:err}). We implicitly 
assumed here that the abundance of an element is the same  at all 
temperatures, an assumption that is not necessarily supported 
from solar observations \citep{jordan98}. 
The data quality at hand does not allow for further discrimination, however. 

Finally, we once more iterated the adjustment of the continuum level to obtain the absolute Fe abundance, as described above, now using the abundances determined from our procedure. The updated values closely agreed with the previously obtained Fe abundances.
The error of the absolute Fe abundance was derived 
by varying it around its best fit value and requiring that the continuum fit be acceptable within one sigma.
The final abundance values we report in this paper refer to the solar photospheric abundances given by \citet{anders89}, except for Fe for which we adopt the value given by \citet{grevesse99}.  

\subsubsection{Errors}\label{sec:err}

Errors arise from different sources. First, they are due to uncertainties in the atomic databases; these are not easily 
quantifiable, but are likely to be in the range of several percent to perhaps  20\%, depending on the line under consideration. 
For method 2, we chose mostly bright, well-studied lines, and for the sake of definition we have assumed  
systematic uncertainties of 10\% for each line. 
Statistical errors also arise from the fit of the $\delta$-line model used 
to extract the line flux; these essentially originate from photon 
statistics.
For each line flux, these two errors were summed in quadrature, and we call them line-errors $\sigma$. 

The errors in the EMD were estimated by statistically varying the fluxes of the 
line blends according to their $\sigma$, and repeating the EMD reconstruction 
for 19 different, perturbed line-flux lists. We thus derived formal upper and lower 
1-sigma ranges of the EMD solutions by using standard formulae from
Gaussian statistics, although we mention that the various solutions 
are not necessarily normally-distributed in $\log$ EM at any given temperature. 
The standard deviations thus derived, however, provide a well-defined characteristic width of
the distribution, and we verified  that the ranges containing 68\% of the solutions and 
the 1-sigma ranges are very similar. Because the EM values scatter considerably in any temperature bin,
we performed the statistics using logarithms of the EMs in each 
temperature bin in order to avoid the average being biased by one or a few
large values. We extended the error analysis to include up to 100 
perturbed line lists but the error ranges did not significantly change.
We caution that the EMD slopes on both sides of the peak temperature could be 
slightly dependent on the initial EMD guess, where we assume a slope of 
$\pm 2$ (see Sect.~\ref{sec:emdrec}). We study the EMD results starting from 
different initial conditions for the slopes in Sect.~\ref{sec:flares}. The final EMD
slopes converged to similar values. These effects are not taken into account in 
the EMD errors reported here. We note that these errors are only given as an indicator for the
uncertainty in the EMD, but they are not explicitly used further in our
error analysis. 

The error of the abundance $A$ is proportional to the line-error, 
$\sigma_{A}= A \cdot\sigma(x)/F(x)$. 
If the abundance is a weighted average, then the final error is 
the larger of i) $(\sum 1/\sigma_{A}^2)^{-1/2}$ and ii) the error of the 
weighted means of all abundance values.
Moreover, there is an error in the abundances arising from the variation in the 20 different EMD reconstructions: 
for each of these reconstructions, we found slightly different abundances with new errors. We defined
the error from this variation as the larger of i)  the average error found in each reconstruction,
and ii) the standard deviation of the twenty abundance values per element. We note, however, that we adopted the
abundance derived from the best-fit solution, with no perturbation applied.

Finally, the error in the adjustment of the continuum (required to determine the line fluxes, Sect.~\ref{ab})
also  affects the abundance errors. As our final error for a given abundance, we summed in quadrature the
error related to the line-error (or the average if multiple lines were used, as defined above), the error arising from the 
variation of the EMD, and the error from the continuum adjustment. We emphasize, however, that this procedure
can provide no more than a simulated estimate of realistic errors. The unknown systematic deviations
in the atomic physics parameters prevent us from obtaining better estimates.

\section{Results}\label{results}

\subsection{Emission Measure Distributions}\label{sec:emd}

\begin{figure}

\clearpage

\begin{center}
\centerline{\hbox{
\includegraphics[width=0.26\linewidth]{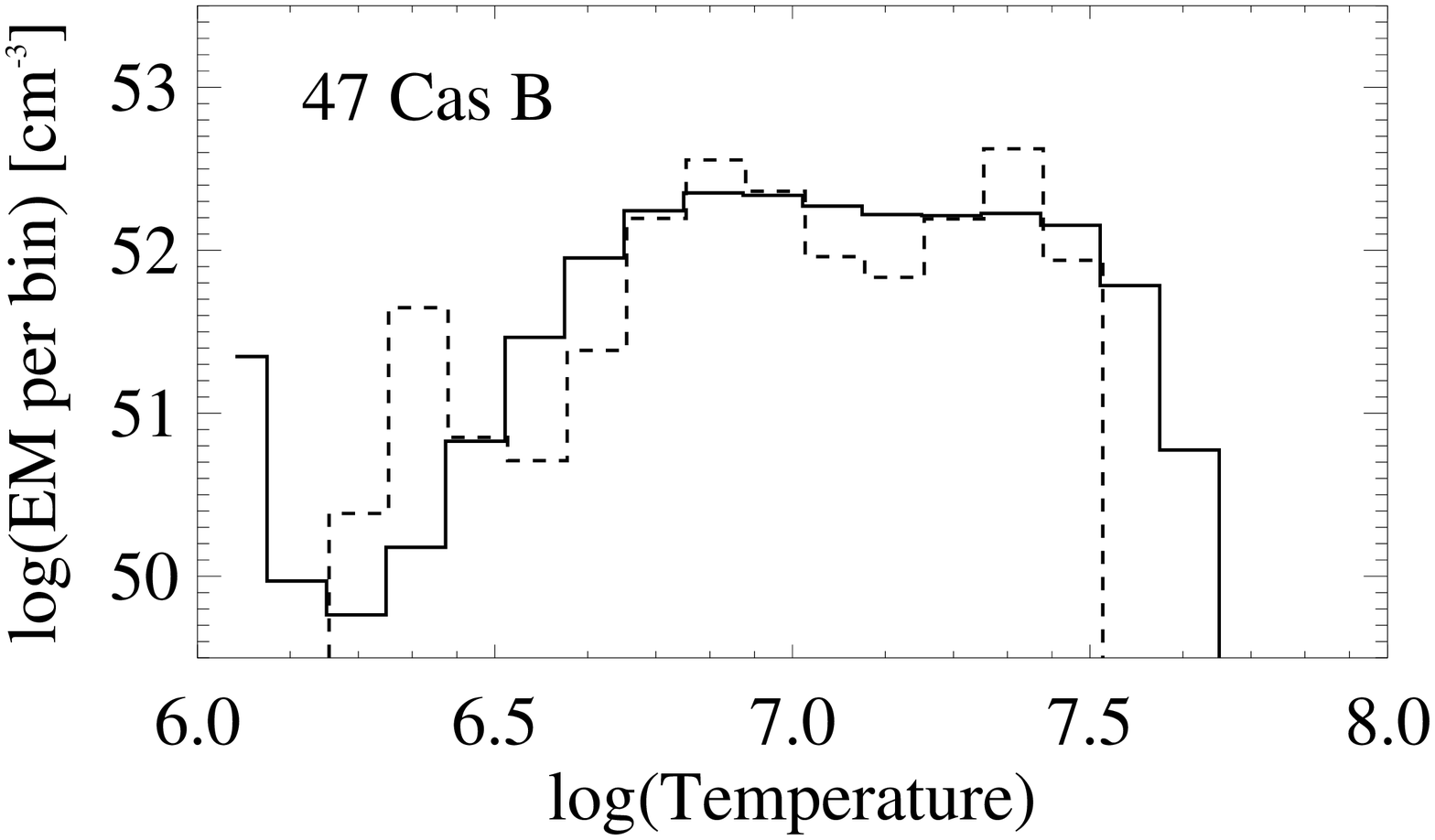}
\includegraphics[width=0.26\linewidth]{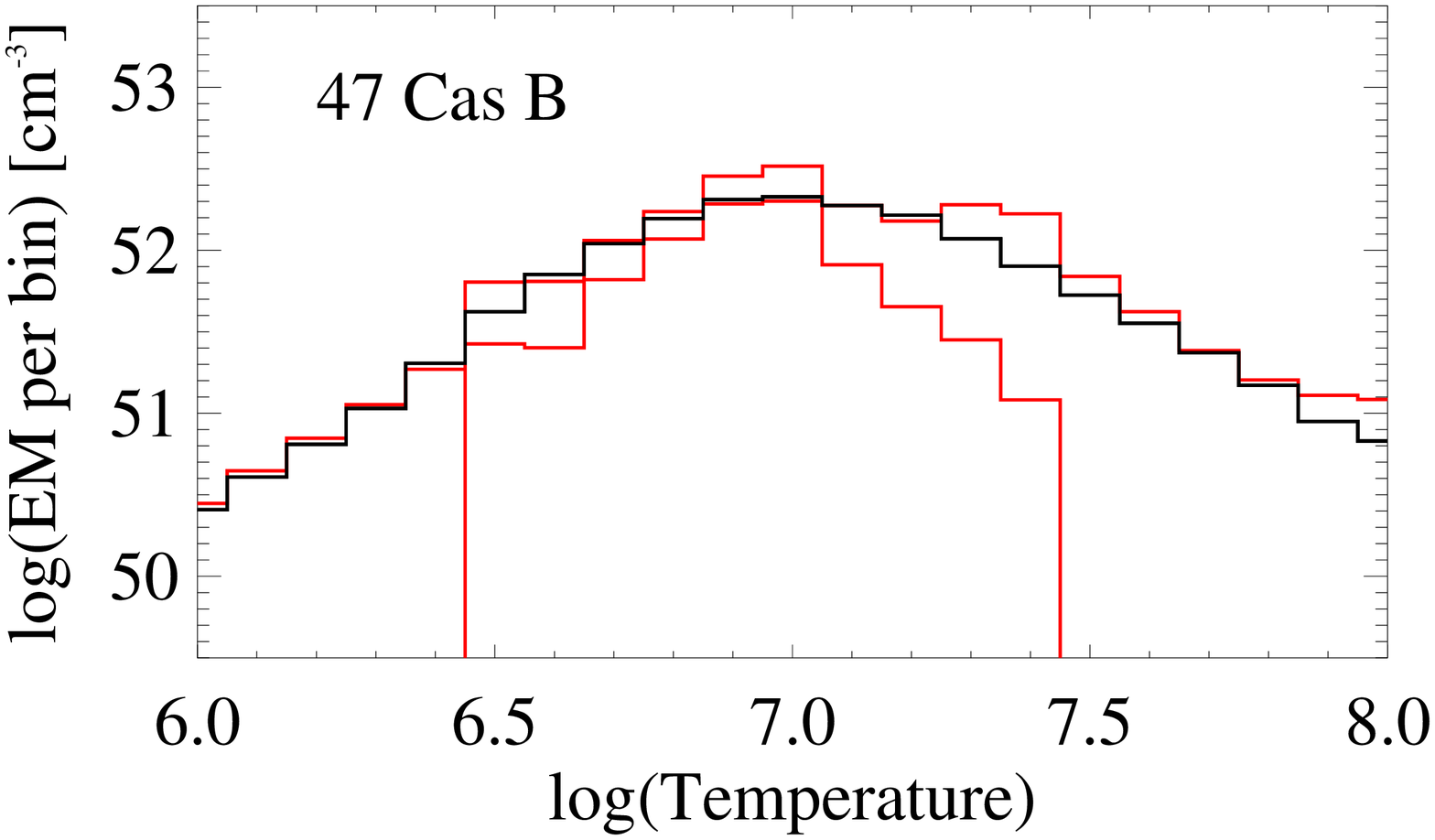}
\includegraphics[width=0.26\linewidth]{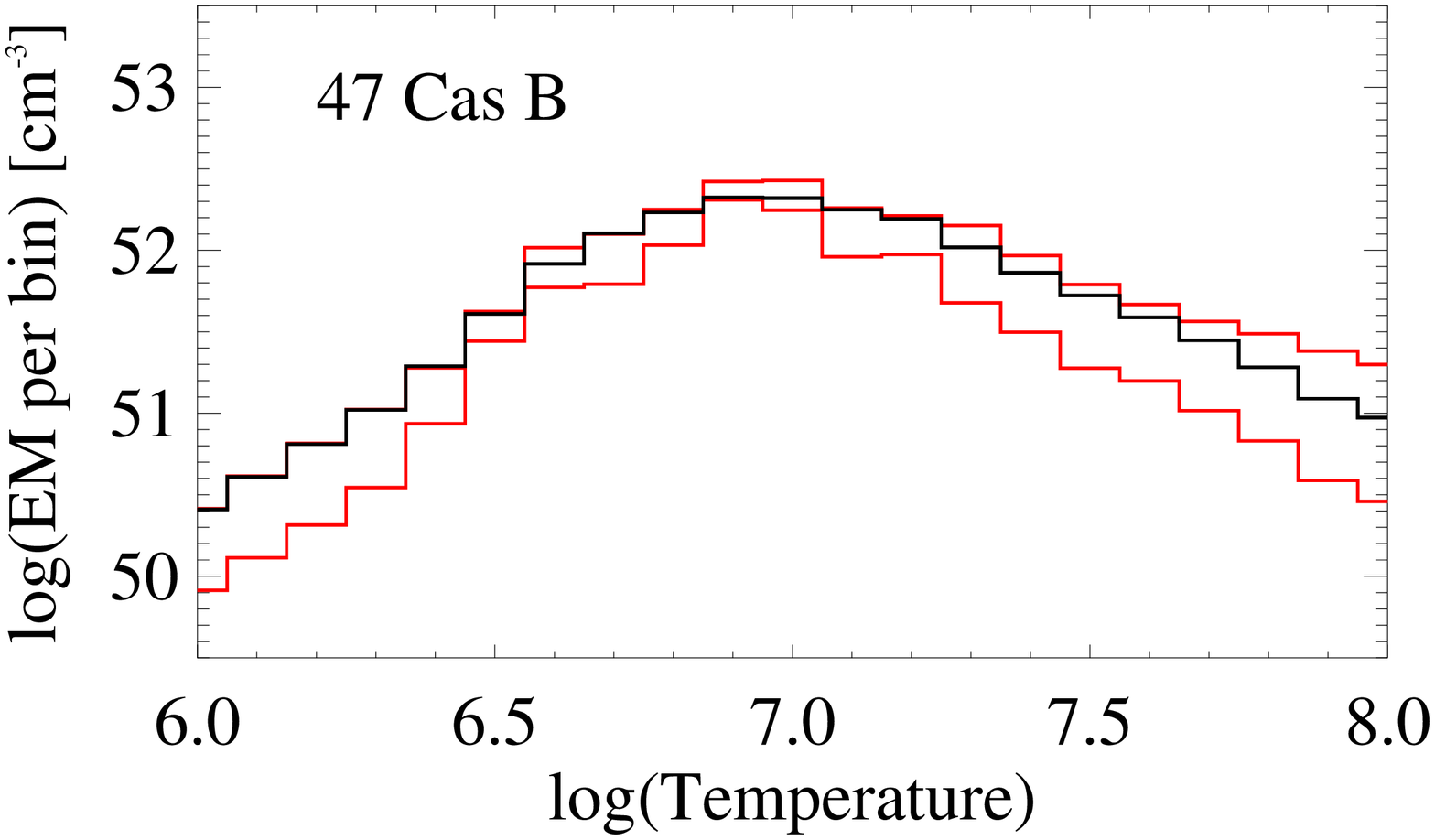}
}}
\centerline{\hbox{
\includegraphics[width=0.26\linewidth]{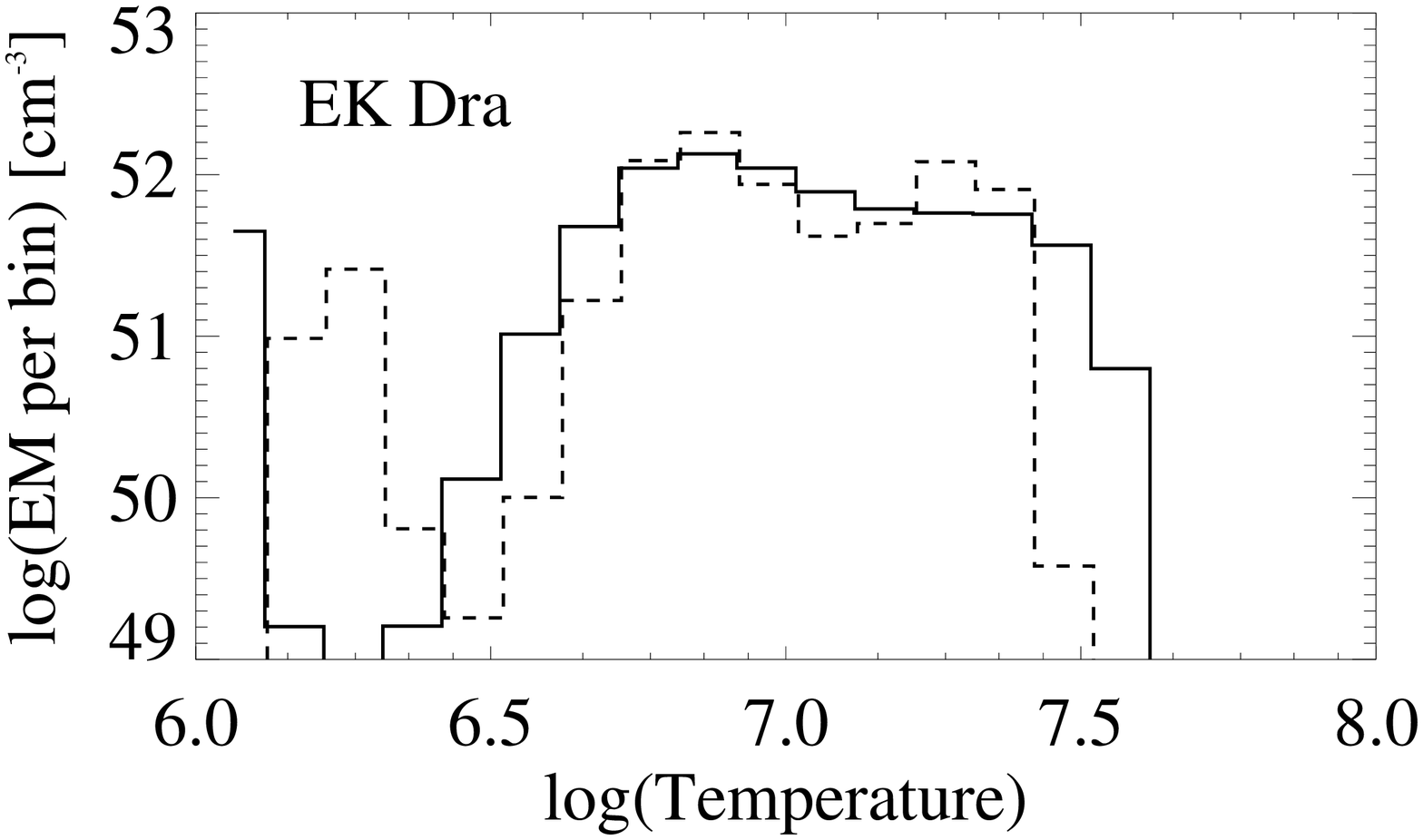}
\includegraphics[width=0.26\linewidth]{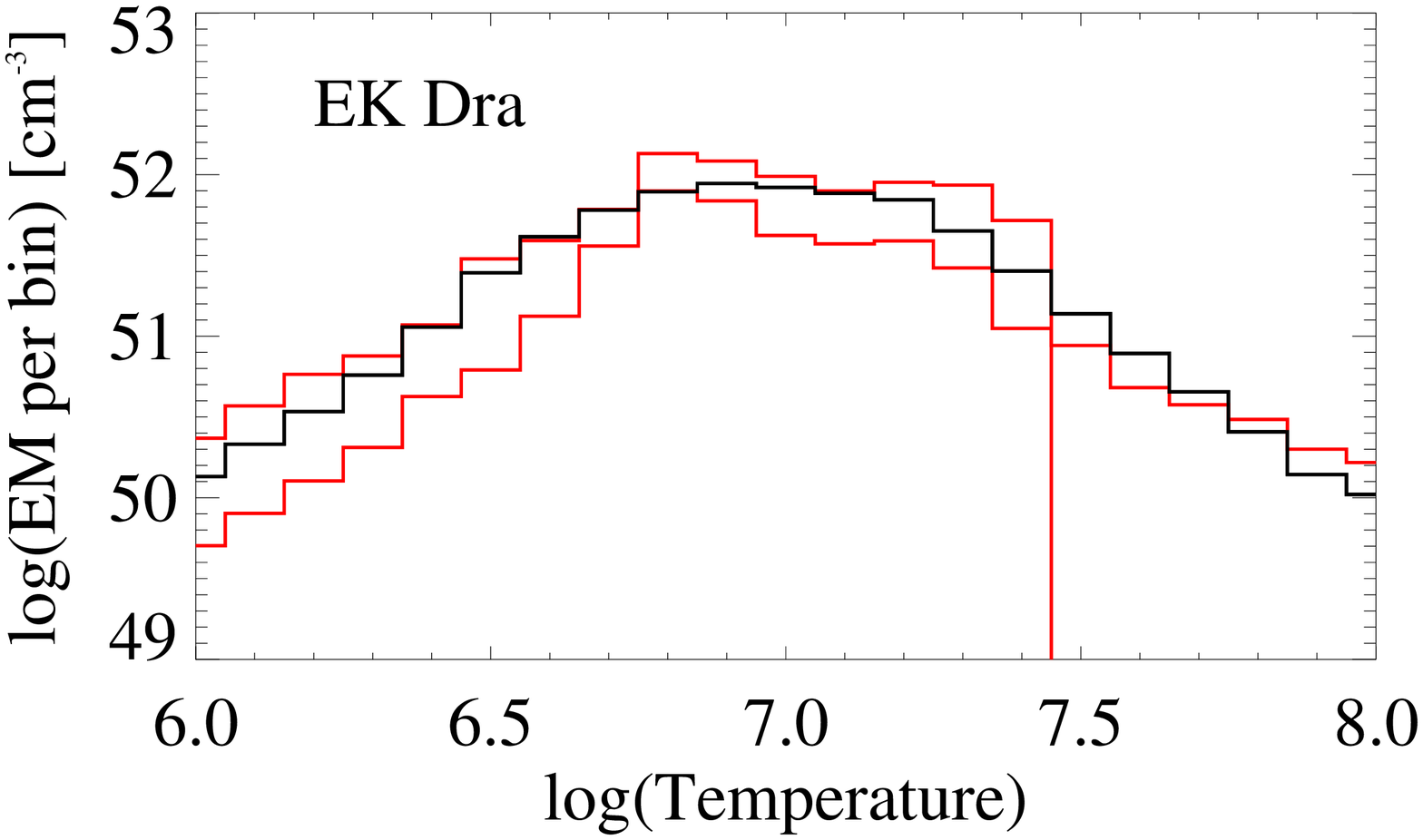}
\includegraphics[width=0.26\linewidth]{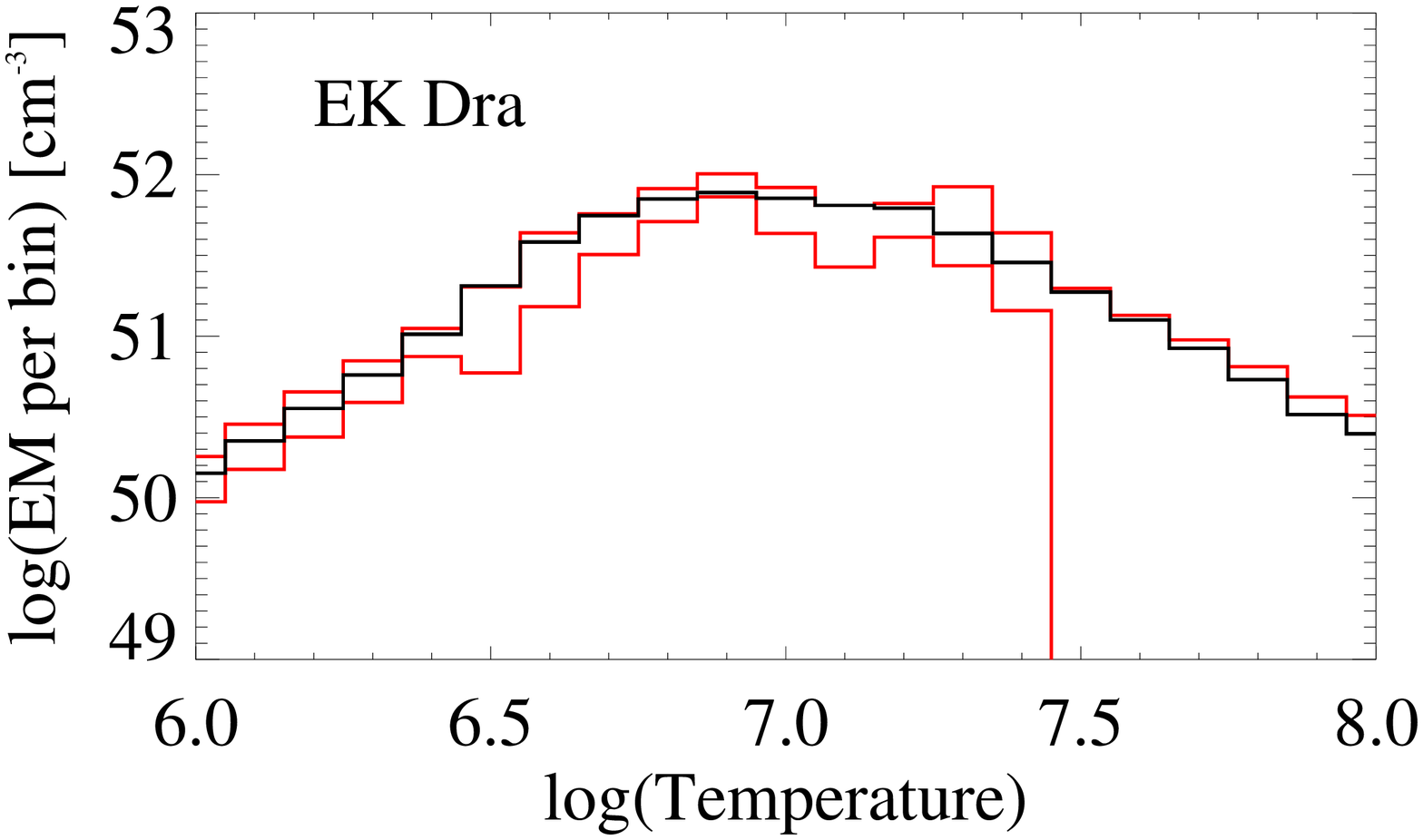}
}}
\centerline{\hbox{
\includegraphics[width=0.26\linewidth]{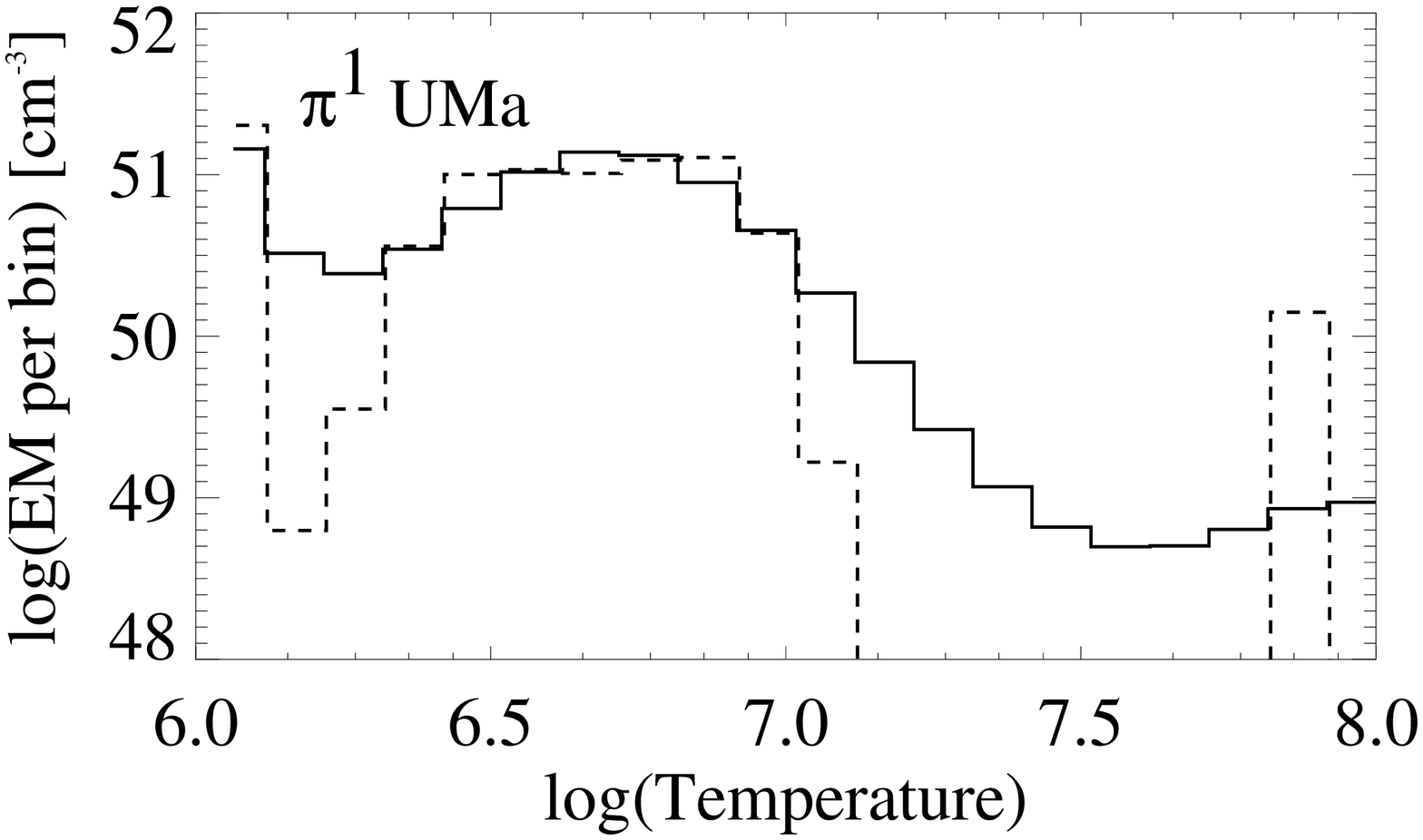}
\includegraphics[width=0.26\linewidth]{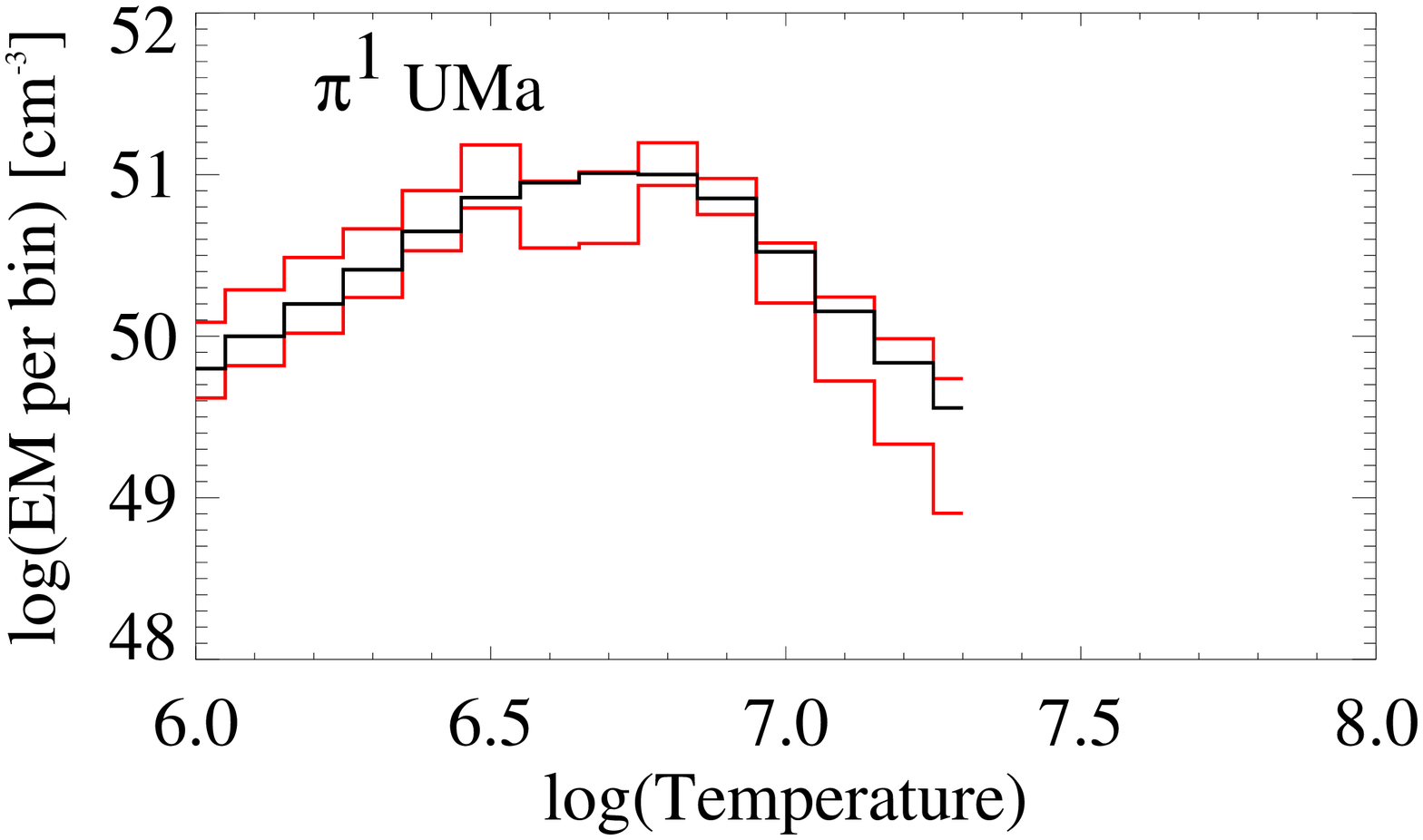}
\includegraphics[width=0.26\linewidth]{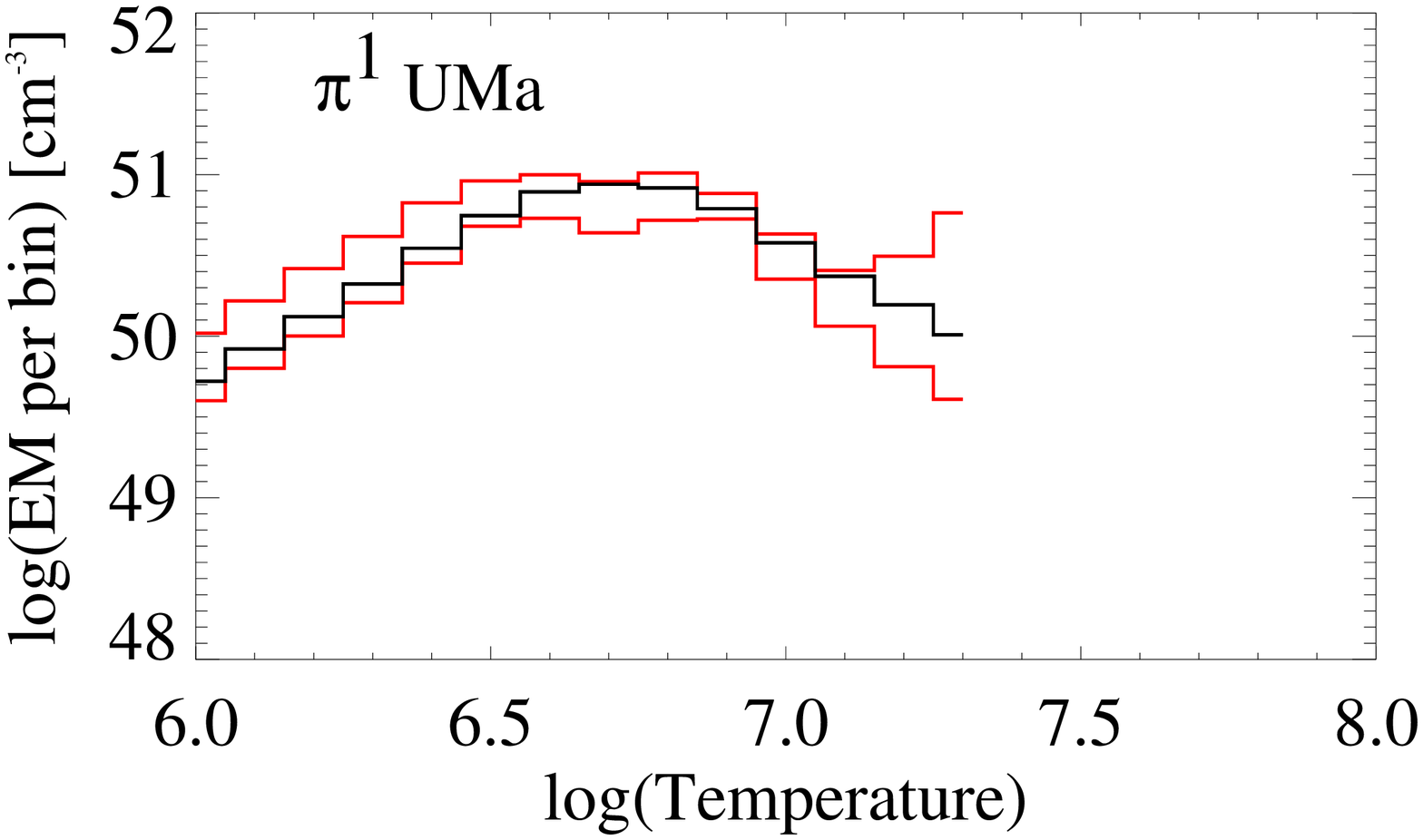}
}}
\centerline{\hbox{
\includegraphics[width=0.26\linewidth]{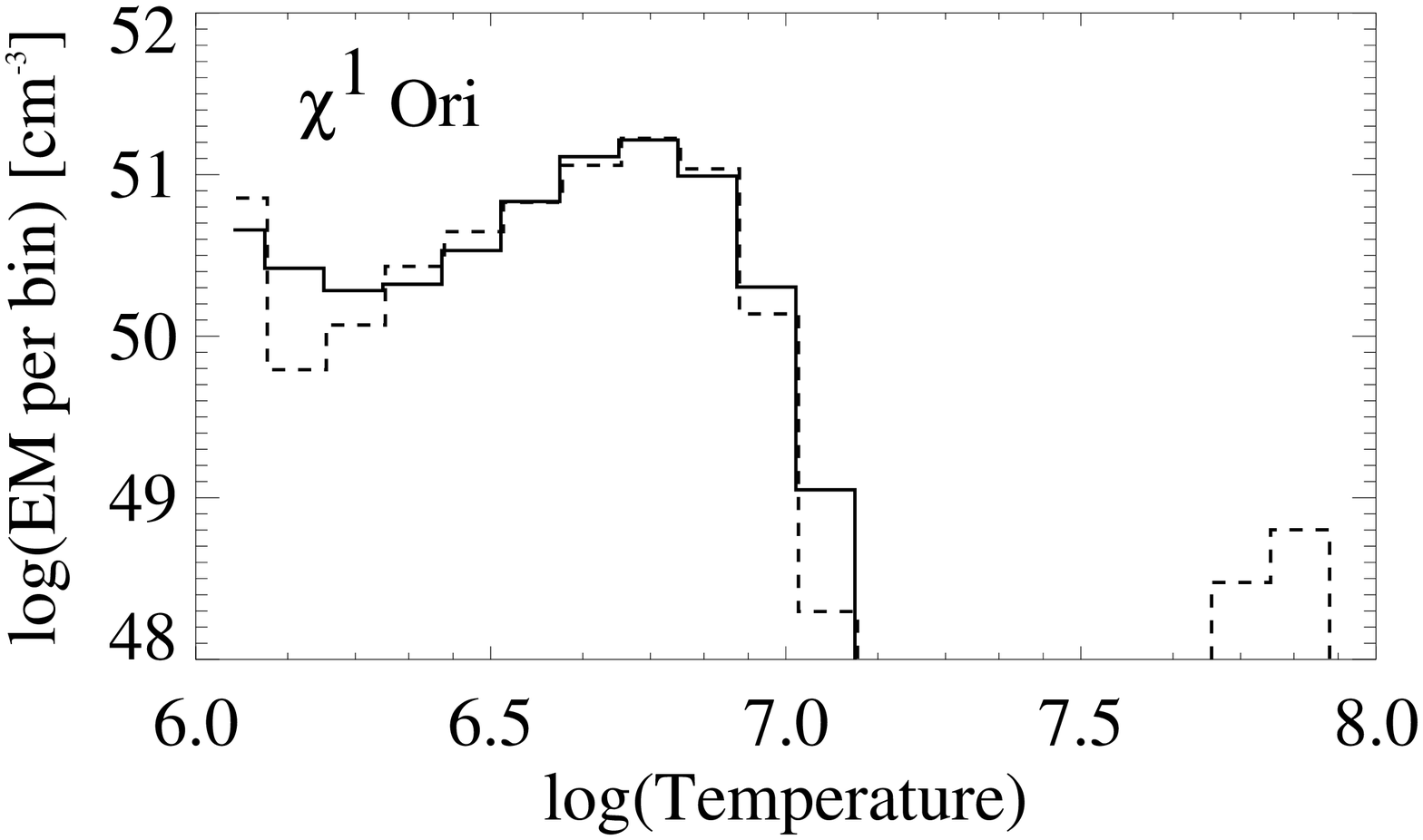}
\includegraphics[width=0.26\linewidth]{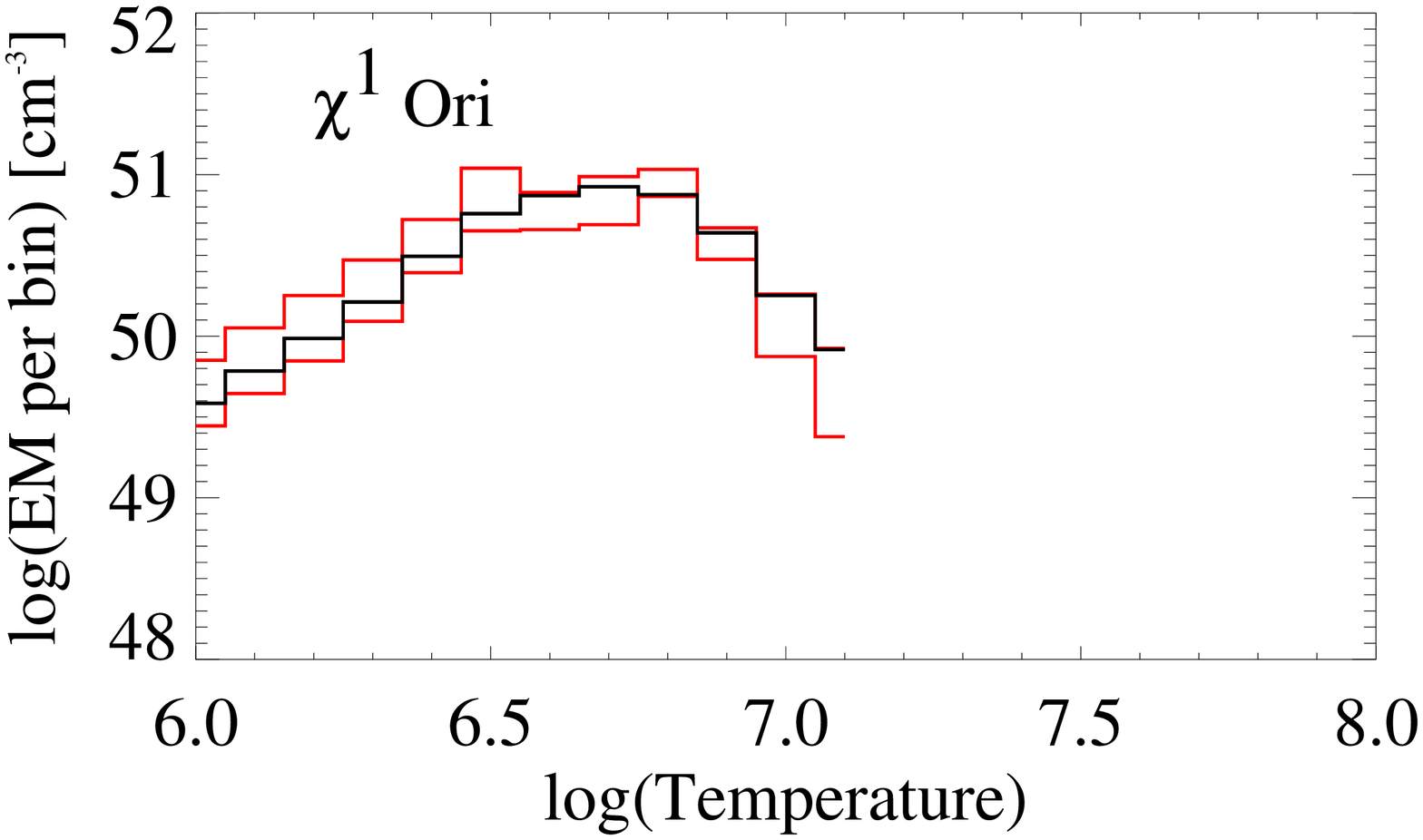}
\includegraphics[width=0.26\linewidth]{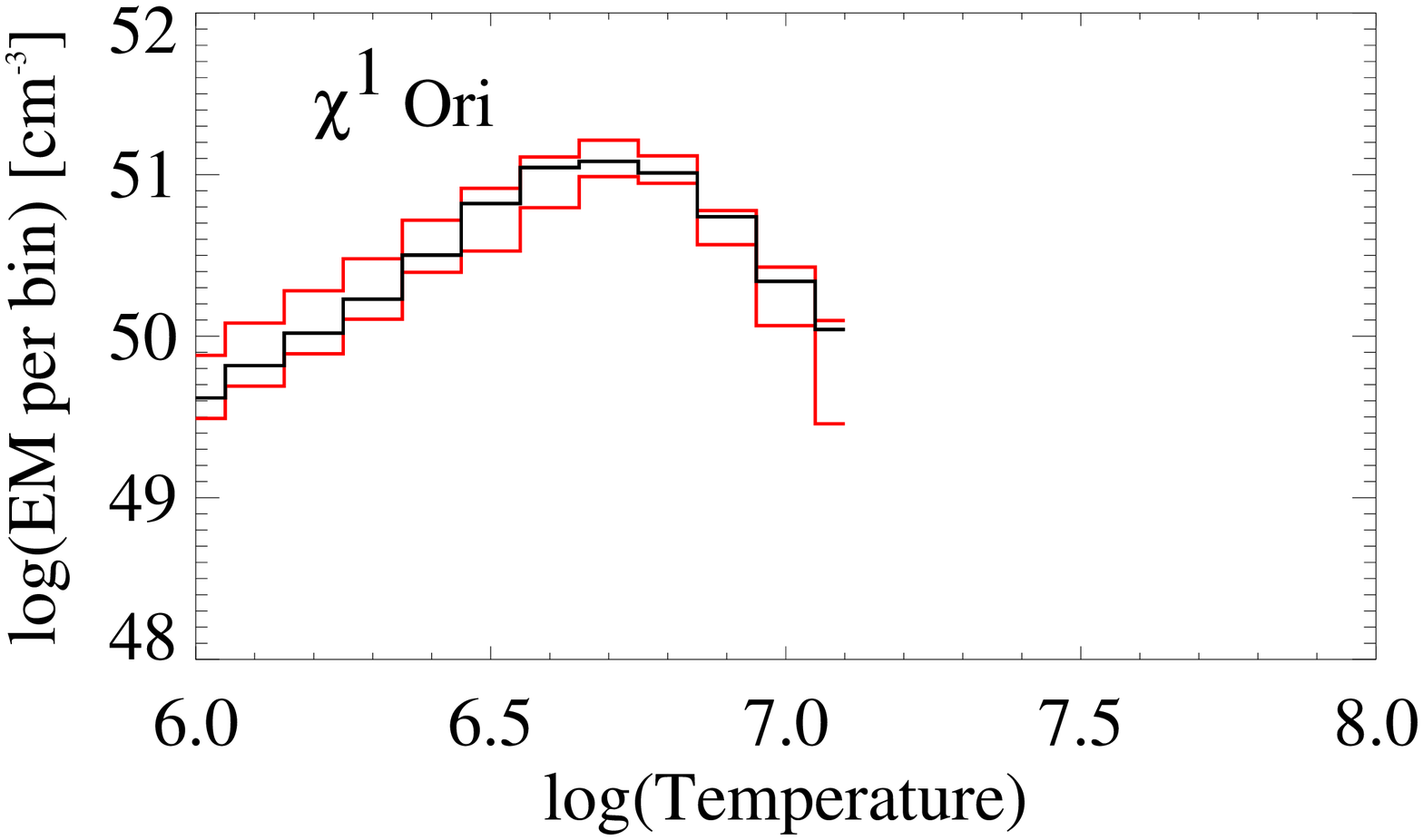}
}}
\centerline{\hbox{
\includegraphics[width=0.26\linewidth]{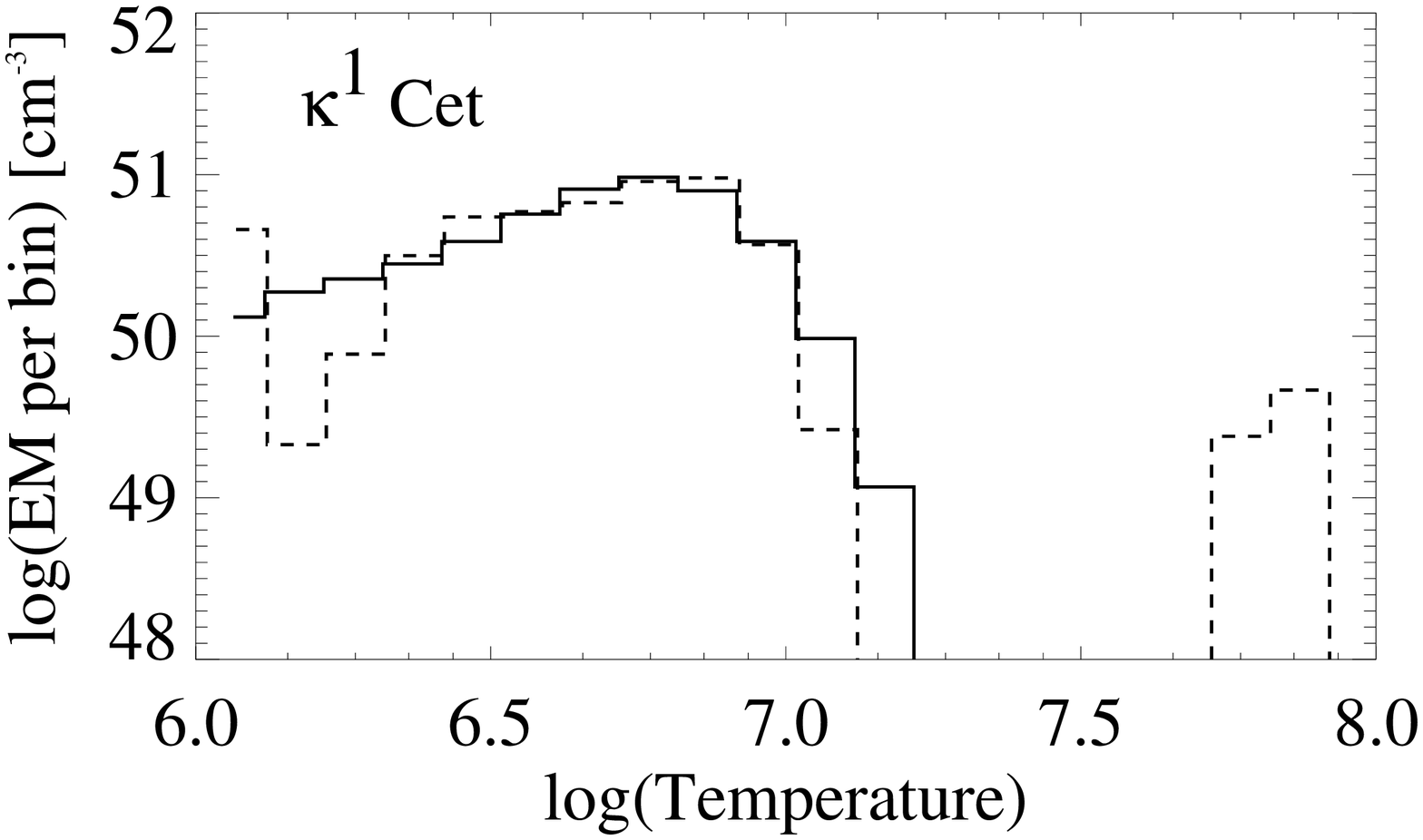}
\includegraphics[width=0.26\linewidth]{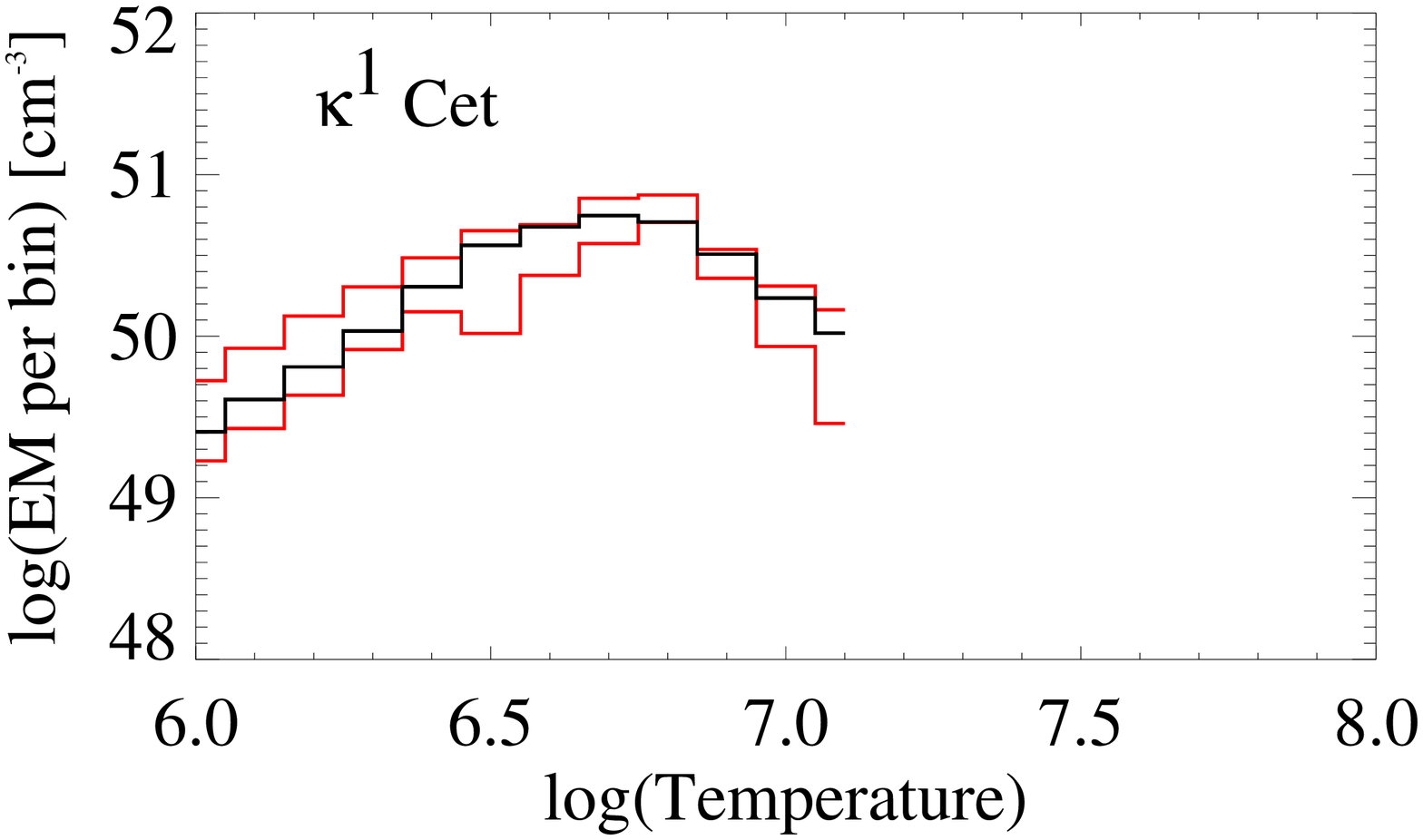}
\includegraphics[width=0.26\linewidth]{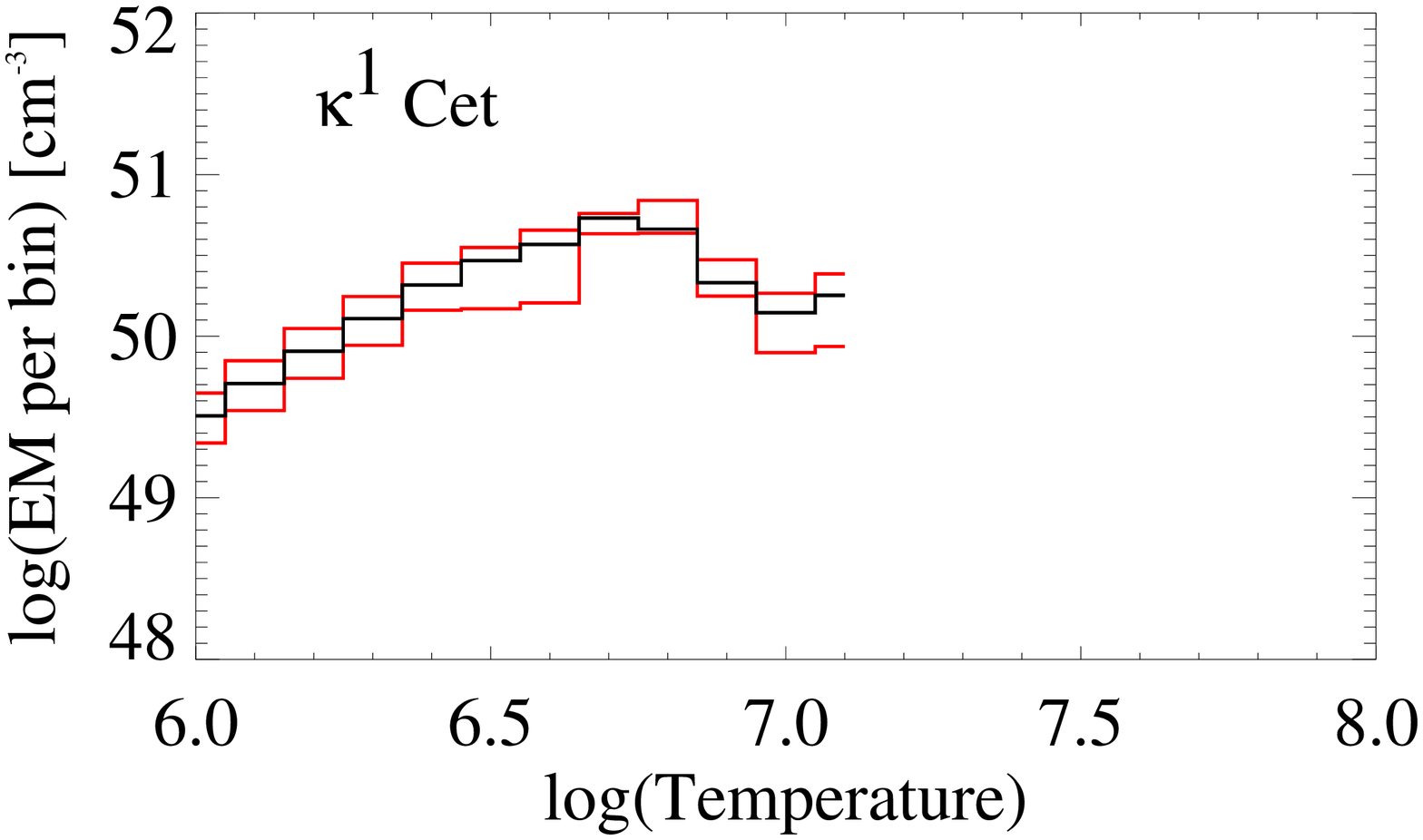}
}}
\centerline{\hbox{
\includegraphics[width=0.26\linewidth]{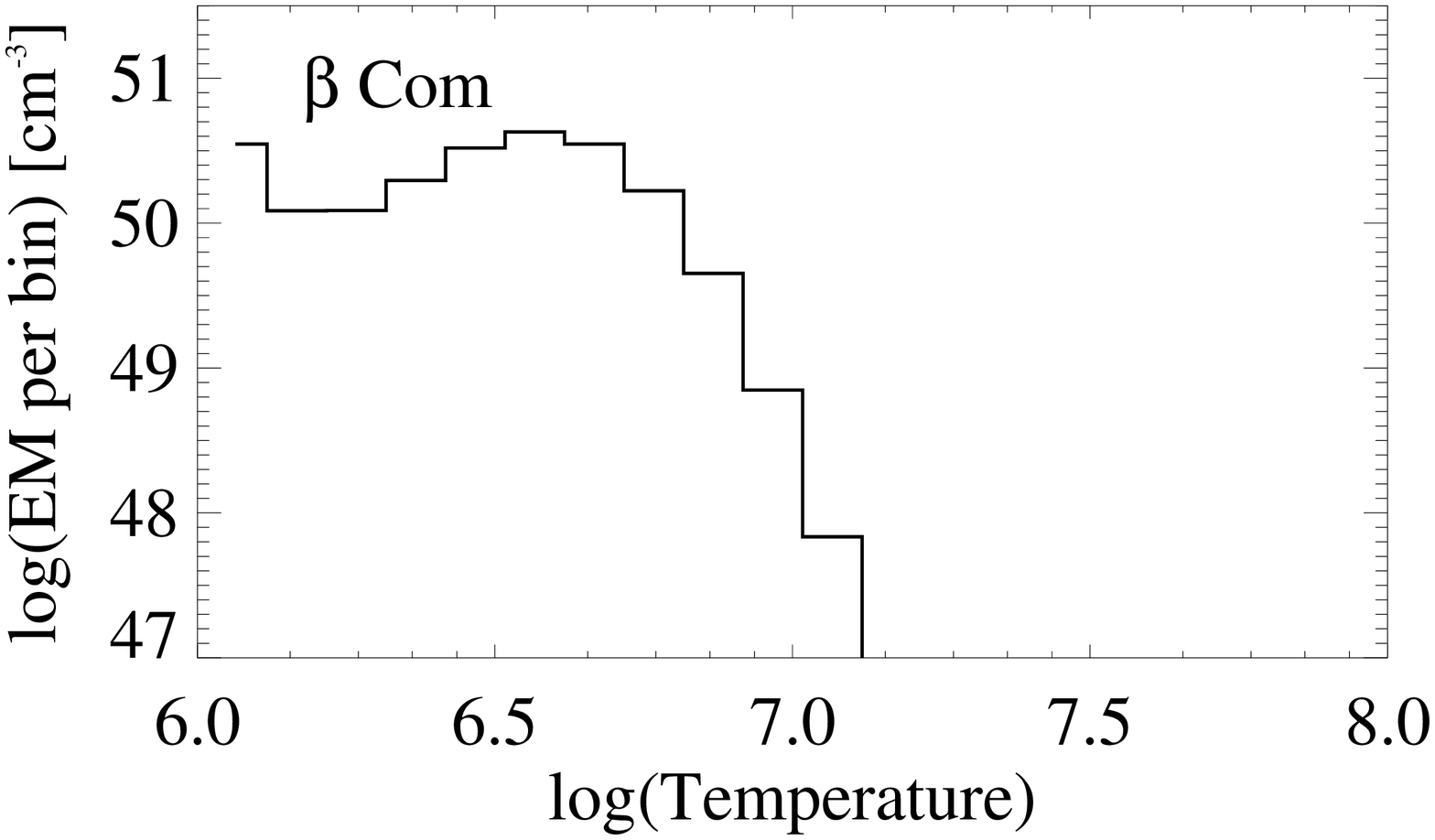}
\includegraphics[width=0.26\linewidth]{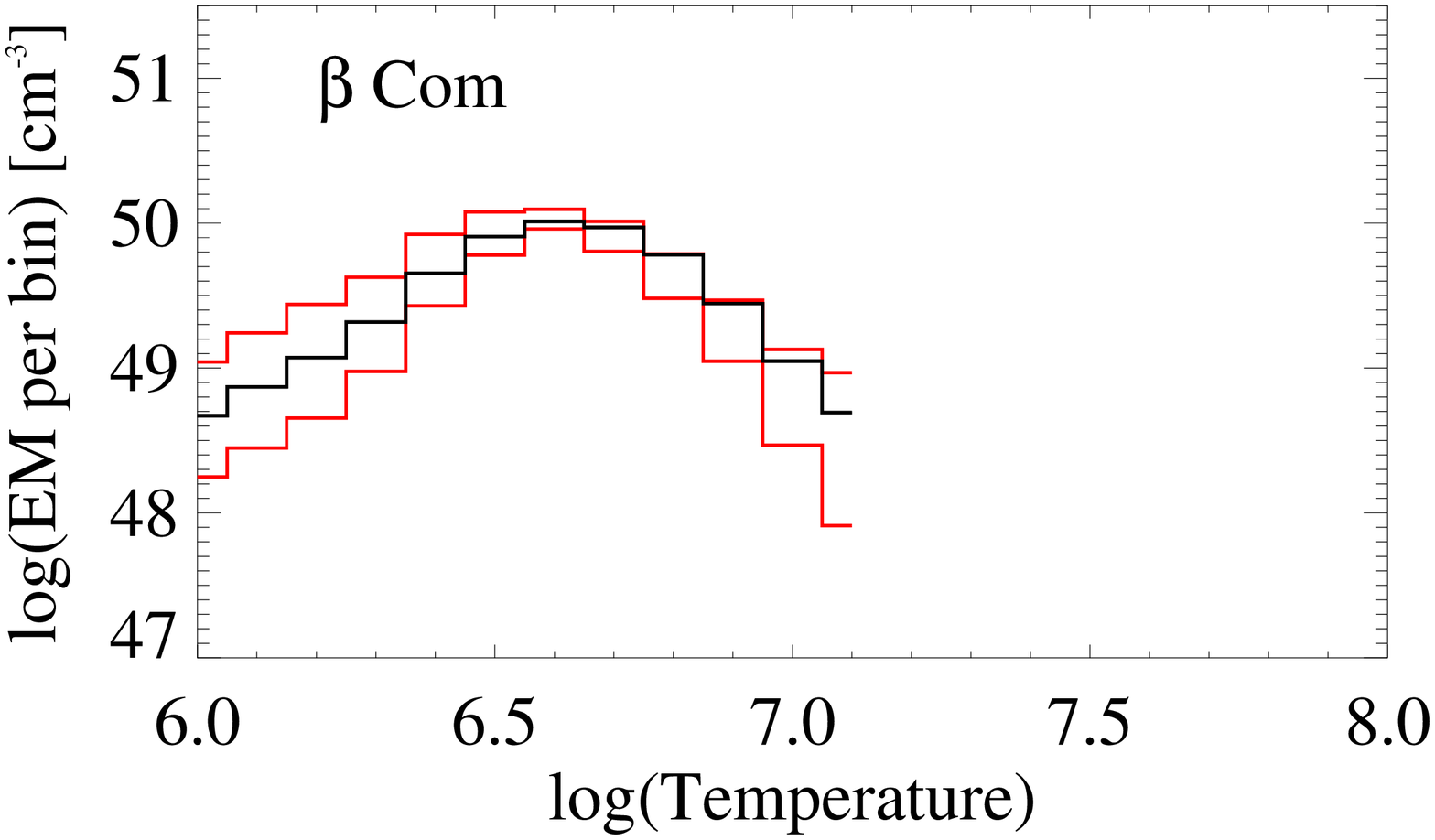}
\includegraphics[width=0.26\linewidth]{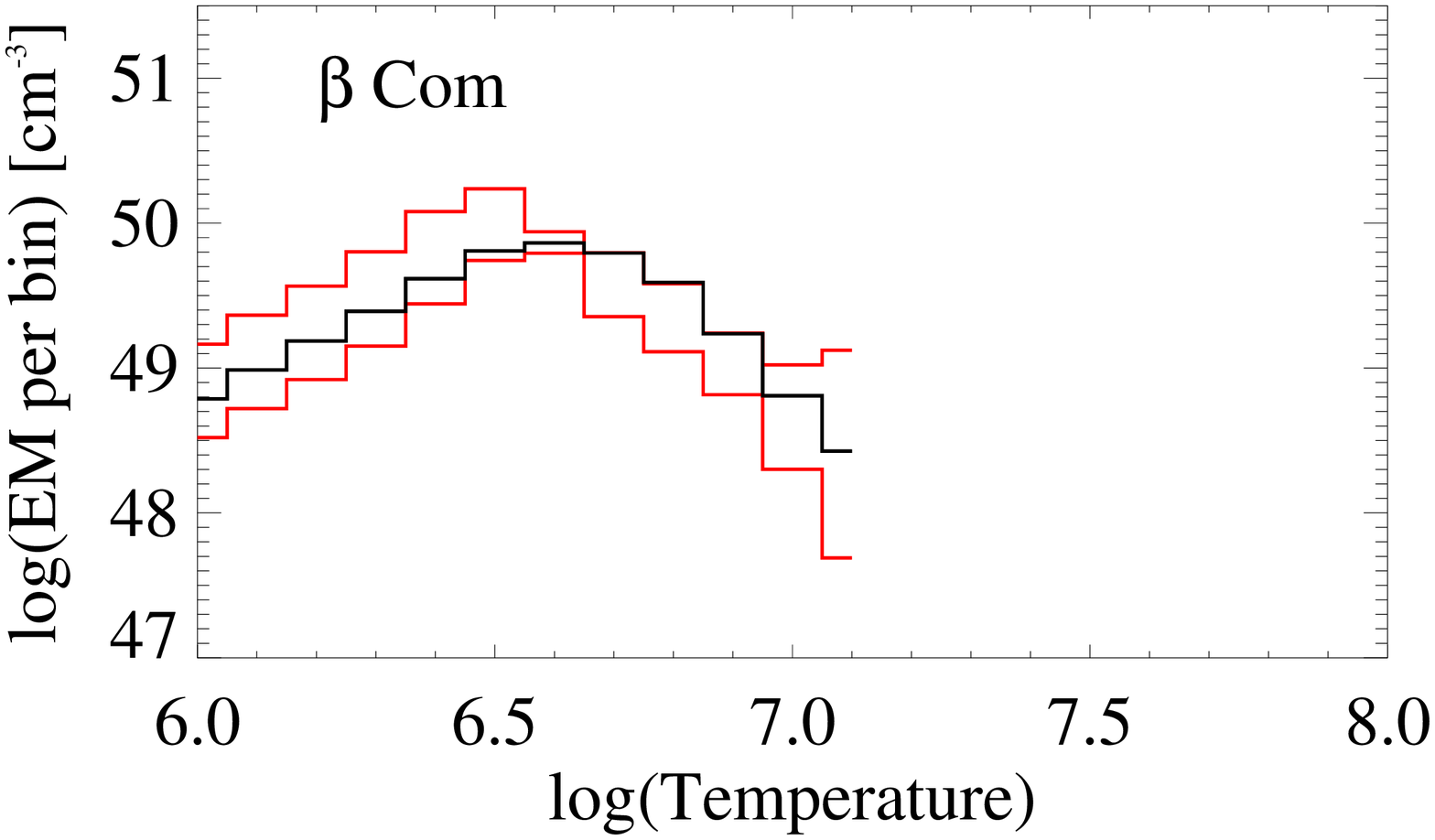}
}}
\end{center}
\caption{Reconstructed emission measure distributions.  
{\bf Left:} From method 1 using SPEX, based on Chebychev polynomials of order 6 (solid line) and order 8 (dashed).
{\bf Middle:} From method 2, based on MEKAL emissivities.    
{\bf Right:} From method 2, based on APEC emissivities. In the middle and right plots, the 
red histograms illustrate the $\pm$ 1$\sigma$ range of solutions from the average of 20 EMDs reconstructed from the original and from the perturbed line lists. The black histograms illustrate the best-fit EMDs, derived from the unperturbed line-flux list.}\label{DEMplots}
\end{figure}

The EMDs derived from the two different methods are shown in Figure~\ref{DEMplots}. In the left column, 
EMDs from method 1 using SPEX in combination with a fit based on Chebychev polynomials of 
degree 6 and, where possible, 8 are plotted. The middle and the right columns show EMDs reconstructed 
with our method 2, based on MEKAL and APEC emissivities, respectively. 
In the middle and right columns, the black histograms illustrate the best-fit EMDs while the red 
histograms mark the $1\sigma$ range at each temperature, derived from the  
perturbed flux lists. As the best-fit EMDs are derived from unperturbed fluxes, 
and they are not equal to the mean EMDs derived from the perturbed flux lists, these ranges of variation do not 
need to be symmetrically arranged around the best-fit solutions. 
In some cases (EK Dra and 47 Cas B), the lower error ranges drop rapidly to 
very low values at certain temperatures. Although this is a consequence of the 
increasing uncertainty in the EMD at the lowest and the highest temperatures,
we note that the error ranges are given on a logarithmic scale; once the ranges
becomes large, the precise level of the lower bound is of little importance.

\begin{figure}[!h]
\begin{center}
\centerline{\hbox{
\includegraphics[width=0.49\linewidth]{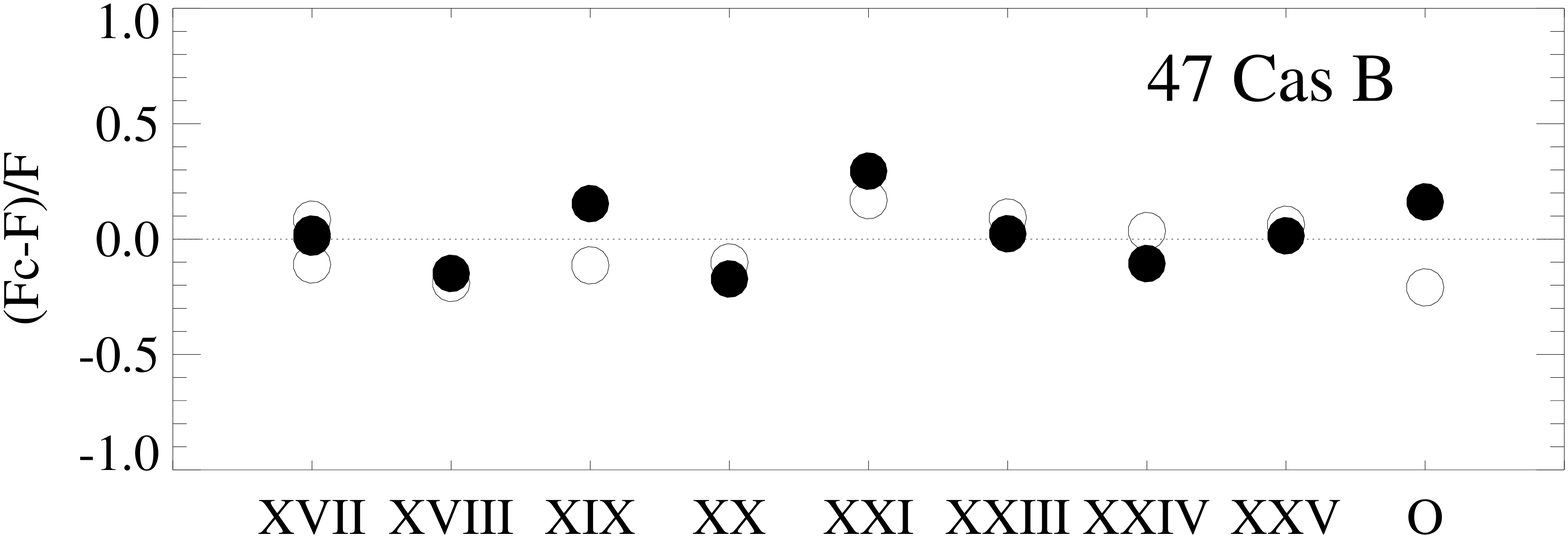}
\includegraphics[width=0.49\linewidth]{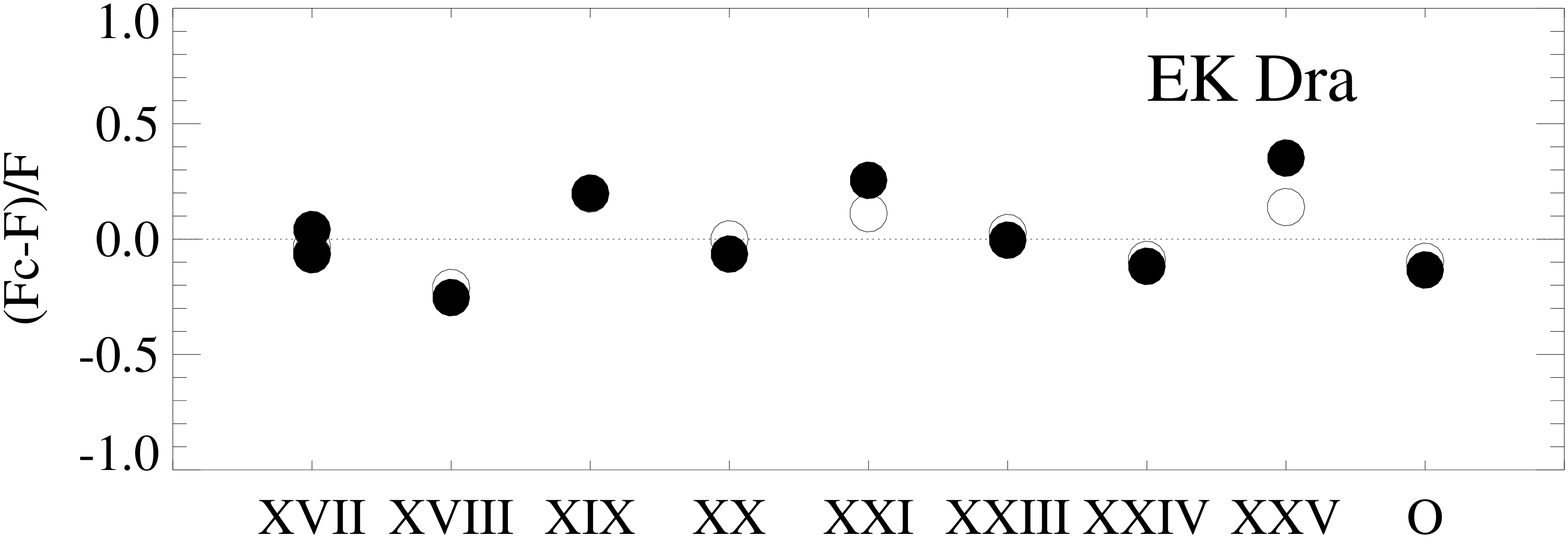}
}}
\centerline{\hbox{
\includegraphics[width=0.49\linewidth]{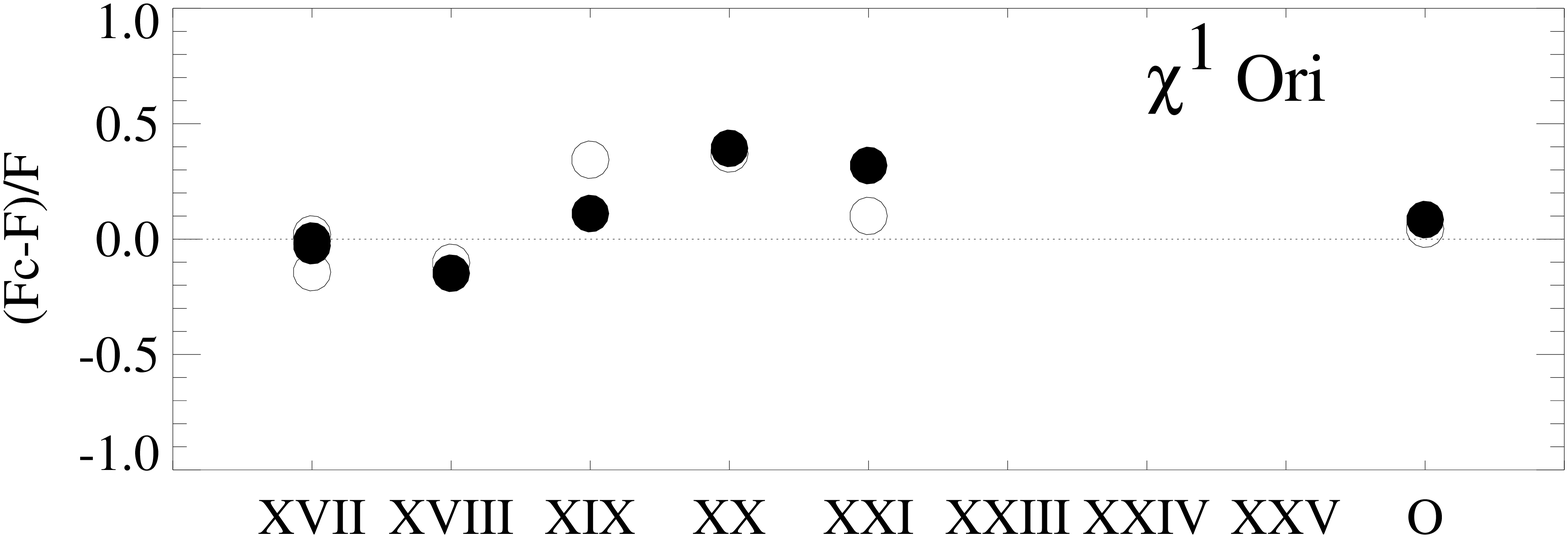}
\includegraphics[width=0.49\linewidth]{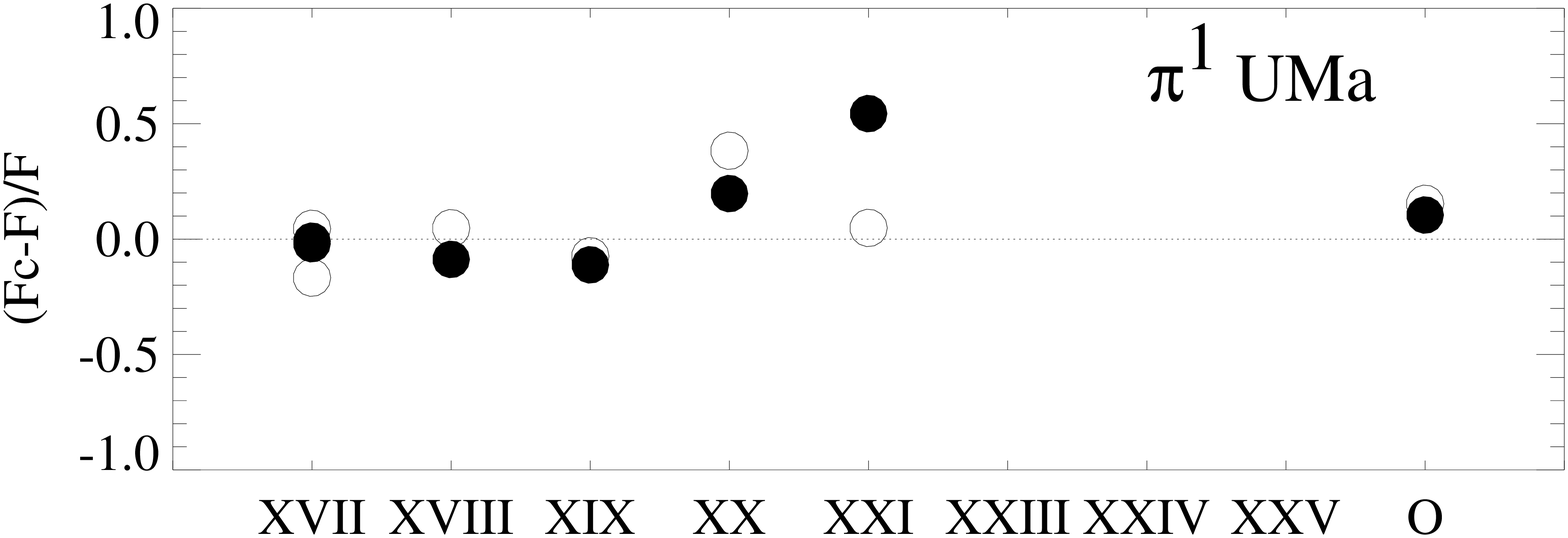}
}}
\centerline{\hbox{
\includegraphics[width=0.49\linewidth]{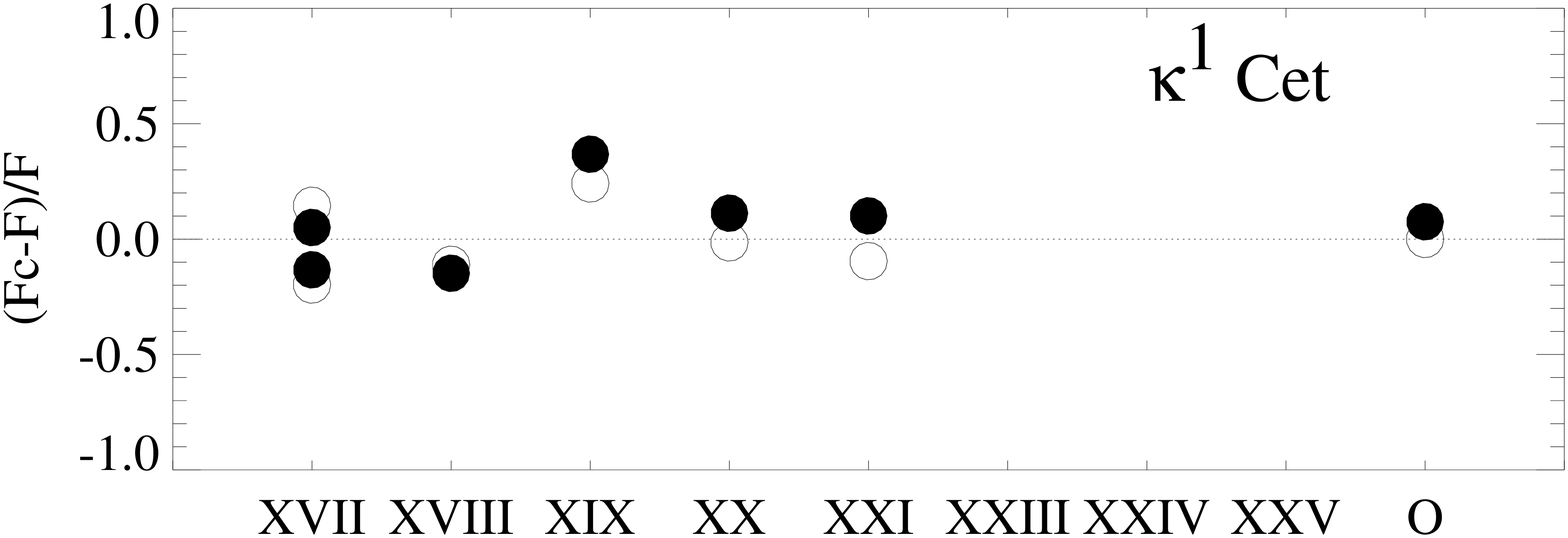}
\includegraphics[width=0.49\linewidth]{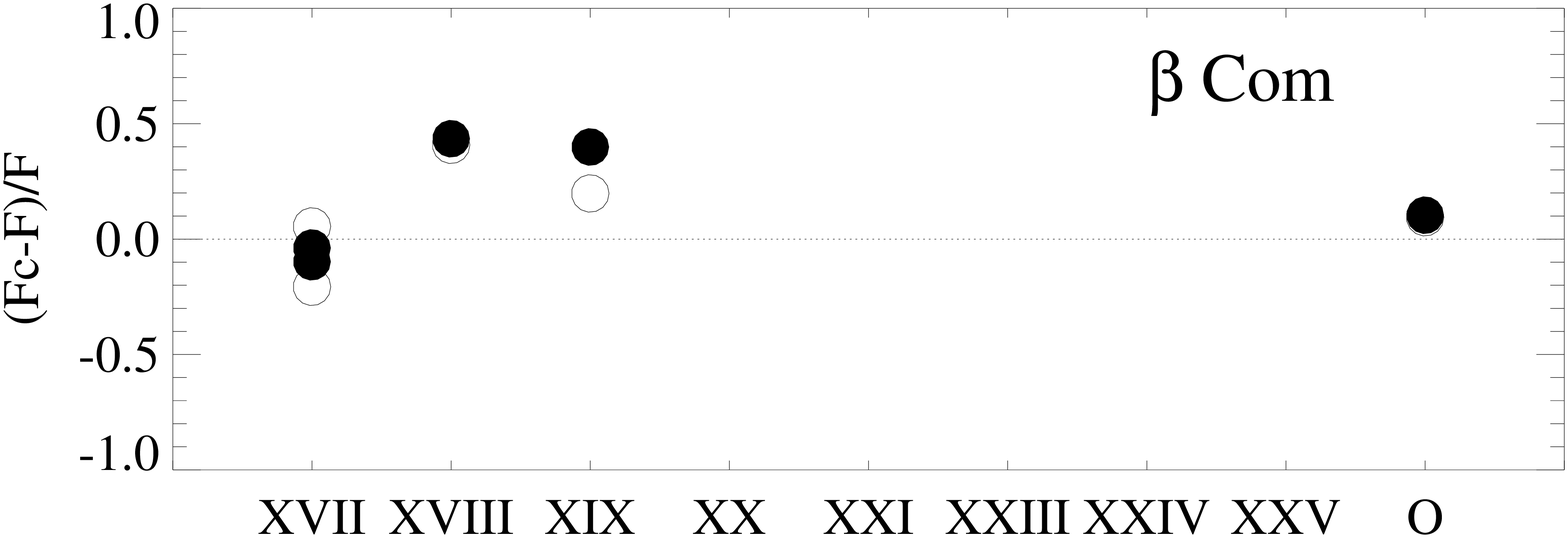}
}}
\end{center}
\caption{Fractional deviation of the predicted Fe line fluxes and the O\,{\sc viii}/O\,{\sc vii} flux ratio from the observed values for method 2. Filled circles: SPEX. Open circles: APEC.}\label{residual}
\end{figure}

For method 2 the quality of the EMD can be 
measured by comparing the predicted and the observed line fluxes. 
The final agreement between predicted and observed line fluxes is illustrated in Fig.~\ref{residual}, where we
show the fractional deviation of the predicted line fluxes from the observed values, $(F_C - F)/F$ for Fe and 
O\,{\sc viii}/O\,{\sc vii} flux ratio. Most line 
fluxes agree within 10--20\%, with the larger deviations mainly relating to the weakest lines, i.e., the lines formed
at high temperatures in the least active stars (e.g., Fe\,{\sc xx} for $\pi^1$ UMa).

The EMDs derived from the different methods show rather similar characteristics. We see that the 
temperature where the EMD peaks decreases toward older, less active stars, namely from
about 10~MK for 47 Cas B and EK Dra to 5~MK for $\pi^1$ UMa, $\chi^1$ Ori and 
$\kappa^1$ Cet, and to  $\la$4~MK for our oldest target, $\beta$ Com. Characteristic values for the Sun are 1--3~MK,
depending on the phase of its activity cycle \citep{peres00}. The 
average temperatures derived from  method 2,  $\log \bar{T}$,  
are listed in Table~\ref{temp}. To obtain these values, we calculated the mean of $\log T$, using
the EMs in each bin as weights. Also given are the lower and upper threshold temperatures that comprise
90\% of the total EM (on each side of the EMD peak).

With method 2, we generally obtain a smoother and 
flatter EMD than with method 1 and the Chebychev polynomial 
approximation. At temperatures above $\log T \approx 7.5$, the EMD is not well constrained.
This is obvious for the cooler coronae which do not provide any useful
spectral lines at those temperatures, and we therefore did not
extend our DEM analysis to this range.  Considerable scatter is still
also found for the EMDs of 47 Cas B and EK Dra, despite the availability 
of Fe\,{\sc xxiii}-Fe\,{\sc xxv}  lines. The reason resides in the fact that 
Fe\,{\sc xxiii} and Fe\,{\sc xxiv} show very faint lines, and Fe\,{\sc xxv} is 
the only blend complex that covers the temperatures above $\log T \approx 7.5$. 
In the more active stars, the EMDs from method 1 
seem to be composed of two peaks at about 6 and 20~MK, well separated
by a  local minimum. 
Also, in most EMDs derived from method 1, we find a deep decrease in 
the EM below about  $\log T \approx 6.2$, combined with a local EM peak 
around $\log T = 6.0$. Both features become stronger if higher polynomial degrees
are used. 

\begin{figure}
\begin{center}
\centerline{\hbox{
\includegraphics[width=0.45\linewidth]{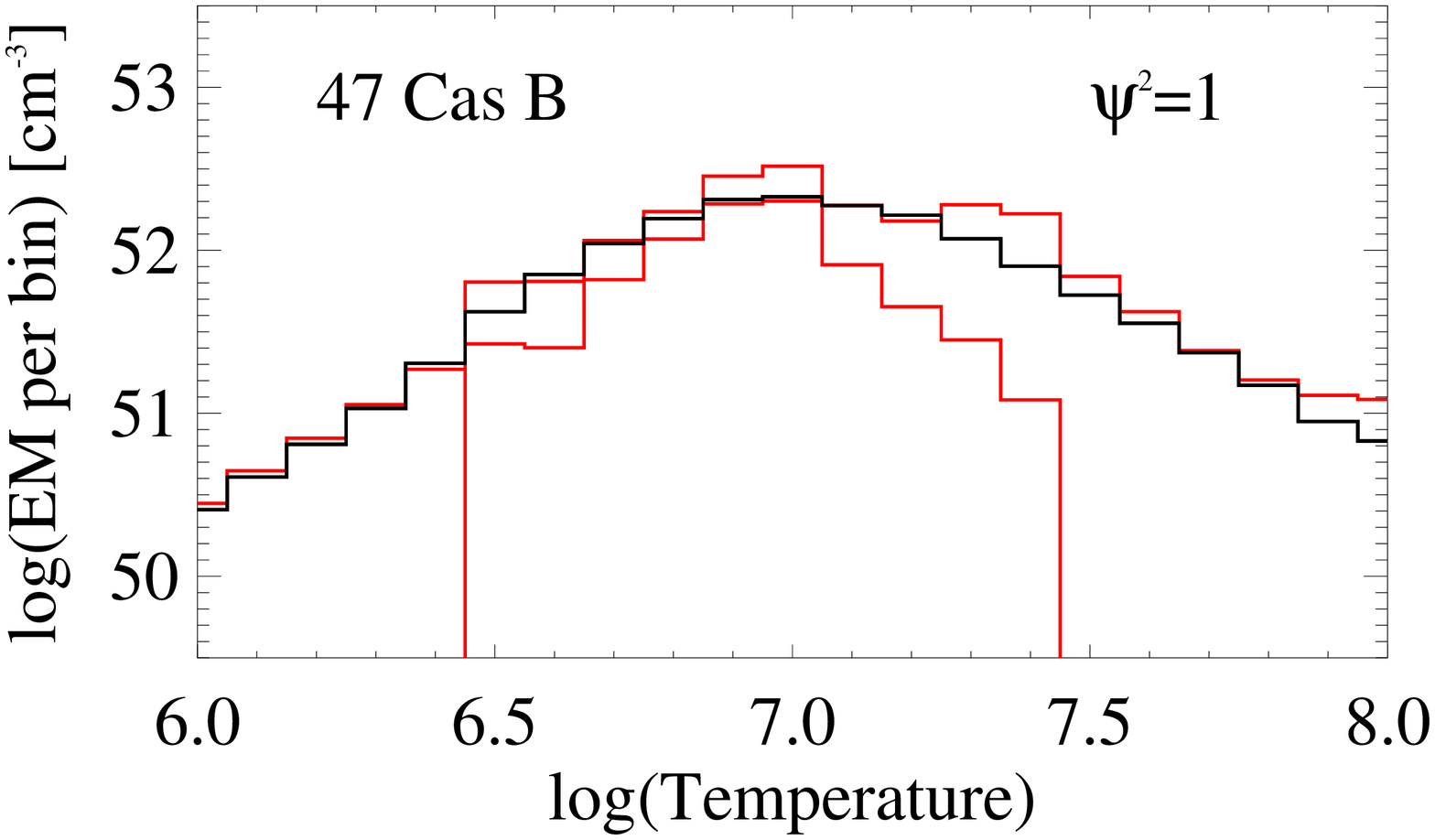}
\includegraphics[width=0.45\linewidth]{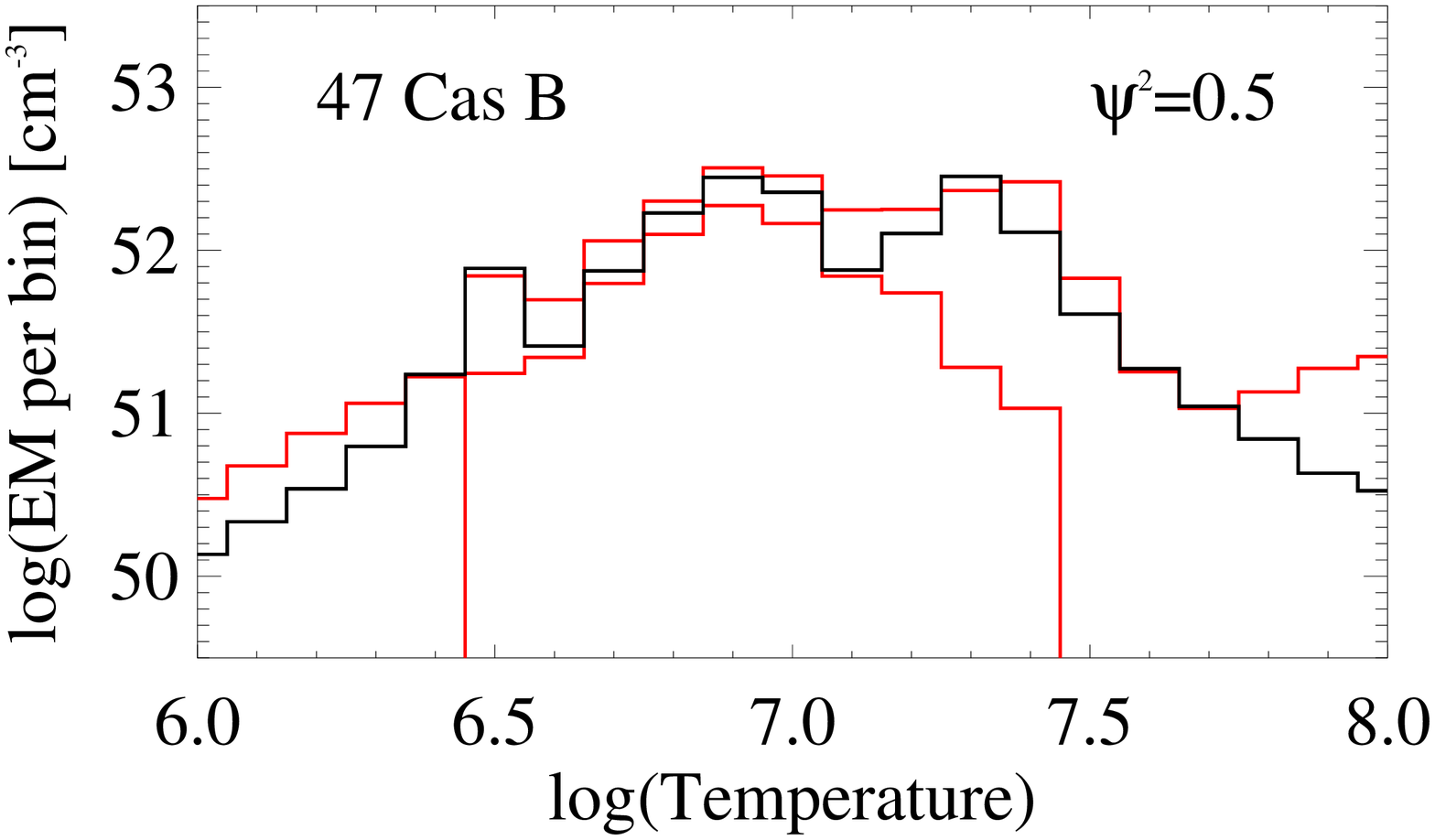}
}}
\centerline{\hbox{
\includegraphics[width=0.45\linewidth]{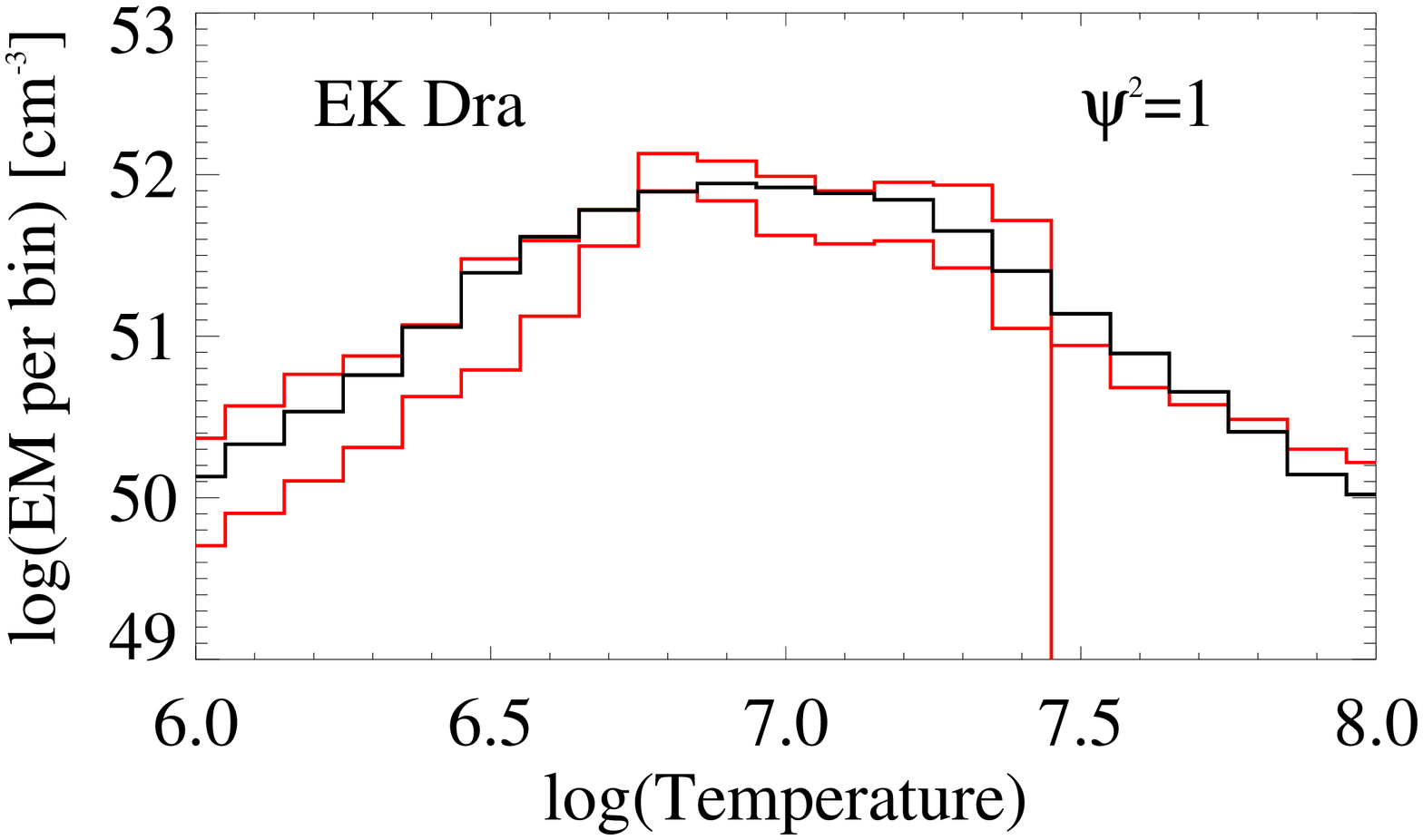}
\includegraphics[width=0.45\linewidth]{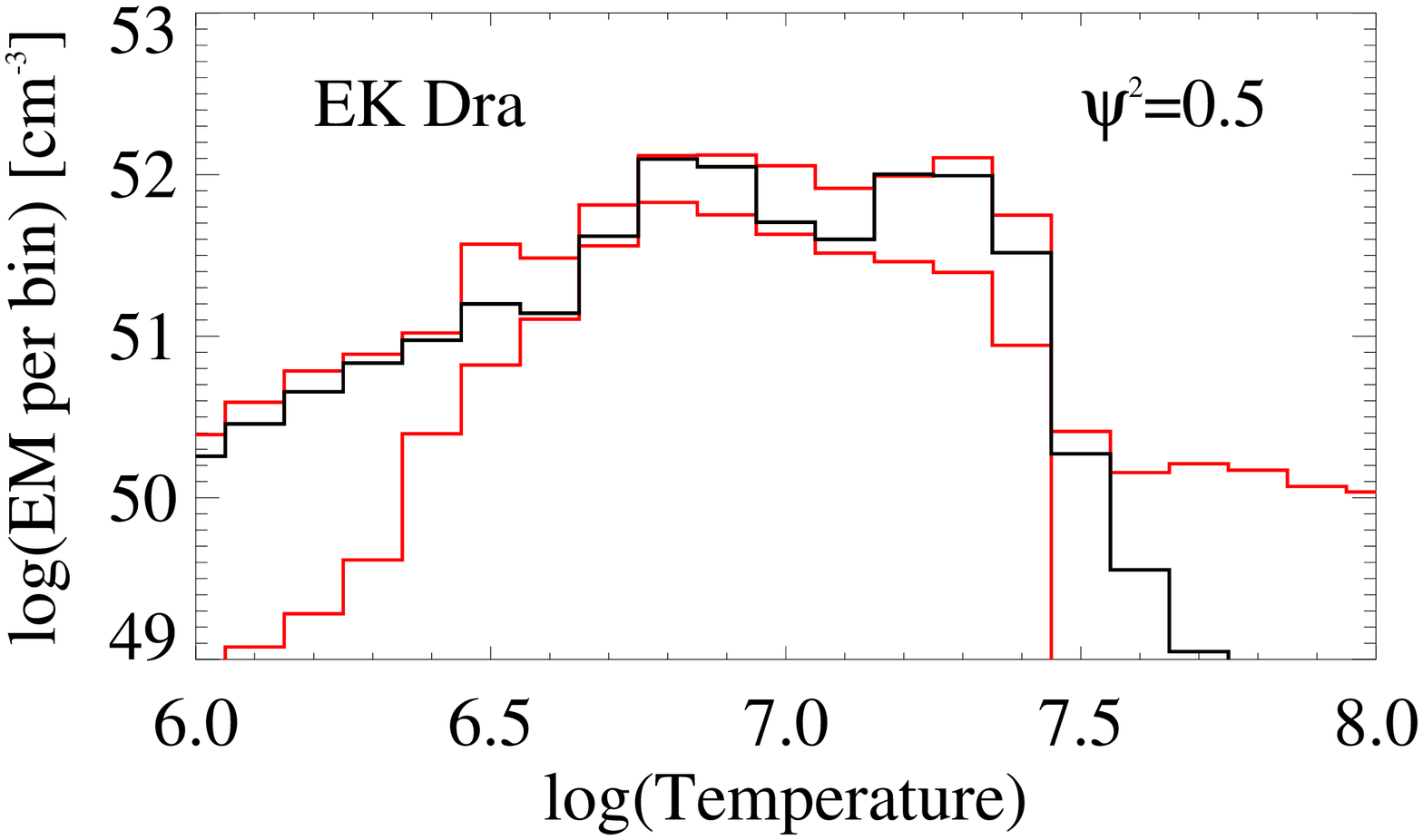}
}}
\end{center}
\caption{Emission measure distributions (using SPEX with method 2). 
{\bf Left:} The iteration was stopped when $\psi^2 = 1.0$  was reached.
{\bf Right:} The iteration was stopped when $\psi^2 = 0.5$ was reached. Note the additional oscillations
in the EMD.}\label{chi_0.5}
\end{figure}

Several effects may contribute to this: First, a large range of solutions may
in fact be compatible with the spectra, given that the spectral inversion is
ill-conditioned, in particular in temperature regions where few constraints are
available. Second, the fit of method 1 iterates to minimum $\chi^2$, which considers 
only the Poissonian errors in the count spectrum and which may introduce 
EMD features of little relevance given the systematic uncertainties in the
atomic physics, while method 2 has been terminated according to $\psi^2$ 
(see Eq.~\ref{for:chi}), which approximately considers the atomic physics uncertainties as well.
Third, method 1 uses many lines that may
introduce uncertainty to the spectral fit, while the result of method 2 almost uniquely relies
on the O\,{\sc viii}/O\,{\sc vii} flux ratio for the coolest portion of the EMD.
And finally, method 1 imposes polynomial constraints on the solution,
which favours the appearance of peaks and valleys in the EMD, while method 2 starts with
a smooth EMD that is changed only in so far as the spectrum requires. 
For example, if a line requires excess EM due to an underestimation of its
emissivity at a given temperature, then the reconstruction process may
compensate by lowering the EM at adjacent temperatures as dictated by lines dominating 
there. To test this hypothesis, we iterated method 2 excessively, to $\psi^2 = 0.5$.
As the two examples in Fig.~\ref{chi_0.5} show, very similar features also evolve in these
examples. The low-temperature slopes appear to become partly steeper as well.

We also note that the amplitudes of the oscillations are compatible with the error ranges   
from perturbing the line-flux lists used in method 2 (cf. Fig.~\ref{DEMplots}). 
It is conceivable that the oscillations found after a deep iteration of method 2
correspond to those seen in method 1 although this cannot be explicitly
proven, given the largely different approaches. However, the magnitudes of 
the oscillations appear to be similar. It is also possible that the oscillations
are present in the stellar EMD; we cannot reliably discriminate between this hypothesis 
and a numerical effect as long as we include statistical errors and assume the presence of systematic 
uncertainties of the magnitude adopted here (see Sect.~\ref{sec:err}).

In contrast, the abundance ratios turn out to be robust, with no
significant change when deeper iterations are applied. In fact, the synthesized spectra
for the two cases are very similar, i.e., the two EMDs represent the spectra almost equally well. 
Comparing the synthesized spectrum with the observations in the wavelength intervals illustrated
in Fig.~\ref{specfit},  we find, for 47 Cas B, a reduced $\chi^2$, $\chi^2_{\rm red} = 1.28$ (for 1091 d.o.f) for the deeper integration 
and $\chi^2_{\rm red} = 1.37$  (for 1091 d.o.f) for our standard convergence criterion. Although at first sight 
this difference appears significant, the important line features and the continuum in the RGS are fitted well in both cases.

We note that the EPIC MOS portion of the synthetic spectrum from method 2 is not very well fitted: the fit 
lies systematically below the data. This feature is probably to be ascribed to a cross-calibration inaccuracy in the effective
areas of the RGS and the EPIC MOS instruments. As a matter of fact, a related effect is present in the spectrum from method 1 
as well, but there, the continuum level of the synthetic spectrum is slightly but systematically too high in the RGS compared to the data,
while the fit in the MOS spectrum is better. This is one important source for the somewhat higher $\chi^2_{\rm red}$ 
for our method 2. We note, however, that the analysis based on method 2 corrects for the continuum discrepancy before line-flux extraction by adjusting the continuum level individually both for the RGS and the EPIC spectra (see Sect.~\ref{sec:emdrec}).

In Table~\ref{redchi}, we compare the $\chi^2_{\rm red}$ 
values with respect to the observations, for the synthetic
spectra constructed from the EMDs that were obtained 
with method 1 and with method 2, respectively. 
Again, only the  regions listed in Table~\ref{region} are used.
The degrees of freedom are also listed in Table~\ref{redchi}. 
Although formally the same wavelength intervals were used, the number of 
degrees of freedom are somewhat different in SPEX and XSPEC. 
This discrepancy comes from two different sources: first, partial bins at the 
beginning and the end of each interval are considered differently (XSPEC
ignoring all partial bins). Although this discrepancy could be reduced 
by minor adjustments, we prefer to keep with the simple prescriptions of 
Table~\ref{region} for easy reproduction of our results. The partial bins occur
in relatively low-flux, shallow regions of the spectrum where essentially continuum
is fitted, hence a difference by a single, usually well fitted continuum bin is of little relevance. 
Second, the standard spectral software packages treat grouped bins
that may contain bad data channels differently (SPEX breaks bins up into
partial bins, while XSPEC does not). This is a feature of the standard software
packages that we test here. 

The $\chi^2_{\rm red}$ from method 2 are slightly larger but 
the differences are not very substantial despite the  
systematic uncertainties in the emissivities adopted in method 2 but not in method 1. 
Beside the fact that in the reconstructed spectrum of  method 1 the MOS spectrum is 
better fitted (given its higher S/N ratio), the slightly better $\chi^2_{\rm red}$  for method 1
could also be a result of excessively deep iterations that aim at fitting 
poorly-described line fluxes at the cost of smoothness
in the EMD. It is noteworthy that the $\chi^2_{\rm red}$ 
values from the deep iterations in method 2 closely approach the 
$\chi^2_{\rm red}$ values of method 1.

There is ample literature on EMDs of active stars available. Several 
other authors have also found EMDs with features resembling ours.
In particular, double-peaked EMDs have previously been reported, e.g., by
\citet{mewe96} for AB Dor, \citet{kaastra96a} for RS CVn binaries, \citet{guedel97a}  for solar analogs, \citet{sanz01} for the RS CVn binary
$\lambda$ And, and \citet{huenemoerder03} for AR Lac, using entirely
different EMD reconstruction methods. In the light of our discussion on errors 
and iteration depths in Sect.~\ref{analysis} and~\ref{results}, we cannot be certain on the actual reality 
of bi-modal DEMs, and given that other authors use similar atomic
physics databases, the same caveat may apply to other EMD reconstructions 
as  well.  \citet{guedel97a} argued, based on solar-flare data, that a bi-modal structure
can arise from the rapid decay of the EM of a population of flares as they are cooling.
The present quality of our EMD inversions does not allow us to make more
definitive conclusions at this point.

\subsection{Abundances}\label{sec:abun}

\begin{figure}
\begin{center}
\centerline{\hbox{
\includegraphics[width=0.43\linewidth]{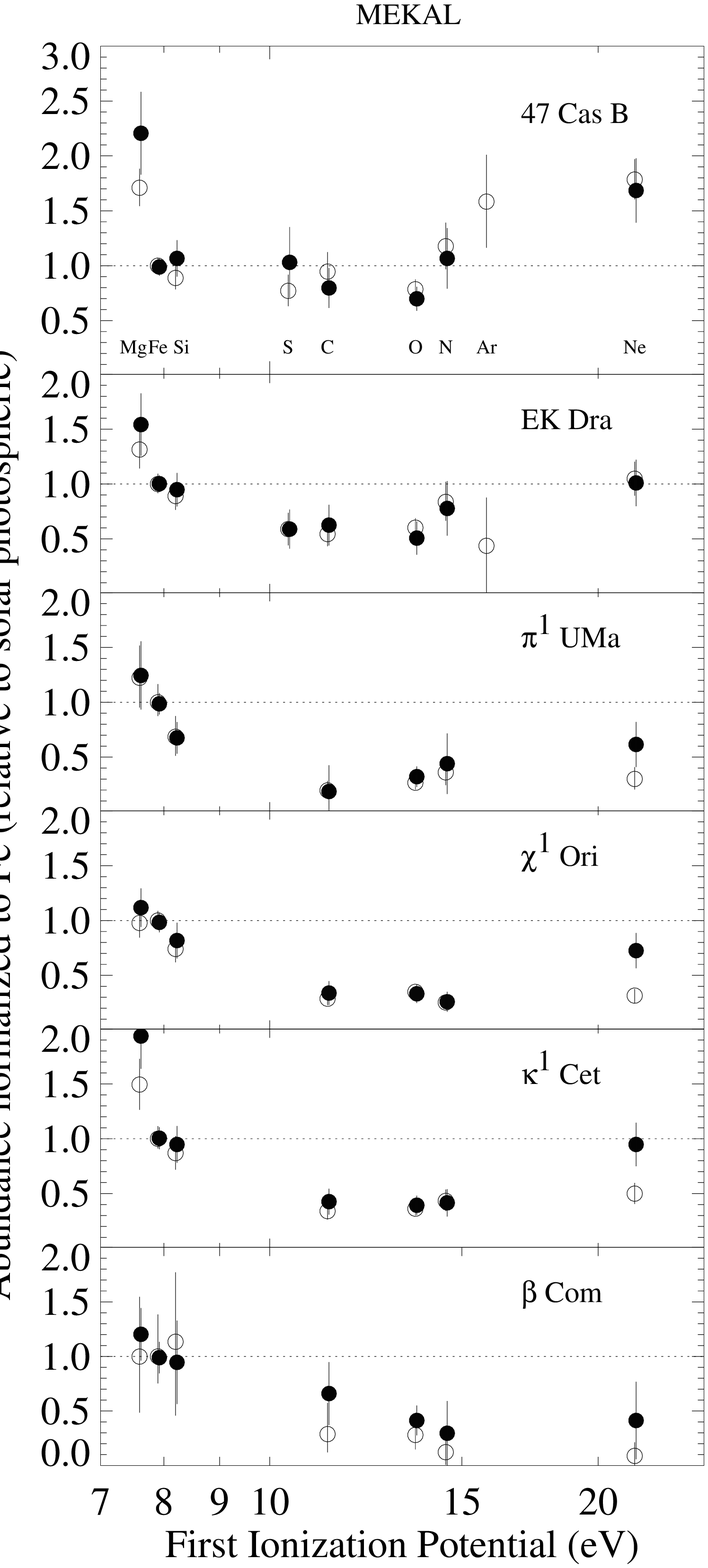}
\hspace{0.5cm}
\includegraphics[width=0.43\linewidth]{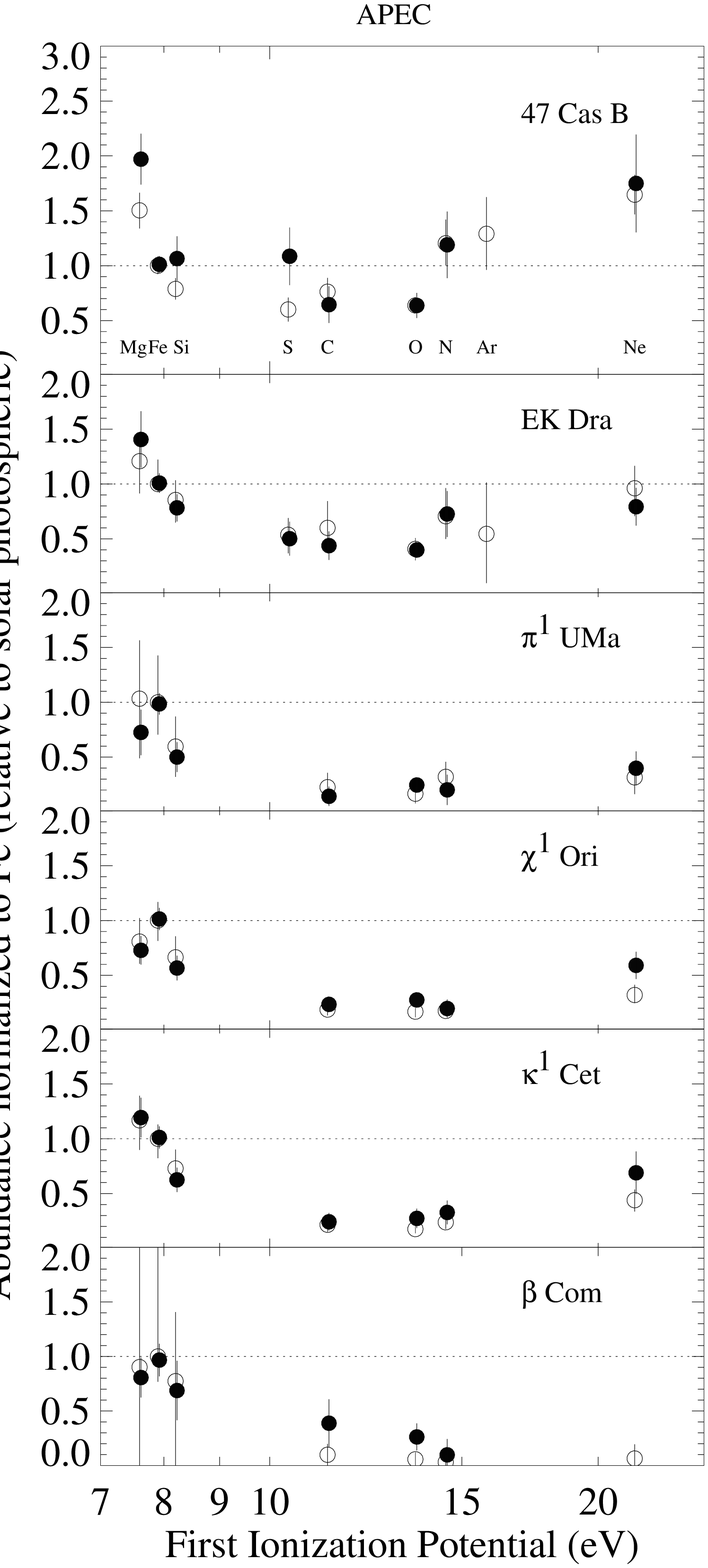}
}}
\end{center}
\caption{Abundances relative to Fe as a function of FIP, normalized to solar photospheric ratios \citep{anders89,grevesse99}. Open circles: method 1; filled circles: method 2. \label{abfig}}
\end{figure}

The abundances found with the different methods are listed in Table~\ref{abuntable} and plotted as a function of the FIP in Figure~\ref{abfig}.
We plot the abundance ratios $A$/Fe with respect to the solar photospheric ratios as a function of the FIP. The open circles 
represent the coronal abundances derived from method 1 whereas the filled circles show the abundances derived from method 2. We find a good agreement between the abundance sets, and an acceptable overall agreement between the results using the MEKAL database and the APEC database.

Nevertheless, some differences can be noted in Table~\ref{abuntable}. In the older stars $\chi^1$ Ori, $\kappa^1$ Cet, and $\beta$ Com, some systematic differences occur for the C/Fe abundance ratios, the abundances derived with APEC being smaller than those derived with SPEX although the error ranges are large. For the same stars, some differences are also present in the N/Fe abundance ratio. In both cases, the abundances are determined from only one faint line (C\,{\sc vi} and N\,{\sc vii}, respectively), making the measurements of the line fluxes difficult (see Tables~\ref{tab:lines} and~\ref{tab:lines_xspec}). 
Differences also occur in the four older stars for the Ne/Fe abundance ratio. This is 
partly due to the lack of strong and reliable (i.e., unblended) Ne lines. While 
in hot coronae, the strong Ne\,{\sc x} line serves as a reliable indicator for the
Ne abundance, this line is much weaker in cooler coronae and strongly blended with Fe
lines. The Ne\,{\sc ix} lines at 13.55--13.7~\AA\ are always blended with Fe lines, but
also become quite faint in the less active stars. Further, the inferred Ne flux in the line feature
depends on the absolute Fe abundance. The latter is poorly determined in particular in the
less active stars where almost no continuum is present. We emphasize that we cannot attribute the
discrepancy to any of the methods. The available data quality simply makes the determination
of the Ne/Fe abundance ratio in cooler coronae ambiguous. 

Fe blending leads to some differences in the Mg/Fe 
abundances as well (the largest deviations are for $\pi^1$ UMa using APEC method 2, and $\kappa^1$ Cet 
using SPEX method 2). Finally the weakness of the S lines in the more active stars leads to some 
differences in the S/Fe abundances.

The absolute Fe abundance was systematically higher when the APEC database was used. Note that in the spectrum of $\beta$ Com, the continuum is almost nonexistent and the derivation of the Fe abundance is difficult. For the latter target, the $\delta$-fit did not converge for the Ne line with APEC. For this reason, these two points are missing in Table~\ref{abuntable} and Figure~\ref{abfig}. 

However, we recognize many of these features to be due to limitations of the data and the reconstruction methods. On the other hand, these systematic differences are small compared to the general trends. As shown in 
Figure~\ref{abfig}, the abundances resulting from the two methods and the two databases agree mostly quite well within the errors, and the general trends are the same, regardless of the method used.

\subsection{Light curves}\label{lc}

\begin{figure}
\begin{center}
\centerline{\hbox{
\includegraphics[width=0.44\linewidth]{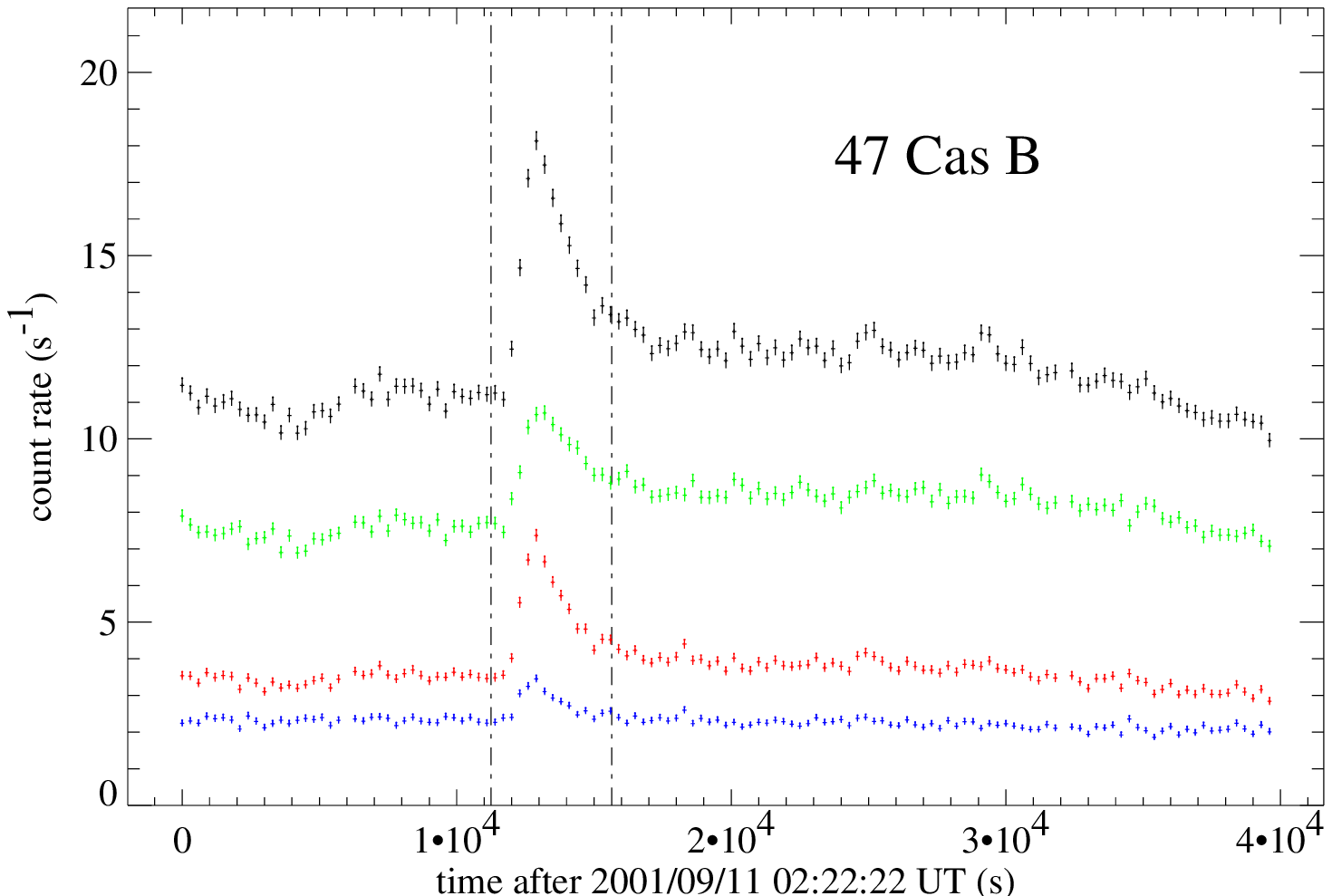}
\includegraphics[width=0.44\linewidth]{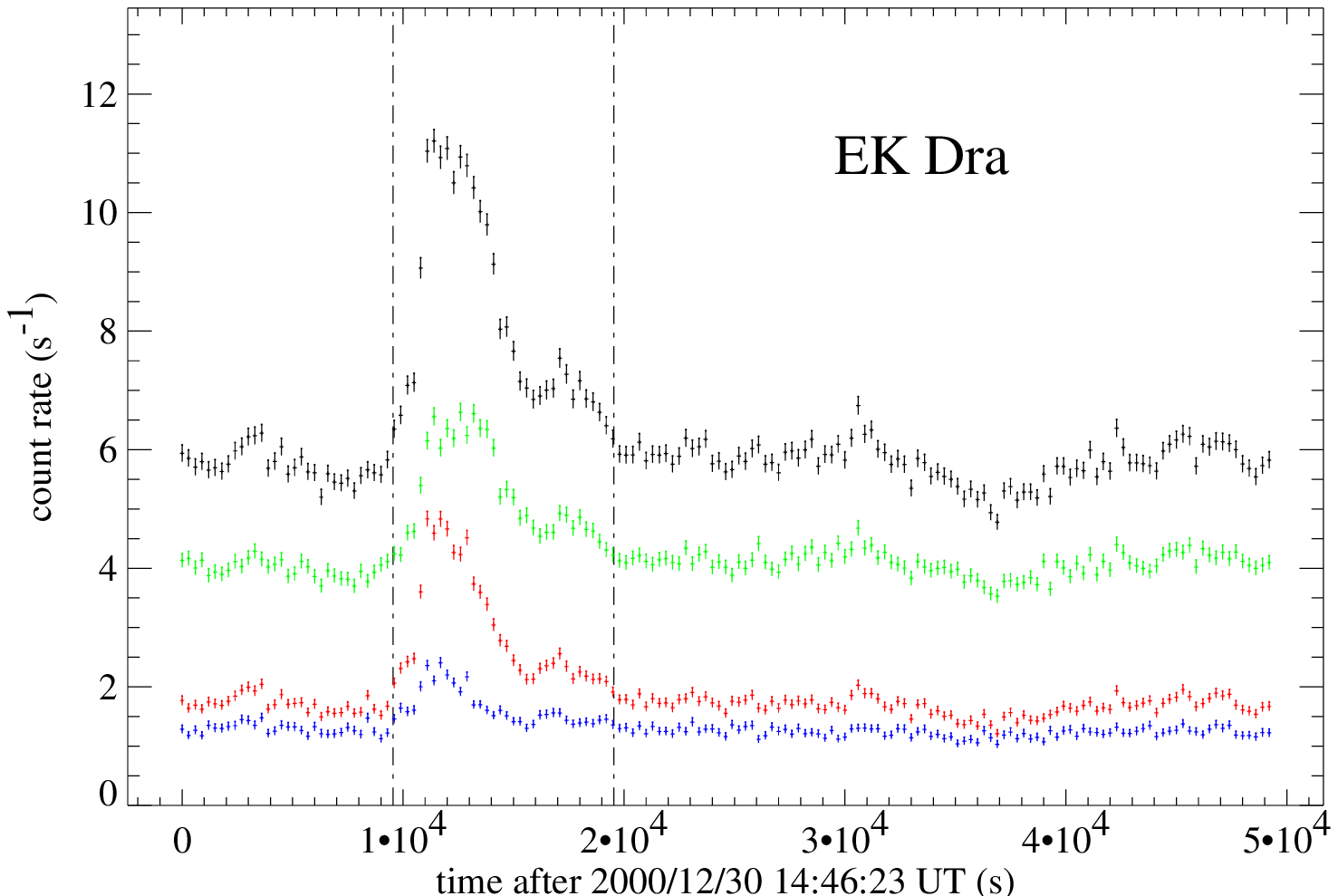}
}}
\centerline{\hbox{
\includegraphics[width=0.44\linewidth]{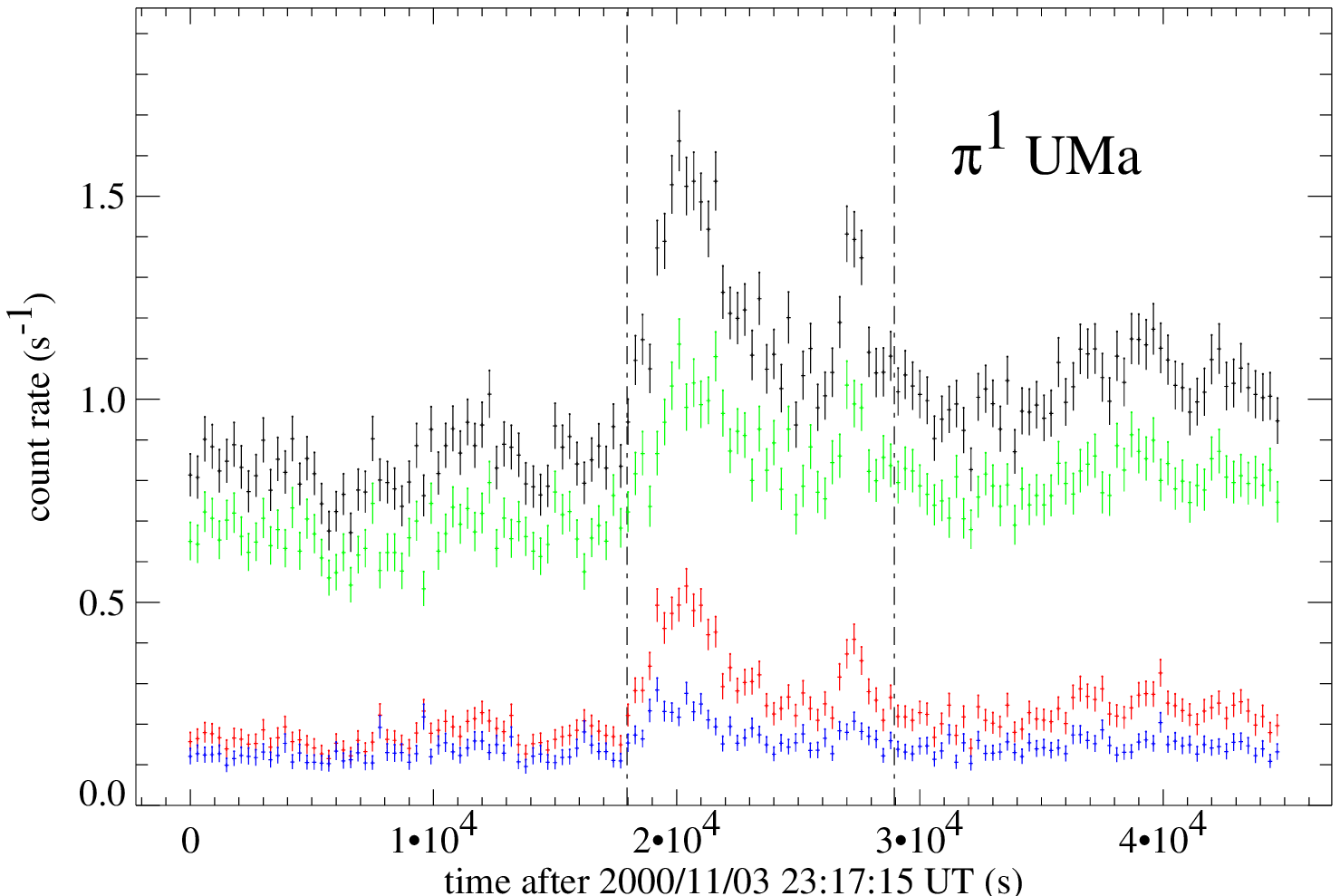}
\includegraphics[width=0.44\linewidth]{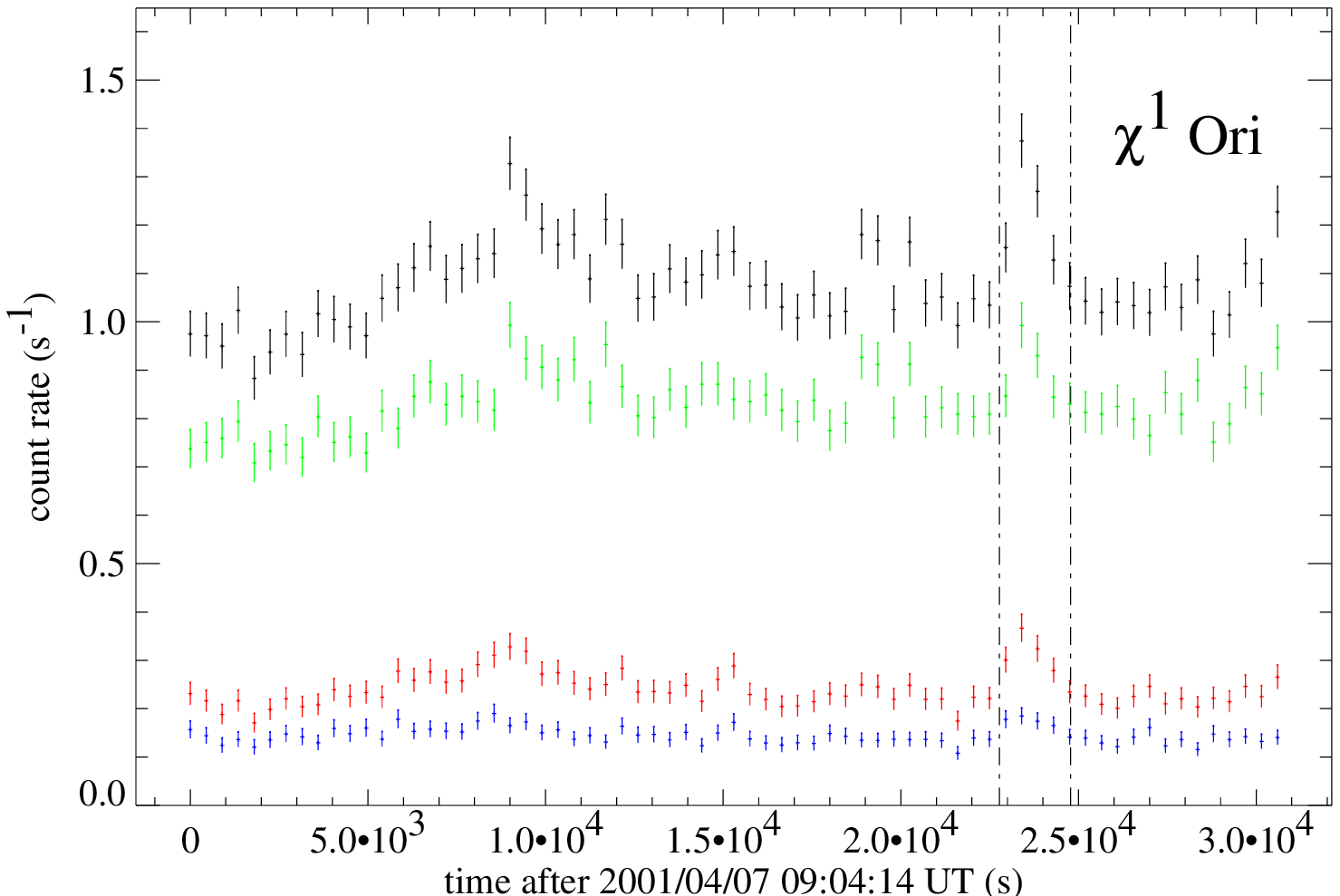}
}}
\centerline{\hbox{
\includegraphics[width=0.44\linewidth]{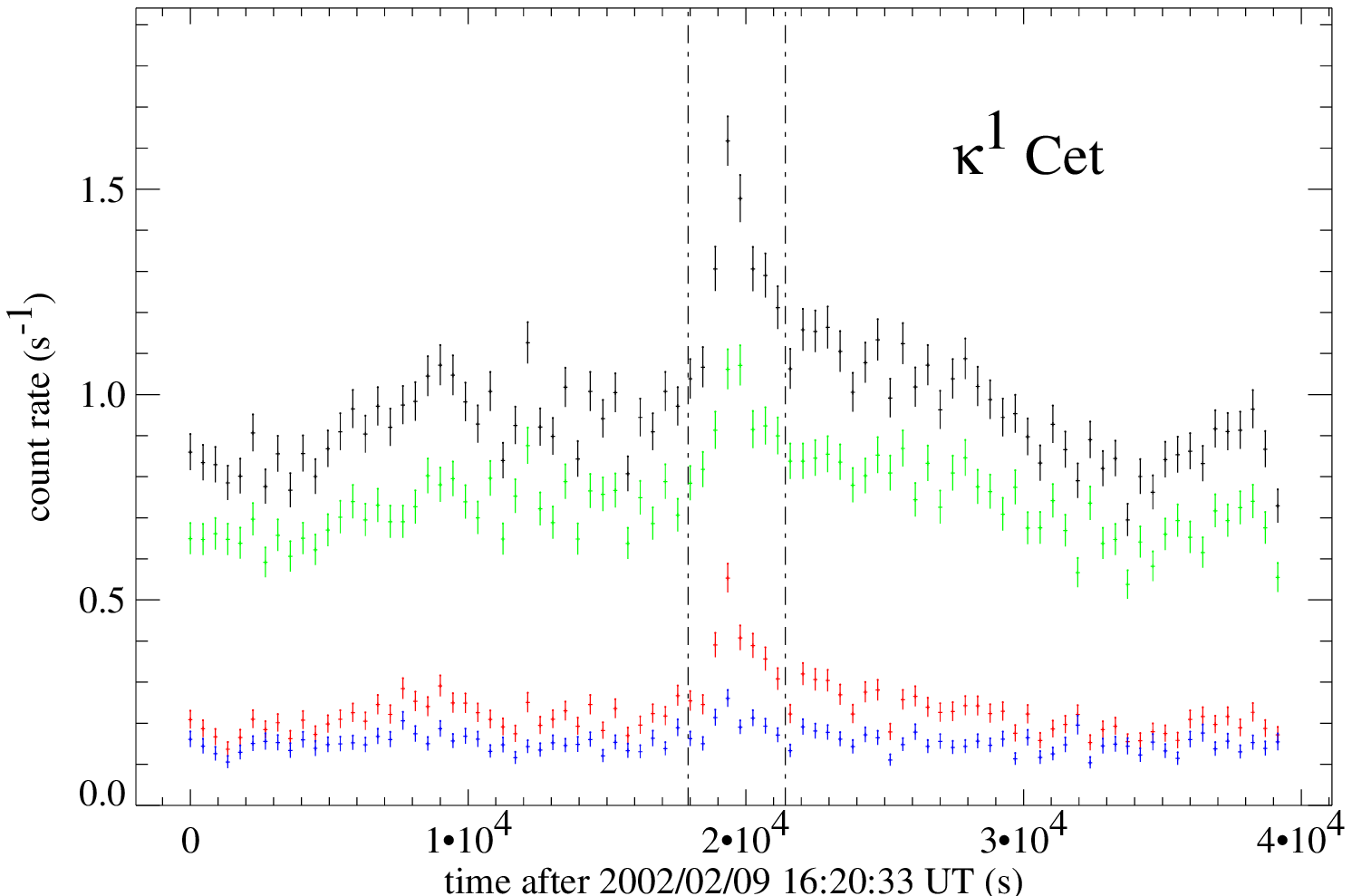}
\includegraphics[width=0.44\linewidth]{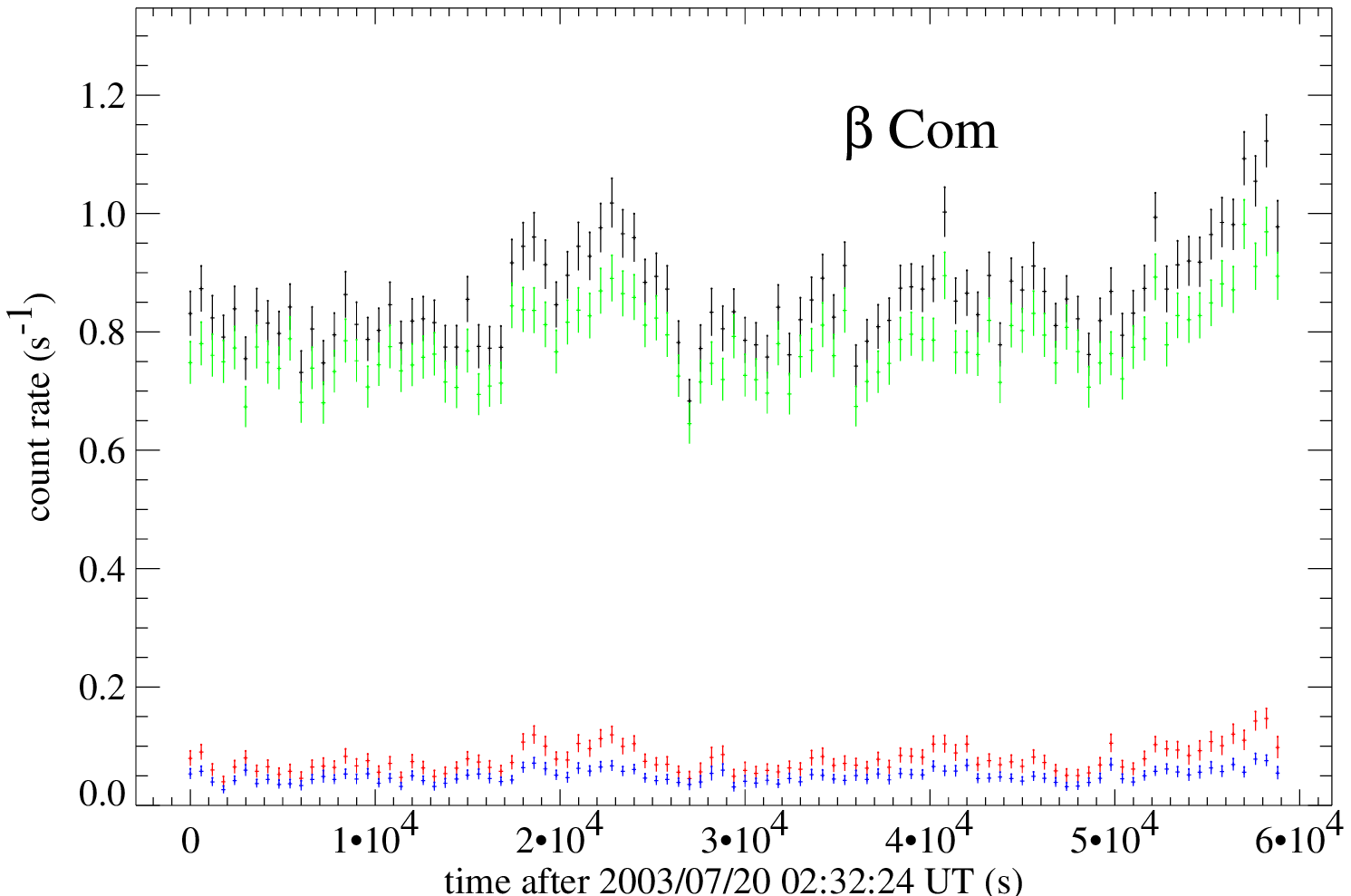}
}}
\end{center}
\caption{Light curves of our targets. The four light curves
in each panel show, from top to bottom, the total count rate in the $0.2-10$~keV
range (black), in the  soft  band ($0.2-1$~keV, green), in
the hard band ($>1$~keV, red, where the upper energy limit is
reported in the caption of Table~\ref{region}),  and the ratio of hard/soft (blue).
For illustration purposes, the hardness ratio has been multiplied by 5, 3,
0.5, 0.5,  0.5, and  0.5 for 47 Cas B, EK Dra, $\pi^1$ UMa, $\chi^1$ Ori,
$\kappa^1$~Cet, and $\beta$ Com, respectively. The bin size is,
for the stars as listed above, 300, 300, 300, 450, 450, and 600~s, respectively.
Only data from detectors that were operated in imaging mode were used, i.e.,
data from the PN camera were not used for $\pi^1$ UMa, $\chi^1$ Ori,
and $\kappa^1$~Cet. For the latter two stars, only one MOS camera was
available in imaging mode (see Table~\ref{log}). The time ranges of the
largest flares excluded from the spectral analysis are also shown by dash-dotted 
vertical lines.
}
\label{lightcurves}
\end{figure}

The light curves of the six stars are shown in Figure~\ref{lightcurves}. For each star, four light curves are plotted. 
They describe, from top to bottom, the total count rates in the  $0.2-10$~keV range (black), in the soft band 
($0.2-1$~keV, green), in the hard band ($>1$~keV, third plot), and the ratio 
between the hard and the soft count rates (blue). The upper energy thresholds used for the hard band vary from star to star and are listed in Table~\ref{region}. Only data from detectors that operated in imaging mode were considered (for 47 Cas B, EK Dra and $\beta$ Com, the data from MOS1, MOS2 and PN were used; for $\pi^1$ UMa, data from MOS1 and MOS2, and for $\chi^1$ Ori and $\kappa^1$ Cet, only data from one MOS camera, MOS2 and MOS1 respectively, were used).

The light curves display considerable variability. We observe the presence of large flares on all stars in our sample except $\beta$ Com. However, even after excluding these flares, the light curves still show considerable 
variability that cannot be described by steady, quiescent coronal emission. 
We also note that the hard emission becomes weaker toward older stars, in agreement with the decline of the average coronal temperature described earlier.

\section{Discussion}\label{discussion}

\subsection{Correlation between the Parameters}

The results of our analysis of six solar analogs clearly show a number of
trends that we wish to quantify below using our results from method 2. Before doing so, we note that
the X-ray results from the observation of $\kappa^1$ Cet are rather similar to
results from $\pi^1$ UMa and $\chi^1$ Ori, despite the former's significantly 
longer rotation period and, hence, higher age. The reason for this discrepancy is
not entirely clear, but we note that $\kappa^1$ Cet has a slightly later spectral type
and therefore a somewhat lower mass than the other targets. It is  known
that later-type stars evolve more slowly \citep{soderblom93} and that they remain
in a state of maximum X-ray luminosity (the saturation limit) for longer rotation periods
\citep{pizzolato03} than stars of earlier spectral type. Both effects make
$\kappa^1$ Cet look somewhat younger than inferred from a rotation-activity 
relation that is appropriate for early G stars.

\begin{figure}
\begin{center}
\includegraphics[width=0.95\linewidth]{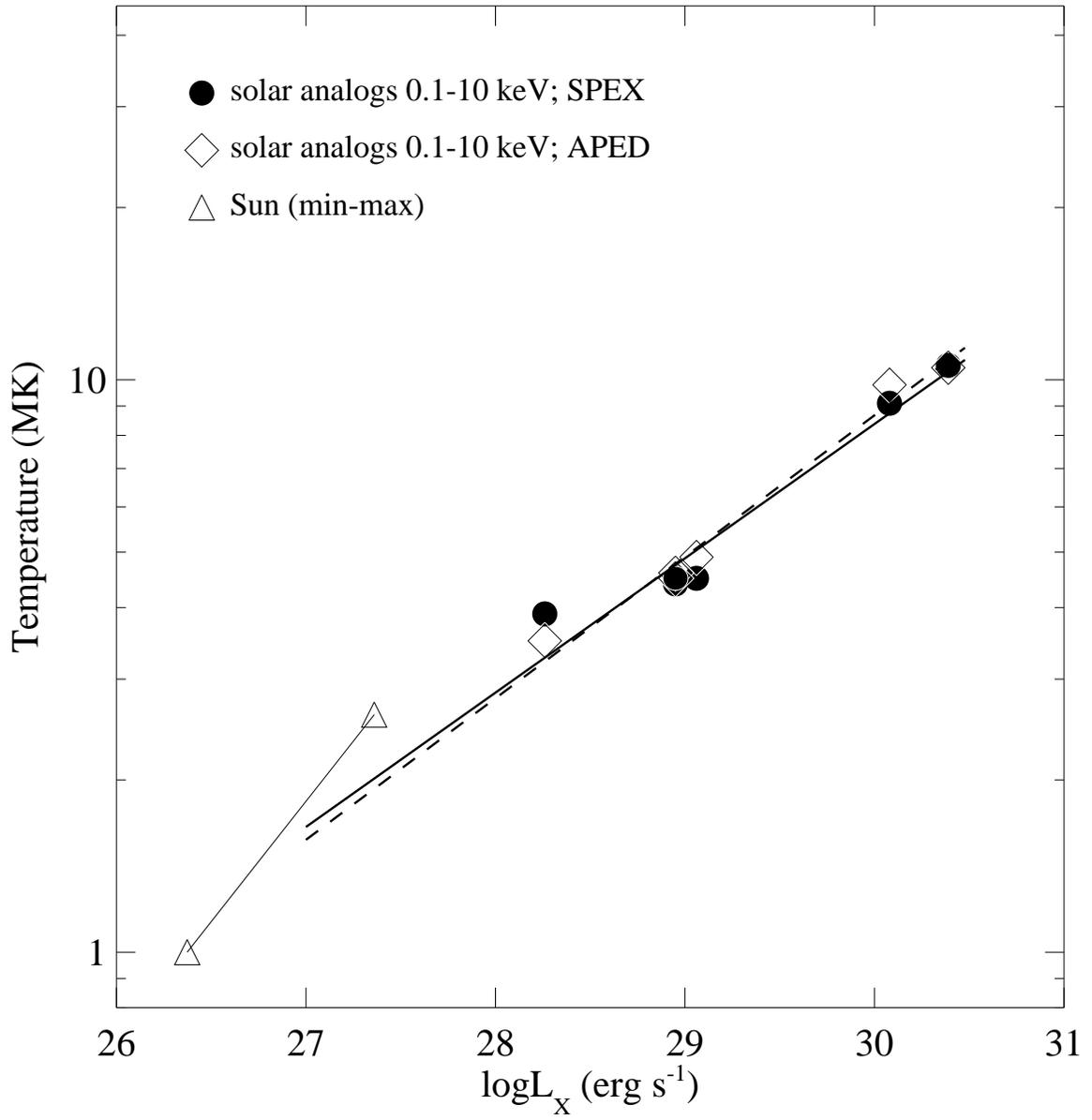}
\end{center}
\caption{Mean coronal temperature as a function of the X-ray luminosity. The dashed and solid lines are the regression fits to the results based on APEC and MEKAL, respectively. }
\label{templx}
\end{figure}

In Figure~\ref{templx}, we plot the mean coronal temperature, $\bar{T}$, (see Table~\ref{temp}) as a function of the total luminosity 
$L_\mathrm{X}$. We fitted the data with a power-law. Because both variables to be correlated are
likely to be affected by systematic scatter around any power-law, 
we use the ordinary least squares bisector method as described by \citet{isobe90}.
Clearly, the two parameters are correlated. We find, for our results from MEKAL and APEC, respectively, 
\begin{eqnarray}
L_{X} & \approx & 1.17 \times 10^{26} \bar{T}^{4.26 \pm 0.41} \rm \qquad erg\,s^{-1} ~  (MEKAL), \label{eq:lx_t}\\
L_{X} & \approx & 1.61 \times 10^{26} \bar{T}^{4.05 \pm 0.25} \rm \qquad erg\,s^{-1} ~   (APEC),  
\end{eqnarray}
where $\bar{T}$ is in MK. These relations are consistent with the results of \citet{guedel97a}, except 
that the latter authors used the higher temperature for a model with two thermal components fitted to {\it ROSAT} spectra.
Two points for the Sun at minimum and maximum activity level are also plotted for comparison (after \citealt{peres00}).  

\begin{figure}
\begin{center}
\includegraphics[width=0.70\linewidth]{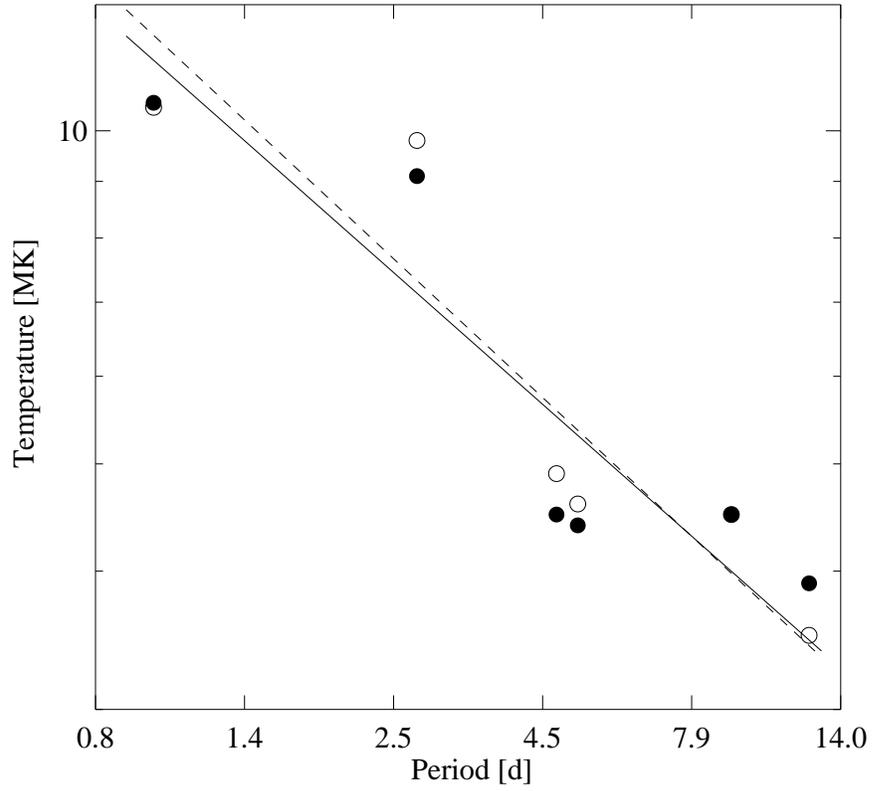}
\end{center}
\caption{Mean coronal temperature as a function of the stellar rotation period. Open circles refer to the APEC values, filled circles to the MEKAL values, both based on method 2. 
The dashed and solid lines are the regression fits to the APEC and MEKAL values, respectively.} 
\label{age_t}
\end{figure}

In Figure~\ref{age_t} the temperature as a function of the period is shown. Again, we fitted the data
with a power-law.   
In this case, the relations calculated from the two power-laws are given by 
\begin{eqnarray}
\bar{T} & \approx &11.6  P_{\rm rot}^{ -0.48\pm 0.07} \rm \qquad MK~  (MEKAL),  \\
\bar{T} & \approx &12.2  P_{\rm rot}^{ -0.50\pm 0.08} \rm \qquad MK~   (APEC), \label{eq:t_prot} 
\end{eqnarray}
where $P_{\rm rot}$ is the rotation period in days. Similar results were obtained for the {\it
ROSAT} data by \citet{guedel97a}, again considering the higher fit-temperature instead of 
the mean temperature. As we averaged the temperature with the EM in each bin used as weights,
the steady decrease of the mean temperature with period is consistent with a decrease
of the amount of hot plasma as the star spins down. 

Equations~\ref{eq:lx_t}--\ref{eq:t_prot} allow us to check for consistency with published relations between $L_{\rm X}$ and 
$P_{\rm rot}$. By combining formula (6) with (8) or, respectively,
(7) with (9), we find 
\begin{eqnarray}
L_{\rm X} & = & 4.01\times 10^{30} P_{\rm rot}^{-2.04\pm  0.36} \rm \quad  erg~s^{-1} ~ (MEKAL), \\
L_{\rm X} & = & 4.04\times 10^{30} P_{\rm rot}^{-2.03\pm  0.35} \rm \quad  erg~s^{-1} ~ (APEC),
\end{eqnarray}
which is consistent with previously reported dependences of this type
\citep{pallavicini81, guedel97a}. A linear regression for $\log L_{\rm X}$ and $\log P_{\rm rot}$ yields the
same result, with a power-law index of --2.03.

\begin{figure}
\begin{center}
\centerline{\hbox{
\includegraphics[width=0.70\linewidth]{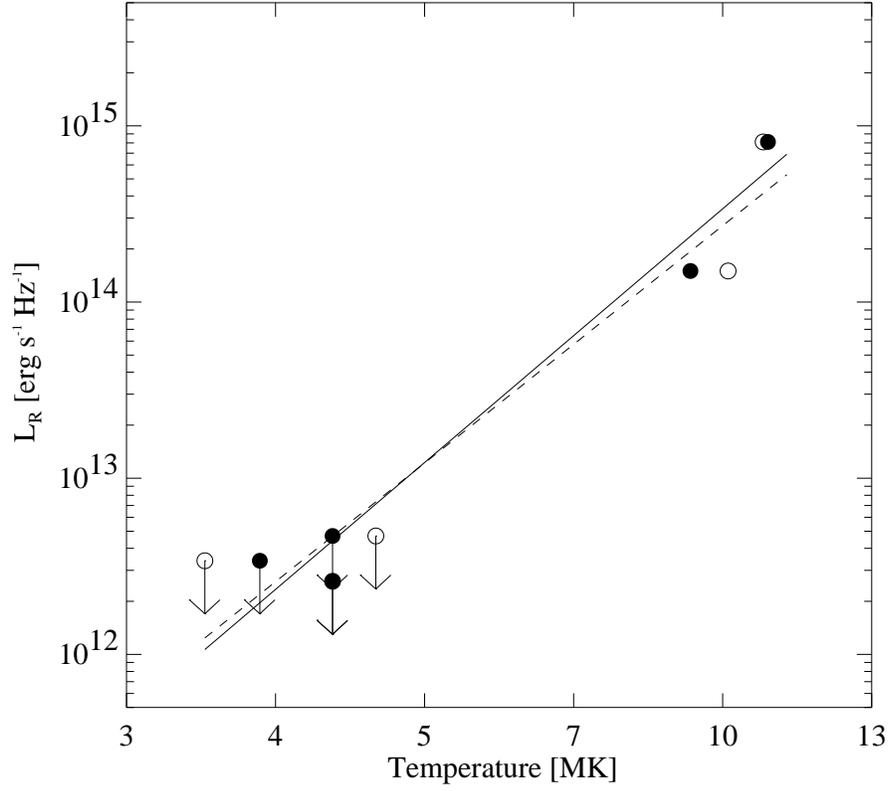}
}}
\end{center}
\caption{Radio luminosity as a function of the temperature. Filled circles give mean temperatures derived from the MEKAL database, while open circles are temperatures derived with APEC, both based on method 2. The luminosity values for the coolest stars $\pi^1$ UMa, $\kappa^1$ Cet and $\beta$ Com, displayed with an arrow, are upper limits.}
\label{t_lr}
\end{figure}

We also studied the relation between radio luminosity and the temperature. Radio luminosities or upper limits thereof are available \citep{guedel01b}
for five out of the six targets and are plotted in Figure~\ref{t_lr}. They refer to low emission levels outside obvious flares. We use the values of the upper limits in the regression analysis. The slopes of the power-laws are therefore lower limits, given by
\begin{eqnarray}
L_R &  \approx & 0.86 \times 10^{9}  \bar{T}^{5.65 \pm 0.46} \rm \qquad erg\,s^{-1}\,Hz^{-1}~ (MEKAL) \label{radio1}, \\
L_R &  \approx & 1.69 \times 10^{9}  \bar{T}^{5.29 \pm 0.74} \rm \qquad erg\,s^{-1}\,Hz^{-1}~ (APEC) \label{radio2}. 
\end{eqnarray}
These relations suggest a relation between the nonthermal electron population, responsible for radio gyrosynchrotron emission, and coronal heating.

\subsection{Abundances}

In the solar corona, the so-called FIP effect has been observed, 
in which the elements with a FIP lower than 10 eV are overabundant relative to the solar 
photospheric composition, whereas the elements with a higher FIP show the same abundance as 
the solar photosphere \citep{feldman92,laming95,feldman00}. Recent spectroscopic analysis with
{\it XMM-Newton} and {\it Chandra} has shown that in very active stars, an inverse effect
is present, in which the low-FIP elements are depleted relative to the high-FIP elements  
\citep{brinkman01}.
In our sample, we observe an evolutionary trend from an inverse FIP effect for the most active
star 47 Cas B to a solar-like FIP effect in the oldest stars (Figure~\ref{abfig}). 
We note, however, that the absolute abundances of low-FIP elements such as Fe do not
reach values as high as in the solar corona, where overabundances by factors of a few are common \citep{feldman92}.

\begin{figure}
\begin{center}
\centerline{\hbox{
\includegraphics[width=0.40\linewidth]{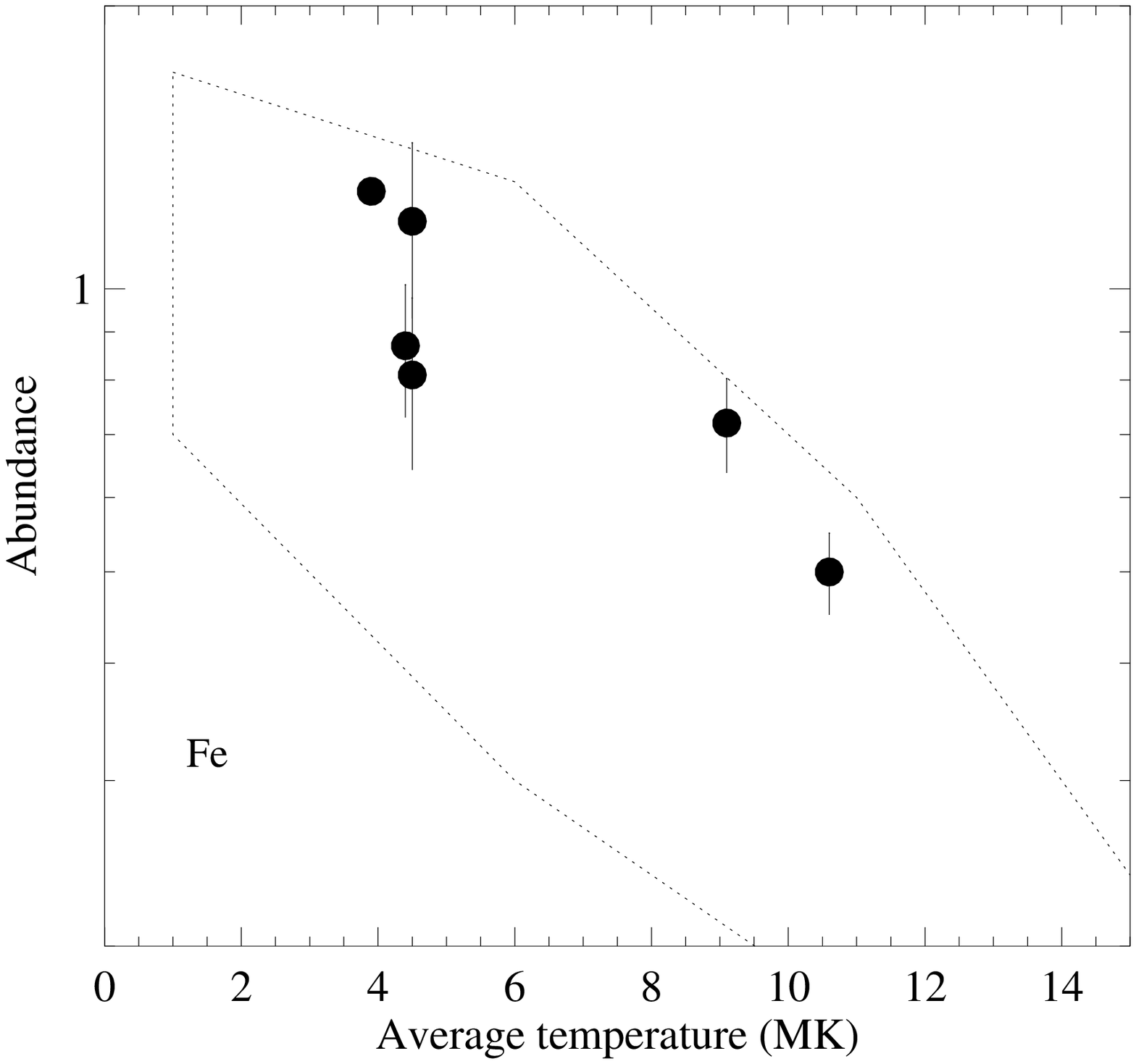}
\includegraphics[width=0.40\linewidth]{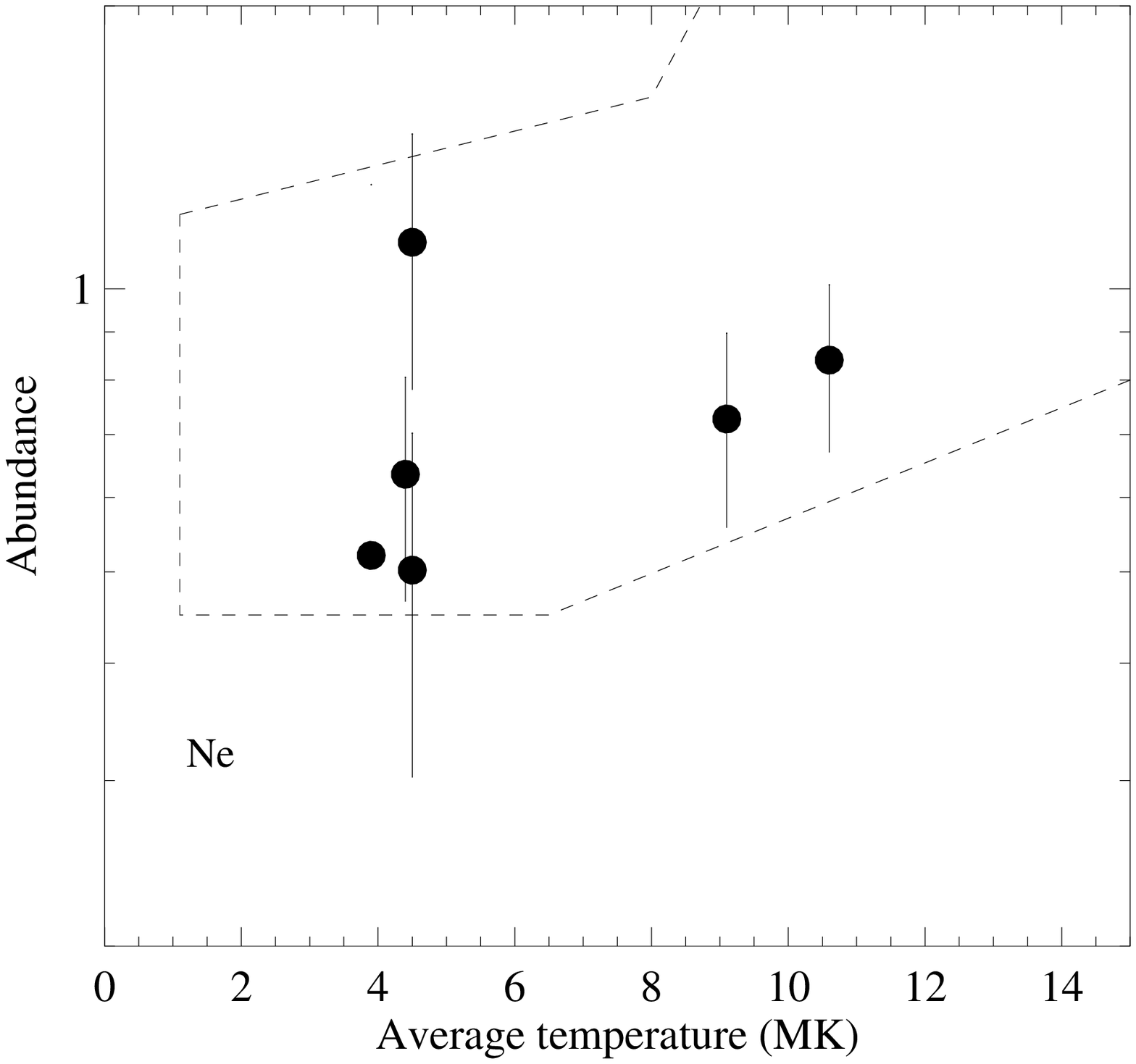}
}}
\centerline{\hbox{
\includegraphics[width=0.40\linewidth]{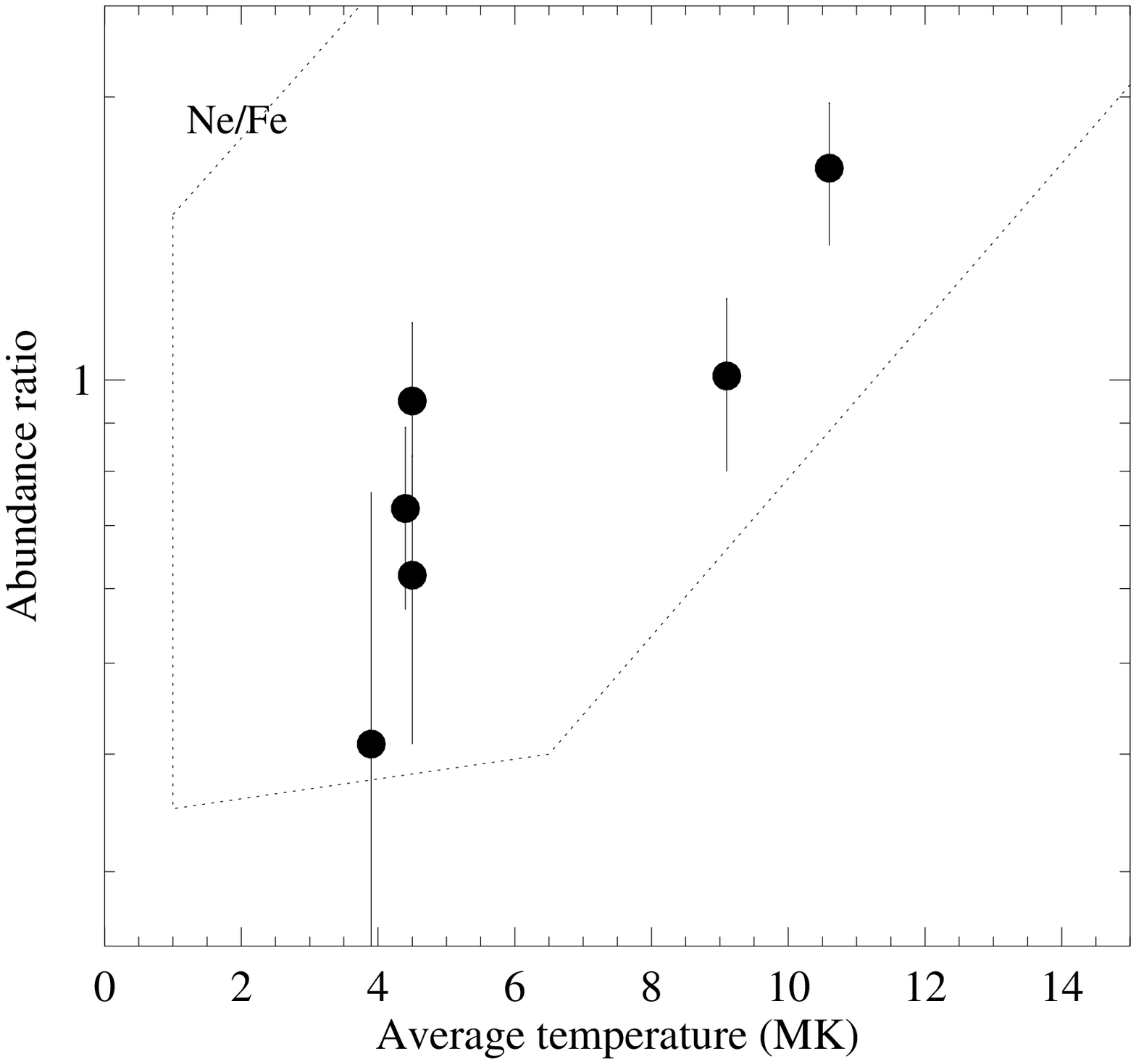}
\includegraphics[width=0.40\linewidth]{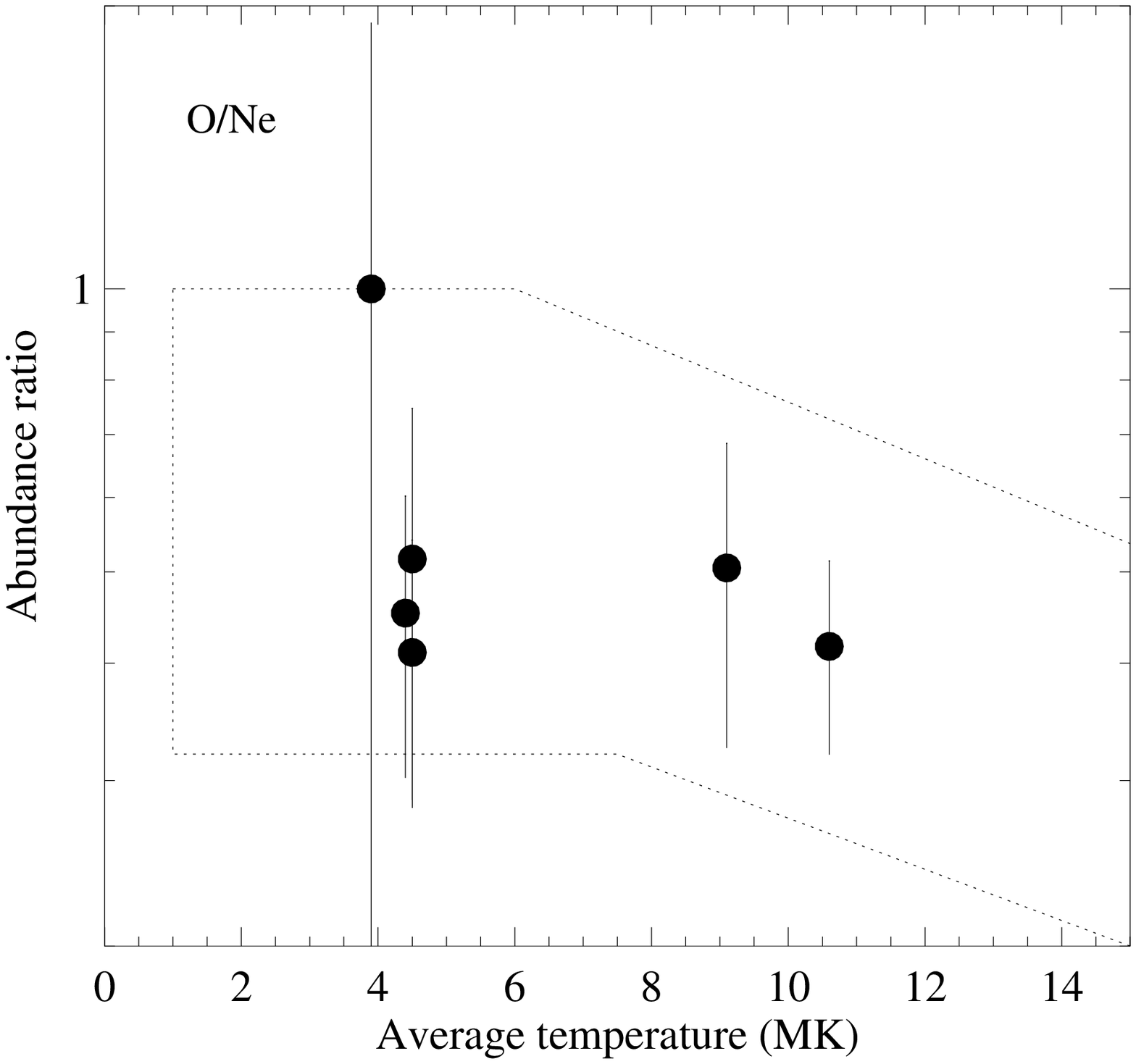}
}}
\centerline{\hbox{
\includegraphics[width=0.40\linewidth]{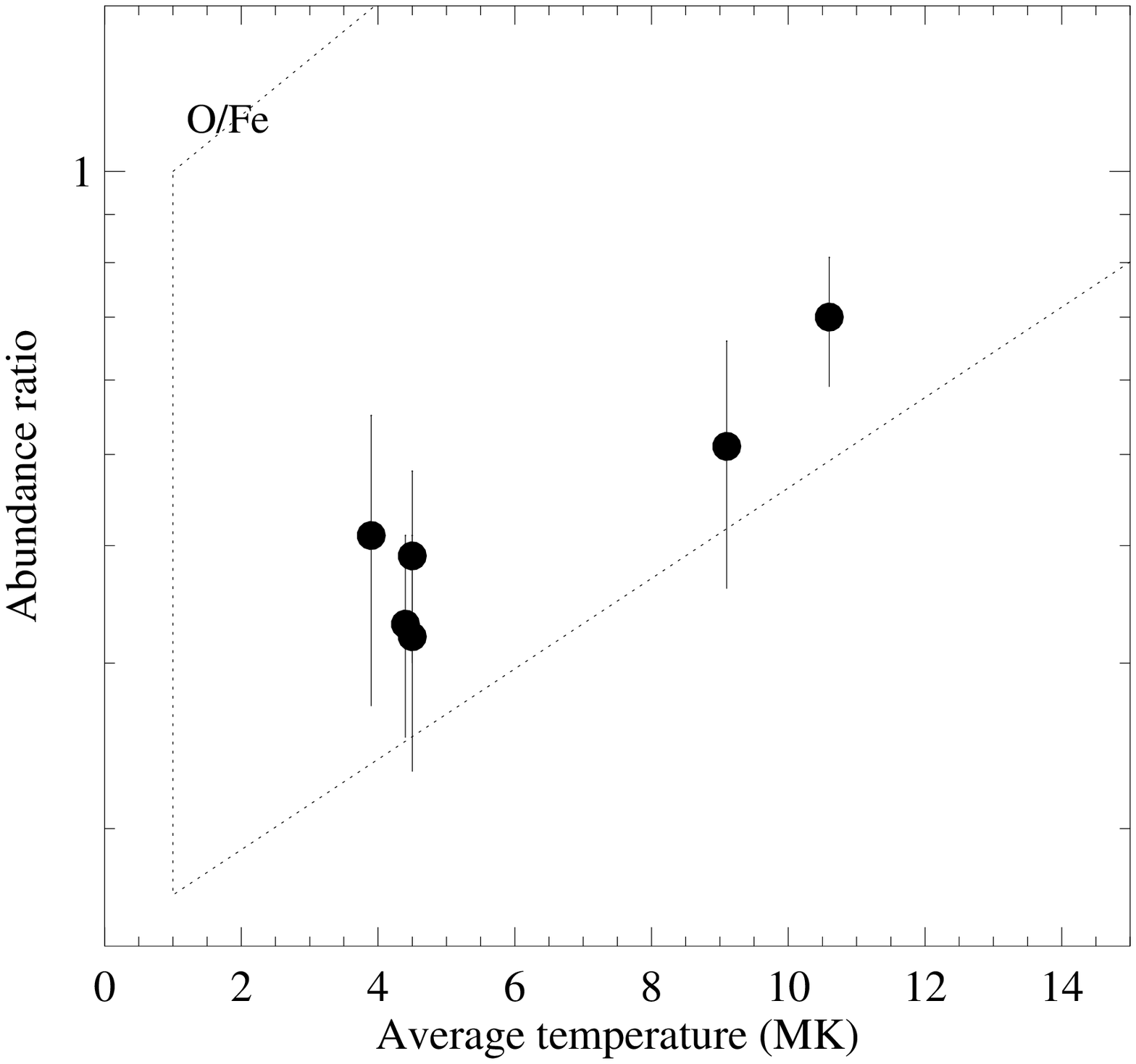}
\includegraphics[width=0.40\linewidth]{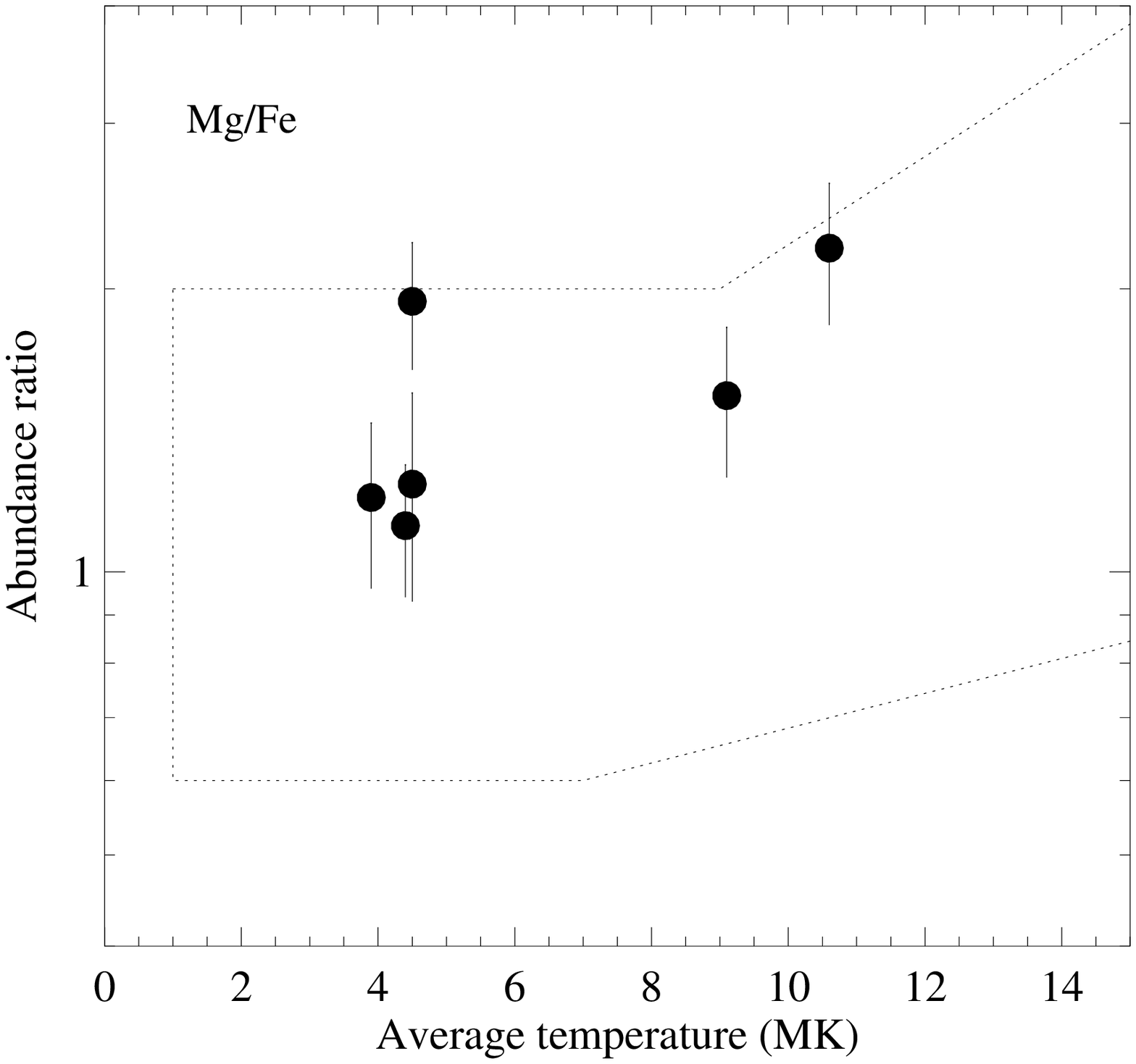}
}}
\end{center}
\caption{Abundances of Fe, Ne, and abundance ratios of Ne/Fe, O/Ne, O/Fe, and Mg/Fe are plotted as a function of the mean coronal temperature. Values from method 2 (MEKAL) have been
used. The dotted contours delimit the regions derived from a larger stellar sample \citep{guedel04}. Note that the displayed range of the abundances or abundance ratios is 1 dex in each of the six plots. \label{abt}}
\end{figure}

In Figure~\ref{abt}, the abundances of Fe, Ne, and the ratios $A$(Ne)/$A$(Fe), $A$(O)/$A$(Ne), $A$(O)/$A$(Fe), and $A$(Mg)/$A$(Fe) are plotted as a function of the
temperature (based on MEKAL/SPEX, method 2). The dotted regions include the ranges of a larger stellar sample \citep{guedel04}. 
The abundance of the low-FIP element Fe tends to decrease from a nearly photospheric value for stars with an average  
temperature of 3 to 5 MK to a lower abundance of $\approx$ 0.5 for the two more active stars EK Dra and 47 Cas B. 
The error bars for  the coolest star ($\beta$ Com) are not plotted for Fe and Ne, since they exceed the range illustrated 
in the figures. This is due to the near-absence of a continuum in this star, which makes absolute abundance determinations difficult.
The abundance ratios, however, are robust (see also \citealt{audard04}).

In the middle left and the bottom left panels, the abundance ratio of Ne/Fe and O/Fe, respectively, 
are shown as a function of the average coronal temperature. Because Ne and O are high-FIP elements, 
and Fe is a low-FIP element, the ratios increase with temperature, confirming the evolutionary trend of 
decreasing  low-FIP elements (such as Fe) with increasing activity. A similar
effect was found by \citet{audard03a} for a sample of active RS CVn-type binaries. This is also
confirmed by a larger sample of active stars \citep{guedel04}.

The plots in the middle right and bottom right panels show the abundance ratio of O/Ne, two high-FIP 
elements, and of Mg/Fe, two low-FIP elements. Our sample 
of stars is too small to constrain a trend in these plots. 
However, the larger star sample studied by \citet{guedel04} 
shows a nearly flat distribution for both ratios.

Could it be that the coronal abundance pattern reflects the composition of
the underlying photosphere? This view does not find support from other studies.
Although a full picture would include knowledge
of the abundances of other elements such as C, N, O, we see, from the summary in Sect.~\ref{sec:phcomp},
no support for a photospheric abundance pattern that significantly deviates 
from solar. Finally, we recall that the solar coronal composition does not reflect the
photospheric composition either, hence an agreement between photospheric and coronal
abundances is not a priori anticipated for solar analogs.

\subsection{Flares and Coronal Heating}\label{sec:flares}

In the previous sections, we have highlighted correlations
between observable parameters, and we have found continuous 
variability in all six targets to an extent that hardly any 
time interval is free of fluctuations. Although
conventional interpretation of coronal structure often makes use
of the approximation of static coronal loops (as, e.g., described
by \citealt{rosner78}), the interpretation of the phenomenology
revealed by our light curves cannot be rooted in strictly
static loops, although static loop models may, under certain
circumstances, serve as approximations even under flaring conditions 
\citep{jakimiec92}. 

There is appeal in the alternative, extreme model assuming that
the coronal emission is entirely due to dynamic, flaring loops. 
A number of observed features reported in this paper and in the previous
literature seem to support such a model: i) with the sensitivity
available with {\it XMM-Newton} and {\it Chandra},  X-ray emission 
previously ascribed to a quiescent component is now recognized
to be continuously variable; in the most extreme cases, no steady component
can reasonably be identified in the light curves \citep{audard03b}.
ii) More active stars (i.e., stars with a higher $L_{\rm X}/L_{\rm bol}$ 
ratio) appear to maintain hotter coronae. This is difficult
to explain with a model that assumes a corona composed only
of steady magnetic loops, with the principal determinant of
$L_{\rm X}$ being the magnetic filling factor, up to the
empirical saturation value of $\log L_{\rm X}/L_{\rm bol} \approx -3$.
Such models do not automatically explain why the coronal temperature
increases with increasing $L_{\rm X}$. We discuss this point further in
our conclusions below. iii) Active stars continuously produce radio emission from accelerated 
electrons. The lifetime of the latter is short, probably amounting to
no more than seconds to minutes \citep{kundu87}. In the solar corona,
flare energy-release processes are required to produce such electron
populations. In order to generate the observed stellar radio emission, high-energy 
electrons must be replenished frequently.

This question has previously been addressed by studying, for a model 
in which the corona is heated entirely by flares, the EMD
\citep{guedel97c,guedel03}, the average temperature to be ascribed to such a corona
\citep{audard00}, and the light curve characteristics expected from a
superposition of stochastic flares \citep{audard99, audard00, kashyap02,
guedel03, arzner04}. We are now in a position to apply the methodology
to our results.

First, we use the shapes of our EMDs to characterize the underlying flare
population in the framework of this model. \citet{guedel03} derived
an analytic expression for a DEM of a flare-heated corona under the 
assumption that the temperature and the flare density both decay 
exponentially with time constants $\tau_T$ and $\tau_n$, respectively,
and that a relation between flare peak-EM and peak-$T$
holds, EM$_p \propto T_{p}^b$, as reported by \citet{feldman95}. Then, the 
DEM follows power-law relations on both sides of its peak temperature ($T_m$), with
\begin{equation}\label{demslopes}
 Q(T) \propto \left\{ \begin{array}{ll} T^{2/\zeta}\quad\quad\quad\quad\quad\quad\quad\quad\quad\   &, T \le  T_m \\
                                     T^{-(b-\phi)(\alpha-2\beta)/(1-\beta) +2b - \phi} \quad       &, T \ge  T_m \end{array} \right.
\end{equation}
where $b \approx 4.3\pm 0.35$ as derived from a large sample of stellar 
flares \citep{guedel04}. The temperature $T_m$ depends on the energy of the
smallest flares participating in the heating in this simple model \citep{guedel03}.
Further, $\beta$ is a power-law
index for a relation between the flare e-folding decay time and its radiated
energy, $\tau\propto E^{\beta}$. As noted by \citet{guedel03}, $\beta$ 
is probably close to zero 
although an extreme case of $\beta = 0.25$ was also studied. The variable
$\phi$ gives the slope of the cooling function (radiative power per unit
emission measure) for a power-law approximation  in the
temperature  interval of interest.
We use $\phi = -0.3$ as an approximation of the slope of the cooling function
in the logarithmic temperature interval 6.8--7.5, i.e. the temperature interval above the 
EMD peak temperature (see 
Fig.~10 in \citealt{audard04}). The parameter 
$\alpha$ is of primary interest for us: It is the exponent of the distribution
of the occurrence rate $N$ of flares in radiated energy, viz.,
$dN/dE \propto E^{-\alpha}$ as found for a large sample of solar flares
(e.g., \citealt{crosby93}), but applicable also to stellar coronae (see references above).
In the first equation, applicable to the cooler portion of the DEM, 
$\zeta = \tau_n/\tau_T$. This parameter describes the amount of heating 
occurring during the flare decay, with $\zeta=2$ corresponding to free cooling
without heating, and $\zeta\approx 0.5$ corresponding to extreme heating rates
during the decay \citep{reale97}.

We have measured the slopes of our EMDs on both sides of $T_m$. To find the 
possible ranges for the best-fit slopes, we re-analyzed our data using
method 2 by starting with different initial conditions for the slopes in
our iteration, but the DEMs converged to similar values. The ranges
of the resulting best fit slopes (not considering the error ranges in the DEM)
are reported in Table~\ref{tab:lc}. We then used Eq.~\ref{demslopes} to determine the most likely
$\zeta$ and $\alpha$ with their acceptable ranges. The results are also reported
in Table~\ref{tab:lc}.

From the low-$T$ slopes of the EMDs derived with method 2 using SPEX, we find $\zeta$ to be around unity
in all cases. Such values are typical for
individual flares observed on active stars (see \citealt{guedel03} and
references therein) and support our assumption that such flares contribute
significantly to the overall observed emission. Static loops, on the other
hand, normally produce shallower DEMs, with slopes of +1 to +3/2, depending
on the amount of conductive flux at the loop footpoints \citep{rosner78,
vdoord97}. 

From the high-$T$ slope, we derive $\alpha \approx 2.2-2.8$, in excellent
agreement with $\alpha$ values determined from long light curves of 
active stars \citep{audard00, kashyap02, guedel03, arzner04}. If $\alpha
> 2$ and the emitted energy is integrated over the flare-rate distribution 
to obtain the total radiative loss, i.e., $L_{\rm tot} =\int_{E_1}^{E_2} E 
(dN/dE) dE$, then $L_{\rm tot}$ diverges as $E_1 \rightarrow 0$, i.e.,
the smallest flares dominate coronal heating, and a lower energy
cutoff is required for this power-law.

\begin{figure}
\begin{center}
\centerline{\hbox{
\includegraphics[width=0.45\linewidth]{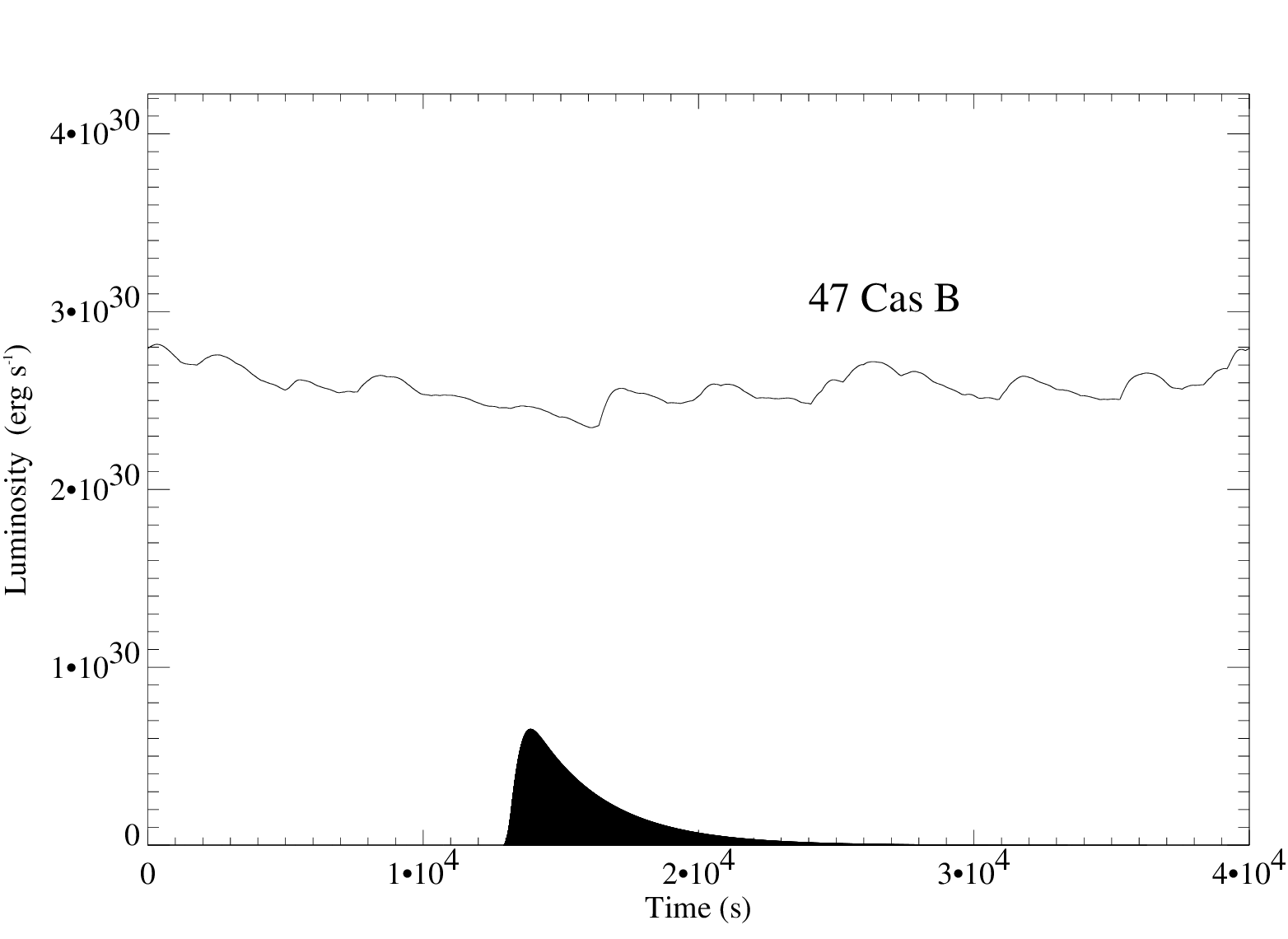}
\includegraphics[width=0.45\linewidth]{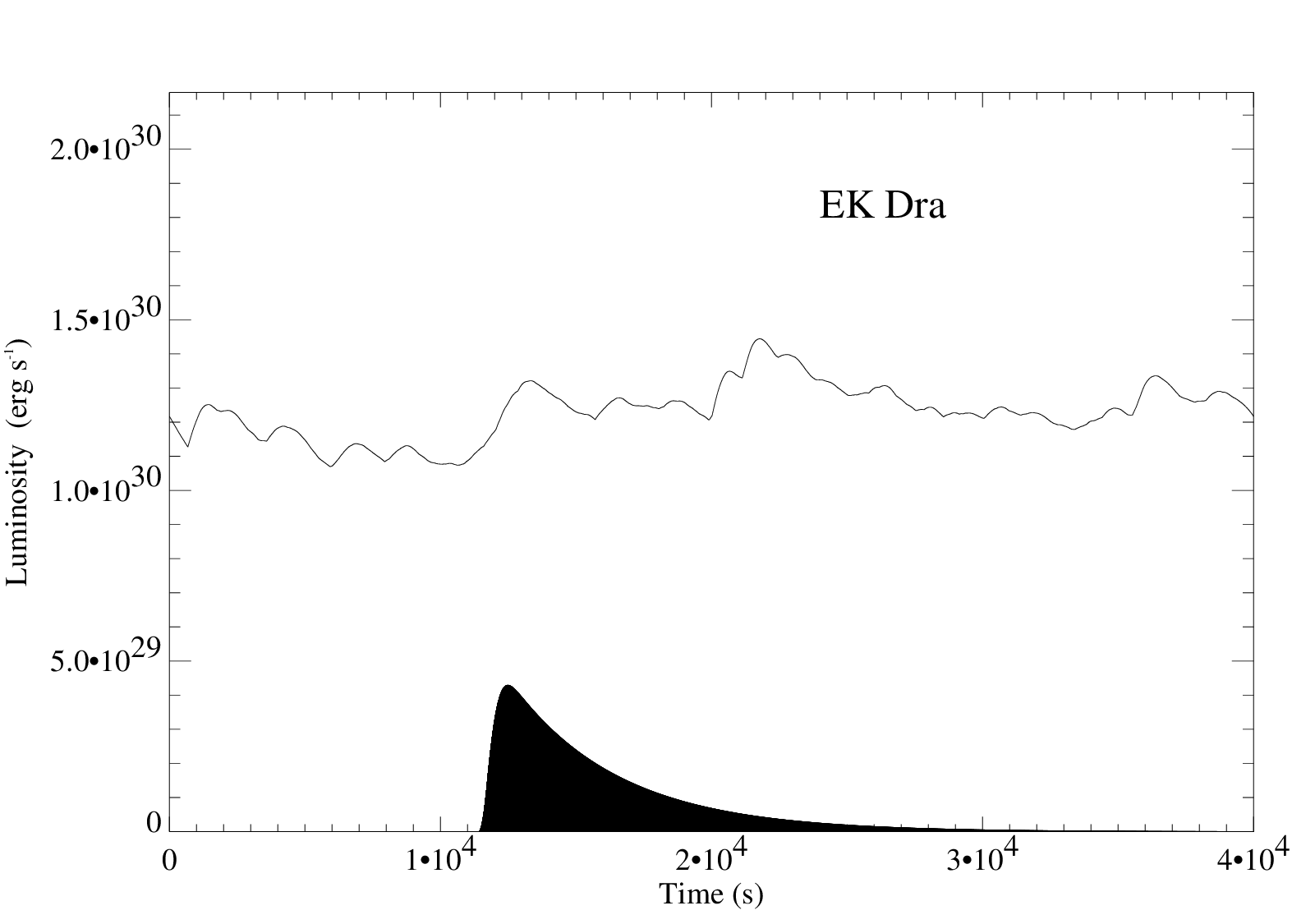}
}}
\centerline{\hbox{
\includegraphics[width=0.45\linewidth]{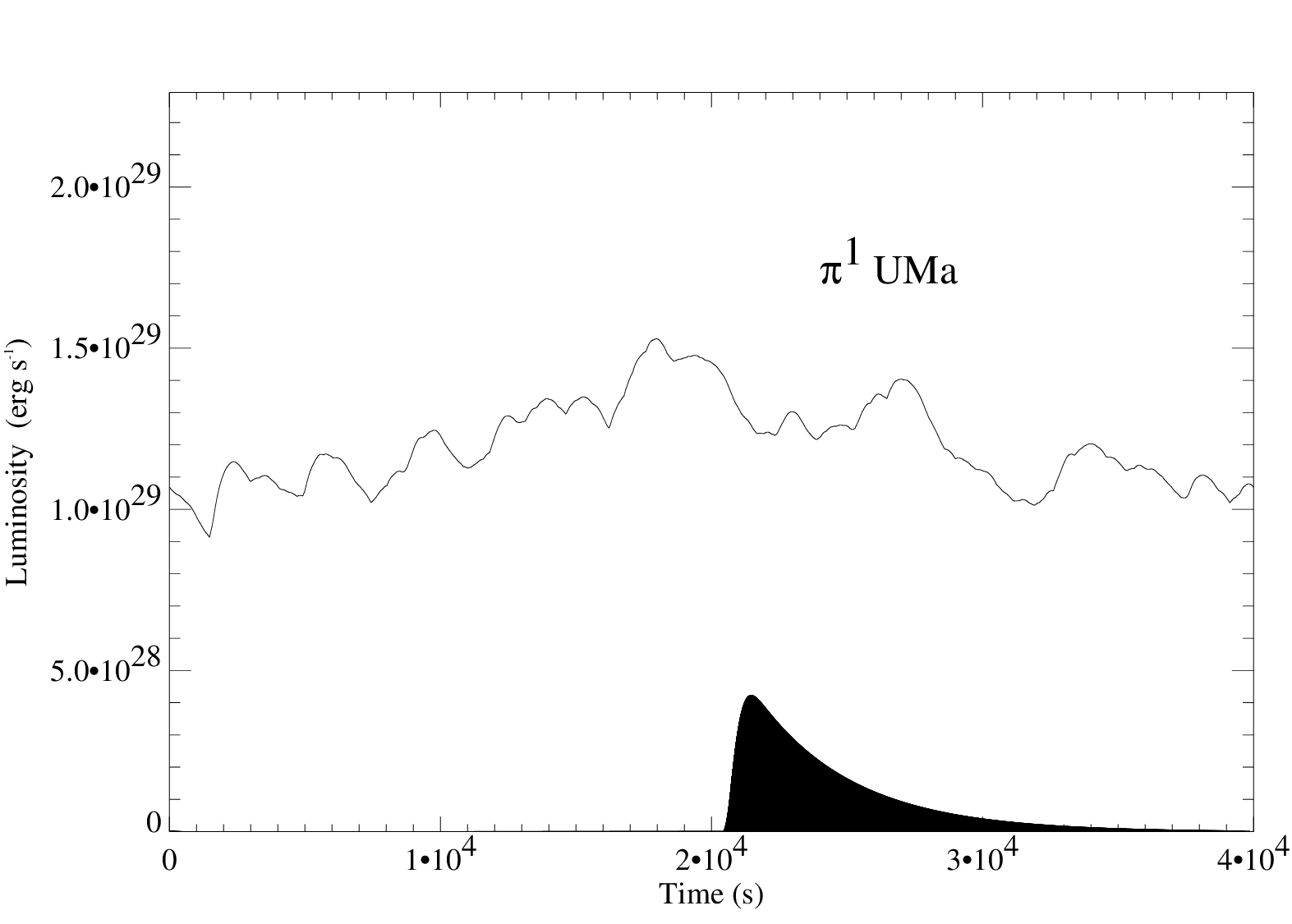}
}}
\end{center}
\caption{Simulated light curves obtained assuming a flare distribution that is based on the $\alpha$ values found from our EMDs. The maximum flare energy and the difference between largest  and smallest flares assumed here are given in Table~\ref{tab:lc}. The black shapes represent the largest flares actually used to synthesize the light curves from the simulations (and this flare may therefore be smaller than the largest flare actually observed in the light curves in Fig.~\ref{lightcurves}, which are not considered for this comparison). \label{simlight}}
\end{figure}

We now continue this consideration by simulating light curves based on 
a characteristic flare shape for the three most active stars, assuming a flare-rate distribution based
on $\alpha$ as determined above. The flare shape was derived from the 
convolution of an exponential function (cut off at t $<$ 0, describing the decay) and a Gaussian 
(important to describe the rise and the peak phase). The shapes of the
largest flares in the light curves were used to estimate the characteristic rise and 
the decay time parameters. We performed two
sets of simulations: one with $\beta$ = 0 and one with $\beta$ = 0.25. 
In each simulation $\alpha$ was chosen within the acceptable range for a given $\beta$.
The flare decay-time was varied according to the scaling 
$\tau \propto E^{\beta}$. 
For the largest flare in each light curve, we measured the amplitude and fitted the decay phase
with an exponential function to find the decay time, and hence the emitted energy.  
We then set the maximum flare energy $E_2$ 
equal to the energy of the largest observed flare 
and constructed a power-law distribution of flares in energy, down to a 
selectable minimum flare energy $E_1$. The flares were
randomly distributed in time (assuming a total of 40 ks), and their light curves
were superimposed. The rate of flares at a given energy, equivalent 
to the probability of the largest flares to occur within the 
simulation time, could be statistically varied.  
We measured the modulation depth, i.e., the ratio between
the root-mean-square scatter of the light curve and the average luminosity, and compared
it with the same measure for the observed light curves outside the outstanding largest one or two
flares. The minimum flare energy $E_1$ and the numbers of large flares occurring was varied until
the modulation depth and the average luminosity level agreed with the observed light curves. 
The modulation depth of the observed light curves is given in Table~\ref{tab:lc}.
It increases with decreasing activity. 
The simulated light curves, for $\beta$ = 0 and  $\alpha$ = 2.25, 2.28, 
and 2.54 for 47 Cas, EK Dra, and $\pi^1$ UMa respectively, are 
shown in Figure~\ref{simlight}. The range $\log E_2/E_1$ (in dex)
of flare energies thus required for the light curve is also given in Table~\ref{tab:lc}.

We see that typically 1.8$-$5.3 orders of magnitude of flare energies for $\beta$ = 0 and 
1.8$-$6.8 for $\beta$ = 0.25, respectively, are required
to ``describe'' our light curves. The interesting point is that
this range is larger for the more luminous stars.
This can be understood because a narrow range of flare energies produces
a more strongly modulated light curve, while a large number of small
flares merely adds a quasi-steady baseline level. However, in Equation~\ref{demslopes},
the turnover in the DEM is determined by the smallest flares participating 
in the statistics. One would thus expect that $E_1$ decreases with decreasing activity,
contrary to the results in Table~\ref{tab:lc}. However, we emphasize that our
light curve model is based on the extreme assumption that all emission
originates from flares. If there is a steady baseline level of X-ray emission 
not directly related to flare decays, as suggested from solar observations, we can obviously not determine $E_1$
from light curve analysis. Also, our model does not take into account any possible 
change in $\alpha$ at lower energies. We know from solar observations that there will be a continuation toward
smaller flares as well, rather than a lower energy threshold. The values in Table~\ref{tab:lc} are therefore
only indicative of the range of flare energies required in the most extreme model
discussed here.

\section{Conclusions}\label{conclusions}

As the rotation rate of a solar analog decreases during its evolution on the 
main sequence, the efficiency of the internal dynamo weakens, resulting
in a decrease of the magnetic activity in the stellar atmosphere. We have
studied systematic trends in the long-term evolution of stellar coronal
X-ray emission for ages in the range 0.1--2~Gyr. As a consequence of the 
stellar spin-down, the X-ray luminosity steadily decreases from levels
that may be close to the empirical saturation limit in the youngest stars
($L_X/L_{\rm bol} \approx 10^{-3}$), to levels approximately two orders of
magnitude lower within the first two Gyr. During the next $\approx$3--4~Gyr,
$L_X$ reduces by another factor of ten to levels as seen in $\alpha$ Cen,
$\beta$ Hyi, or the Sun \citep{guedel97a}. The overall trends reported
here confirm earlier studies based on lower-resolution X-ray spectroscopy
\citep{maggio87, guedel97a}. 

The high-resolution X-ray spectroscopy now available has permitted a more
detailed study of the composition and the thermal structure of solar-like 
coronae than was previously possible. We have studied emission measure
distributions and coronal element abundances for all six targets in a 
homogeneous way, applying two widely differing methods and using two different 
sets of atomic parameters. It is important to recall that both of our
methods include modeling of line blends to the extent possible with
the presently available line emissivities.

There is gratifying agreement between the results
from the two methods, although it appears that the choice of the atomic 
database introduces systematic differences. This is perhaps not surprising
as the presently available compilations of atomic parameters are incomplete
and suffer from systematic uncertainties. A more serious limitation is
set by the mathematical problem of spectral inversion itself.
While even counting statistics at the percent level makes the inversion
problem ill-conditioned \citep{craig76, judge02}, systematic uncertainties in the
line emissivities of perhaps up to 10--20\% may introduce various structure in our
EMDs that may not correspond to coronal features. We have
suggested that a reasonable convergence criterion should be set, although
it is difficult to assess at what level artificial structure is introduced
into a reconstructed EMD. Regardless of these inconsistencies, however, 
we recover element abundances that are rather robust.

With these limitations in mind, we have found systematic trends in the
EMD structure as a star ages. Not only does the total emission measure 
continuously decrease, the temperature where the EMD peaks also decays 
with time. The EM-weighted logarithmic average of the coronal temperature $\bar{T}$  
thus follows a  power-law dependence on the X-ray luminosity, namely 
$\bar{T} \propto L_X ^{0.25\pm0.02}$. 
A similar relation between the dominant coronal temperature and $L_{\rm X}$ was already studied by \citet{schrijver84},
using {\it Einstein} data of a large sample of stars. \citet{schmitt90} found for 1-$T$ model fits 
$T \propto L_X ^{0.4}$ from {\it Einstein} data. Finally, \citet{guedel97a} used
{\it ROSAT} data and a G-star sample similar to ours and found a
relation between the higher temperature of a 2-$T$ model and $L_{\rm X}$
in complete agreement with our results.

The observed trends cannot be explained by a model that is based exclusively
on different filling factors of the surface magnetic field, as an increased
filling factor does not explain per se why the temperature should increase.
We have discussed an extreme case of an alternative model in which a
statistical distribution of flares is responsible for the correlation
between $L_{\rm X}$ and $\bar{T}$. 
The question then is why the flare rate (above any given base-level energy)
is higher in more active (i.e., more rapidly rotating) stars. Our data cannot give 
a conclusive answer. One possibility is that more active stars indeed do
show a larger magnetic surface filling factor, and that the higher density of 
magnetic loops leads to more magnetic reconnection, thus producing a higher
flare rate \citep{guedel97a}; this, then, also includes a more prominent
population of very large flares  that produce both large emission measures and
very high temperatures, thus shifting the average temperature to higher 
values.  Once the magnetic filling dilutes,  the interactions  between neighboring
magnetic loop systems will become less frequent, and both $L_{\rm X}$ and $\bar{T}$
decrease (see \citealt{guedel97a}).
 
Alternatively, instead of increasing the surface filling factor, other ingredients
may systematically change with changing rotation period,
such as the structure of surface magnetic field in active regions or differences
in the convection pattern that jostles the magnetic-loop footpoints. In a statistical-flare
model, then, the way to explain the $L_{\rm X}-\bar{T}$ correlation would
be to reheat the same active regions more frequently in more active stars,
as some process brings non-potential energy into the magnetic fields at a higher rate.
As argued by \citet{audard00}, such frequently-heated loops can be  approximated by
static loops although the heating process is non-static. In this case, approximating the
loop temperature with its apex temperature (where most of the EM is found), the
\citet{rosner78} loop scaling law predicts $L_{\rm X} \propto T^{3.7}$ for loops of given length
below the coronal pressure scale height, and $L_{\rm X} \propto T^{4.7}$ for loops
larger than this limit (see \citealt{audard00} for details). These
predictions are close to our observational finding. On the other hand, static loops
would predict slopes of the DEM on the low-temperature side that are significantly smaller
than those determined by us.

To conclude, we are presently unable to distinguish between a model in which
the flare rate is controlled by the magnetic filling factor, and one in
which constrained active regions flare progressively more frequently as the rotation
rate of a star increases.  Both approaches, however, are compatible with
the hypothesis that much of the coronal heating is induced by flaring, regardless
of the ultimate cause of the increased flare rate in more active stars.

Because larger flares produce hotter plasma \citep{feldman95}, 
more active stars produce hotter coronae. This trend is unequivocally
recovered from our observations and further supports a picture in which
flares contribute significantly to the overall coronal heating (e.g., 
\citealt{audard00, kashyap02, guedel03}).

We note in passing that an alternative view with similar consequences has recently been 
presented by \citet{peres04}. These authors suggest that, based on properties of
coronal structures seen on the Sun, a higher occurrence of very compact, hot features 
including flares make more active coronae hotter.

We have also derived element abundances and found good agreement between
our two methods. Somewhat more systematic deviations can be noted if different
atomic databases are used (MEKAL, APEC), but the trends in the abundance
pattern agree and markedly change with changing activity level. A similar
trend was noted for RS CVn binaries by \citet{audard03a}, but the latter study
referred to extremely active stars in which the abundance pattern
changed from a strong inverse FIP effect to a flat distribution with decreasing
activity (or mean coronal temperature). In our sample, the change from an inverse or flat
distribution to a solar-like distribution occurs at ages of less than 300 Myr or
rotation periods longer than $\approx$ 3 days. Incidentally, a rapid decay of
nonthermal radio emission has been noted for the same activity range. We hypothesize
that the same electrons that are responsible for the observed gyrosynchrotron
emission also induce an inverse-FIP effect in the most active stars, as follows
(see \citealt{guedel02} for further arguments): if electrons are streaming along the
magnetic fields toward the chromosphere, they build up a downward-pointing electric 
field that acts to suppress positive currents from the chromosphere to the corona.
In other words, ions in the chromosphere are prevented from streaming into the corona,
while neutral, predominantly high-FIP elements, are not affected. As the electron
population diminishes in less active stars, the suppression of ion diffusion into
the corona disappears, and a solar-like FIP effect can build up, by whatever (still
unidentified) mechanisms. Recently, \citet{laming04} presented an alternative model
in which both the solar-like FIP and the inverse FIP effect are related to a common
plasma-physical cause.

\begin{acknowledgements}
The authors acknowledge helpful comments by the referee.
This research is based on observations obtained with \textit{XMM-Newton}, an ESA science
mission with instruments and contributions directly funded by
ESA Member States and the USA (NASA). AT and MG acknowledge support from the Swiss National 
Science Foundation (grant  20-66875) and from the Swiss Academy of
Natural Sciences.  MA acknowledges support from NASA to Columbia University for \textit{XMM-Newton} mission support and data analysis. SS acknowledges support from NASA/GSFC grant NAG5-13677. This research made use of the SIMBAD database, operated by CDS, Strasbourg.
\end{acknowledgements}

\vskip 02cm

{\it Note added in proof.} $-$ In a recent publication, Antia \& Basu (2005, ApJ, 620, L129)
suggested that a significant discrepancy between helioseismological
studies and modeling of the solar interior, brought about by
new measurements of a lower solar photospheric abundance of
oxygen, may be resolved by an upward revision of the solar
photospheric neon abundance. They therefore predict a solar
photospheric O/Ne abundance ratio lower than that adopted here. This may
explain the generally low value of the coronal O/Ne
abundance ratio observed in the coronae of stars, including
our sample stars (Fig. 13), regardless of activity level
(see also Fig. 37 in G\"udel 2004).

\clearpage

\begin{deluxetable}{llllllll}
\tabletypesize{\scriptsize}
\tablecaption{Program stars, including a comparison with the Sun \label{stars}}
\tablewidth{0pt}
\tablehead{ 
\colhead{Star}    & \colhead{Spec.}  & \colhead{Distance\tablenotemark{a}} & \colhead{$P_{\rm rot}$}      & \colhead{$\log L_X$\tablenotemark{b}} & \colhead{$\log L_X$\tablenotemark{c}} & \colhead{$\log L_X$\tablenotemark{d}} & \colhead{Age\tablenotemark{e}}             \\
 \colhead{}       & \colhead{type}   & \colhead{(pc)}         & \colhead{(d)}             & \colhead{(erg~s$^{-1}$)}  & \colhead{(erg~s$^{-1}$)}   & \colhead{(erg~s$^{-1}$)}  & \colhead{(Gyr)}}
\startdata
47 Cas B      & G0-2~V &  33.56 & $\approx$1.0       & 30.31    &30.35          &   30.39  &  0.1  \\
EK Dra        &  G0~V  &  33.94 &   2.75          & 29.93    &    30.06      &  30.08  &  0.1  \\
$\pi^1$ UMa   &  G1~V  &  14.27 &   4.7           & 29.10    &   29.05       &  29.06  &  0.3  \\
$\chi^1$ Ori  &  G1~V  &  8.66  &   5.1           & 28.99    &     28.95     &   28.95  &  0.3  \\
$\kappa^1$ Cet&  G5~V  &  9.16  &   9.2           & 28.79    &    28.94      &  28.95   &  0.75 \\
$\beta$ Com   &  G0~V  &  9.15  &   12.4          & 28.21    &     28.26     &  28.26  &  1.6  \\
Sun \tablenotemark{e}          &  G2~V  & $5\times 10^{-6}$ & 25.4 & 27.3     &     27.3      &  27.3    &  4.6  \\
\enddata
\tablenotetext{a}{stellar distances from Perryman et al. (1997)}
\tablenotetext{b}{determined from {\it ROSAT} in the 0.1--2.4~keV band \citep{guedel97a, guedel98a, guedel98b}}  
\tablenotetext{c}{determined from {\it XMM-Newton} in the 0.1--2.4~keV band (this work)}  
\tablenotetext{d}{determined from {\it XMM-Newton} in the 0.1--10~keV band (this work)}  
\tablenotetext{e}{from \citet{guedel97a, guedel98a, guedel98b}}  
\end{deluxetable}

\begin{deluxetable}{ccccccccc}
\tabletypesize{\scriptsize}
\tablecaption{Stellar photospheric abundances \label{tab:photab}}
\tablewidth{0pt}
\tablehead{ 
\colhead{Star} & \colhead{Fe}  & \colhead{Mg} & \colhead{Si}      & \colhead{S} & \colhead{C} & \colhead{O} & \colhead{N} & \colhead{Ref.}}
\startdata
EK Dra        &  $1.20$  &  $-$  & $-$    & $-$    &  $-$   & $-$    &  $-$   &  1  \\
$\pi^1$ UMa   &  0.83-1.02  &  0.65-0.83  &  0.78-1.12  &  $-$    &   $-$       &  $-$   &  $-$   & 2\\
              &  0.93~(0.81-1.07)  &  0.74~(0.58-0.95)  &  0.89~(0.83-0.95)  &  $-$    &  0.85~(0.68-1.07)  &  $-$   &  $-$   & 3\\
              &  $0.87$  & $-$    & $-$    & $-$     & $-$    & $-$    & $-$    & 1 \\
              &  $1.10$  & $-$    &  $-$   &  $-$    & $-$    & $-$    & $-$    & 4 \\
              &  0.83-0.98  &  $-$   & $-$      &  $-$    &  $-$   & $-$    & $-$    & 5 \\
              &  1.09~(1.00-1.19)  &  $-$   &  $-$   &  $-$     &  $-$   &  $-$   &  $-$   & 6 \\
$\chi^1$ Ori  &  0.89-0.93   &  $0.91$  &  $0.98$   &  $-$  &  $-$      & $-$   & $-$   & 7   \\
              &  $0.91$   &  $-$   &  $-$    & $-$   &  $-$   &  $-$  & $-$   & 8   \\
              &  $\approx 1$   &  $-$   &  $-$    & $-$   &  $0.63$   &  $-$  & $-$   & 9   \\
              &  1.14~(1.07-1.22)   &  $-$   &   $-$    & $-$   &  $-$    &  $-$  &  $-$  & 6   \\
              &  $1.35$   &  $-$   &  $-$    & $-$   &  $-$    & $-$   &  $-$  & 1   \\
              &  0.66-1.29   &  $-$     &    $-$  &  $-$  &   $-$   & $-$   &  $-$  & 5   \\
$\kappa^1$ Cet&  0.89-1.04  &  0.85-0.98  &   0.95-1.07  & $-$   &  $-$ &  $-$  & $-$  &   2 \\
              &  1.29~(1.17-1.41)  &  0.91~(0.76-1.10)  &   0.85~(0.78-0.93)  &  $-$  & 0.91~(0.71-1.17)   & $-$   & $-$  &   3 \\
              &  $1.0$  &  $-$   &  $-$    &  $-$  &  $-$   &  $-$  & $-$  &   7 \\
              &  $1.13$  &  $-$   &   $-$   & $-$   &  $-$   &  $-$  & $-$  &   6 \\
              &  $1.66$  &  $-$   &  $-$    & $-$   &  $-$   &  $-$  & $-$  &   1\\
              &  0.98-1.10  &  $-$   &  $-$      &  $-$  &   $-$  & $-$   & $-$  &   5 \\
$\beta$ Com   &  $0.93$  &  $-$   &  $-$   & $1.38$   &  $0.98$  &  1.26~(1.05-1.47) & $1.05$  &  10 \\
              &  1.00-1.07  & $1.17$   &  $1.00$   &  $-$   &  $-$   & $-$   &  $-$  &  7  \\
              &  1.15~(1.07-1.25)   &  $-$   &  $-$     &  $-$   &  $-$   &  $-$  &  $-$  &  6  \\
              &  $1.07$  &  $-$   &  $-$    &  $-$   &  $-$   &  $-$  & $-$   &  1  \\
              &  $1.00$  &  $-$   &    $-$  &  $-$   & $-$    &  $-$  &  $-$  &  8  \\
              &  $1.17$  &  $-$   &   $-$   &  $-$   &  $-$   &  $-$  &  $-$  &  4  \\
              &  0.89-1.17  &  $-$   &   $-$   &  $-$   &  $-$   &  $-$  &  $-$  &  5  \\
\enddata
\tablecomments{Values refer to the solar photospheric composition. If available, error ranges are given in parentheses.}

\tablerefs{ 
(1)\citet{rochapinto04};
(2) \citet{ottmann98}; the original values were transformed to solar
                  abundances as given by \citealt{anders89} = AG89 except for Fe for which we use
                  the value given in \citealt{grevesse99} = GS99;  
(3) \citet{gaidos02};  
(4) \citet{gray01} ; 
(5) \citet{cayrel01};  
(6) \citet{taylor03}, corrected to GS99; 
(7) \citet{edvardsson93};  
(8) \citet{gratton96};  
(9) \citet{tomkin95}, corrected to AG89 and GS99;  
(10) \citet{clegg81}.} 
\end{deluxetable}

\begin{deluxetable}{lllllll}
\tabletypesize{\scriptsize}
\tablecaption{Observation log \label{log}}
\tablewidth{0pt}
\tablehead{
\colhead{Star}&\colhead{Instruments}&\colhead{Filter}&\colhead{Start}&\colhead{Stop}&\colhead{Exposure [s]\tablenotemark{a}}}       
\startdata
47 Cas B        & RGS 1 & -     & 2001/09/10 23:30:24 & 2001/09/11 13:39:00 & 36610     \\
                & RGS 2 & -     & 2001/09/10 23:30:24 & 2001/09/11 13:39:01 & 36610     \\
                & MOS 1 & Thick & 2001/09/11 02:04:32 & 2001/09/11 13:29:30 & 36610     \\
EK Dra          & RGS 1 & -     & 2000/12/30 14:01:58 & 2000/12/31 05:17:04 & 41960   \\
                & RGS 2 & -     & 2000/12/30 14:01:58 & 2000/12/31 05:17:04 & 41960 \\
                & MOS 2 & Thick & 2000/12/30 14:10:36 & 2000/12/31 04:38:02 & 41960 \\
$\pi^1$ UMa     & RGS 1 & -     & 2000/11/03 21:44:48 & 2000/11/03 12:28:18 & 38800   \\
                & RGS 2 & -     & 2000/11/03 21:44:48 & 2000/11/03 12:28:19 & 38800 \\
                & MOS 1 & Thick & 2000/11/03 21:53:16 & 2000/11/03 11:49:04 & 38800 \\
$\chi^1$ Ori    & RGS 1 & -     & 2001/04/07 08:56:49 & 2001/04/07 22:31:53 & 29326 \\
                & RGS 2 & -     & 2001/04/07 08:56:49 & 2001/04/07 22:31:59 & 29326 \\
                & MOS 2 & Thick & 2001/04/07 09:03:11 & 2001/04/07 17:45:10 & 29326  \\
$\kappa^1$ Cet  & RGS 1 & -     & 2002/02/09 16:13:01 & 2002/02/10 03:21:41 & 35920     \\
                & RGS 2 & -     & 2002/02/09 16:13:01 & 2002/02/10 03:21:35 & 35920    \\
                & MOS 1 & Thick & 2002/02/09 16:19:33 & 2002/02/10 03:18:09 & 35920    \\
$\beta$ Com     & RGS 1 & -     & 2003/07/20 02:08:16 & 2003/07/20 20:13:28 & 61320 \\
                & RGS 2 & -     & 2003/07/20 02:08:16 & 2003/07/20 20:13:34 & 61320 \\
                & MOS 2 & Thick & 2003/07/20 02:09:06 & 2003/07/20 19:12:36 & 61320    \\
\enddata
\tablenotetext{a}{Exposure time used for the analysis, in seconds (excluding flares)}  
\end{deluxetable}

\begin{deluxetable}{lc}
\tablecaption{Spectral wavelength ranges used for method 1 \label{region}}
\tabletypesize{\scriptsize}
\tablewidth{0pt}
\tablehead{
\colhead{Instrument} & \colhead{$\lambda $ range (\AA)}}           
\startdata
RGS & $8.30-9.50$        \\
RGS & $12.00-13.95$   \\
RGS & $14.15-15.90$\\
RGS & $16.20-17.15$\\
RGS & $17.80-18.30$\\
RGS & $18.75-19.20$\\
RGS & $20.80-21.10$\\
RGS & $21.40-22.40$\\
RGS & $23.65-24.00$\\
RGS & $24.50-24.90$\\
RGS & $28.50-30.10$\\
RGS & $31.10-32.00$\\
RGS & $33.40-33.85$\\
MOS & $1.70\tablenotemark{a}-6.90$\\
MOS & $7.80-9.35$\\
\enddata
\tablenotetext{a}{For 47 Cas B and EK Dra. For the other stars, we used lower limits as follows: 4.96 \AA\ for  $\pi^1$ UMa and $\chi^1$ Ori, 4.13 \AA\  for $\kappa^1$ Cet, and 5.0 \AA\  for $\beta$ Com}  
\end{deluxetable}

\begin{deluxetable}{lllllllll}
\tablecaption{Lines used for the EMD reconstruction\tablenotemark{a}. Method 2, with MEKAL/SPEX\label{tab:lines}}
\tabletypesize{\scriptsize}
\tablewidth{0pt}
\tablehead{
\colhead{Line} & \colhead{$\log T_{m}$\tablenotemark{b}~ (K)} &\colhead{$\lambda$ (\AA)}  & \colhead{47 Cas B} & \colhead{EK Dra} & \colhead{$\pi^1$} UMa & \colhead{$\chi^1$ Ori} & \colhead{$\kappa^1$ Cet} & \colhead{$\beta$ Com}} 
\startdata
Fe\,{\sc xvii} & 6.7 &15.01    & 483.2$\pm$17.2 & 376.9$\pm$13.9  & 61.5$\pm$2.4 & 52.0$\pm$1.6 & 43.5$\pm$1.4 & 8.5$\pm$0.5 \\
Fe\,{\sc xvii} & 6.7 &16.78    & 217.5$\pm$12.3 & 152.7$\pm$9.4   & 29.3$\pm$1.8 & 24.1$\pm$1.1 & 24.8$\pm$1.1 & 4.3$\pm$0.4 \\
Fe\,{\sc xviii}& 6.8 &14.20   & 292.9$\pm$15.2 & 207.4$\pm$11.7  & 20.5$\pm$1.6 & 17.0$\pm$1.0 & 16.2$\pm$1.0 & 1.4$\pm$0.3 \\
Fe\,{\sc xix}  & 6.9 &13.52    & 222.9$\pm$26.6 & 126.6$\pm$19.6  & 14.5$\pm$2.9 & 8.3$\pm$1.6  & 6.8$\pm$1.5  & 0.8$\pm$0.6 \\
Fe\,{\sc xx}   & 7.0 &12.83     & 289.7$\pm$21.9 & 162.5$\pm$59.0  & 7.4$\pm$1.8  & 4.3$\pm$1.1  & 5.8$\pm$1.0  &  $-$          \\
Fe\,{\sc xxi}  & 7.0 &12.29    & 210.2$\pm$24.7 & 149.1$\pm$16.3  & 8.3$\pm$2.1  & 7.3$\pm$1.3          & 8.7$\pm$1.4  &  $-$           \\
Fe\,{\sc xxiii}& 7.2 &11.74  & 183.3$\pm$21.5 & 116.5$\pm $15.1 & $-$          & $-$          & $-$          &       $-$        \\
Fe\,{\sc xxiv} & 7.3 &10.62   & 95.0$\pm$40.1  & 55.8$\pm $13.1  & $-$          & $-$          & $-$          &      $-$        \\
Fe\,{\sc xxv}  & 7.8 &1.85    & 68.0$\pm$17.8  & 22.6$\pm $81.0  & $-$           & $- $         & $-$          &     $-$        \\
\tableline                      
O\,{\sc viii}  & 6.5&18.97    & 639.8$\pm$17.6 & 298.5$\pm$11.5  & 32.2$\pm$1.7 & 28.2$\pm$1.1 & 29.5$\pm$1.1 & 5.6$\pm$0.4 \\
O\,{\sc vii}   & 6.3&21.60     & 90.2$\pm$11.8  & 50.1$\pm$8.2    & 11.6$\pm$1.6 & 8.6$\pm$1.0  & 10.1$\pm$0.96& 2.4$\pm$0.5 \\
O\,{\sc vii}   & 6.3&22.10     & 48.6$\pm$10.2  & 39.5$\pm$7.7    &  8.4$\pm$1.5 & 6.5$\pm$0.8  & 6.1$\pm$0.9  & 3.3$\pm$0.5 \\
C\,{\sc vi}    & 6.1&33.73      & 68.2$\pm$7.9   & 38.1$\pm $5.2   &  2.6$\pm$2.3 & 3.6$\pm$0.5  & 4.1$\pm$0.6  & 1.3$\pm$0.3 \\
N\,{\sc vii}   & 6.3&24.77     & 62.2$\pm$8.9   & 32.3$\pm $5.1   &  3.8$\pm$1.5 & 1.8$\pm$0.4  & 2.5$\pm$0.4  & 0.4$\pm$0.3 \\
Ne\,{\sc x}    & 6.8&12.13      & 475.0$\pm$33.3 & 185.7$\pm $21.2 &  $-$         & $-$          &  $-$         & $-$\\
Ne\,{\sc ix}   & 6.6&13.45     & 277.8$\pm$27.6 & 113.0$\pm $17.9 &  8.4$\pm$2.3 & 8.9$\pm$1.5  & 10.0$\pm$1.5 & 0.9$\pm$0.5 \\
Ne\,{\sc ix}   & 6.6&13.70     &      $-$       &      $-$        & 5.9$\pm$2.0  & 5.2$\pm$1.2  & 6.4$\pm$1.2  &    $-$\\
Mg\,{\sc xii}  & 7.0&8.42     & 238.7$\pm$13.2 & 93.9$\pm$8.1    & 3.9$\pm$0.7  & 2.5$\pm$0.4  & 6.1$\pm$0.5  & 0.3$\pm$0.2 \\
Mg\,{\sc xi}   & 6.8&9.17      & 320.6$\pm$10.8 & 157.2$\pm$7.1   & 15.0$\pm$0.7 & 10.4$\pm$0.4 & 14.7$\pm$0.5 & 1.5$\pm$0.1 \\
Si\,{\sc xiv}  & 7.2&6.18     & 114.2$\pm$12.0  & 57.3$\pm$7.9    &  $-$         & $-$          & $ -     $    &      $-$        \\
Si\,{\sc xiii} & 7.0&6.65    & 194.0$\pm$11.0 & 105.6$\pm$7.5   & 5.6$\pm$0.6  & 4.9$\pm$0.4  & 5.7$\pm$0.4  & 0.7$\pm$0.2 \\
S\,{\sc xvi}   & 7.2&4.72      & 55.6$\pm$12.0  &      $-$        &  $-$         & $-$          & $-$          &       $-$        \\
S\,{\sc xv}    & 7.2&5.04       & 93.3$\pm$11.3  & 35.7$\pm$6.3    &  $-$         & $-$          & $-$          &       $-$        \\
\enddata
\tablenotetext{a}{For each line, the measured luminosity in $10^{26}$~erg~s$^{-1}$  
          using the MEKAL database is given. Note that the entries for the Fe lines contain blends
          of Fe around the given lines}
\tablenotetext{b}{Maximum line formation temperature}
\end{deluxetable}

\begin{deluxetable}{lllllllll}
\tablecaption{Lines used for the EMD reconstruction\tablenotemark{a}. Method 2, with APEC/XSPEC \label{tab:lines_xspec}}
\tabletypesize{\scriptsize}
\tablewidth{0pt}
\tablehead{
\colhead{Line} & \colhead{$\log T_{m}$\tablenotemark{b}~ (K)} & \colhead{$\lambda$ (\AA)} &  \colhead{47 Cas B} & \colhead{EK Dra} & \colhead{$\pi^1$ UMa} & \colhead{$\chi^1$ Ori} & \colhead{$\kappa^1$ Cet} & \colhead{$\beta$ Com}}
\startdata
Fe\,{\sc vii} & 6.7   &15.01 &   477.5$\pm$19.6 & 378.7$\pm$16.3  & 61.5$\pm$2.9 & 51.9$\pm$1.7 & 44.2$\pm$1.8 &  8.1$\pm$ 0.7 \\
Fe\,{\sc vii} & 6.7   &16.78 &   222.3$\pm$18.6 & 150.5$\pm$15.2   & 30.9$\pm$2.7 & 25.1$\pm$1.8 & 25.5$\pm$1.6 &4.5 $\pm$0.7 \\
Fe\,{\sc viii}& 6.9   &14.20 &  297.4$\pm$18.9 & 207.1$\pm$15.9  & 20.4$\pm$2.7 & 17.2$\pm$1.3 & 16.9$\pm$1.3 &1.4 $\pm$0.6 \\
Fe\,{\sc xix} & 6.9   &13.52 &   235.2$\pm$39.8 & 114.7$\pm$34.0  & 14.4$\pm$2.3 &5.8 $\pm$2.8  & 6.6$\pm$2.2  & 0.7$\pm$0.6 \\
Fe\,{\sc xx}  & 7.0   &12.83 &    246.8$\pm$ 30.9& 144.7$\pm$30.1  & 8.6$\pm$1.8  & 4.4$\pm$1.9  & 7.1$\pm$1.5  &  $-$   \\
Fe\,{\sc xxi} & 7.0   &12.29 &   190.0$\pm$40.3 & 132.2$\pm$29.6  & 12.3$\pm$3.7  &  6.8$\pm$2.3 & 9.7$\pm$1.5  &  $-$  \\
Fe\,{\sc xxiii}& 7.2  &11.74 & 155.4$\pm$20.4& 109.6$\pm $14.3 & $-$     & $-$          & $-$          &   $-$        \\
Fe\,{\sc xxiv}& 7.3   &10.62 &   71.3$\pm$59.4  & 53.5$\pm $33.5  & $-$             & $-$          & $-$          &         $-$        \\
Fe\,{\sc xxv} & 7.8  &1.85 &    72.7 $\pm$17.9  & 40.6$\pm $14.1  & $-$             & $- $         & $-$          &         $-$        \\
\tableline                            
O\,{\sc viii}  & 6.5&18.97 &   651.1$\pm$17.7 & 291.4$\pm$11.4  & 32.9$\pm$1.7 & 28.4$\pm$1.1& 30.1$\pm$8.7 & 5.6$\pm$0.4 \\
O\,{\sc vii}   & 6.3&21.60 &    99.7$\pm$20.8  & 45.5$\pm$13.3    & 11.8$\pm$1.6 & 9.1$\pm$1.4  &10.9 $\pm$1.4& 2.8$\pm$0.9 \\
O\,{\sc vii}   & 6.3&22.10 &     43.8$\pm$10.6 &39.5 $\pm$11.8    &  8.4$\pm$1.5 & 6.4$\pm$1.2  & 6.5$\pm$1.2  & 3.1$\pm$0.9 \\
C\,{\sc vi}    & 6.1&33.73 &     56.9$\pm$7.6   & 30.1$\pm $5.0   & 2.5 $\pm$1.0 & 2.8$\pm$0.5  & 3.5$\pm$0.5  & 1.3$\pm$0.3 \\
N\,{\sc vii}   & 6.3&24.77 &    69.2$\pm$9.2   & 32.4$\pm $5.2   &  2.1$\pm$0.9 & 1.5$\pm$0.4  & 2.7$\pm$0.5  & 0.2$\pm$ 0.2\\
Ne\,{\sc x}    & 6.8&12.13 &     500.2$\pm$35.3 & 167.0$\pm $21.7 &  $-$            & $-$          &  $-$         &  $-$        \\
Ne\,{\sc ix}   & 6.6&13.45 &    285.0$\pm$42.7 & 87.5$\pm $36.2 & 7.9$\pm$4.0 & 10.2$\pm$2.7  &10.6 $\pm$ 2.9 &$-$\tablenotemark{c} \\
Ne\,{\sc ix}   & 6.6&13.70 &            $-$       &      $-$                          &5.1$\pm$2.0  &  6.0$\pm$1.3 & 7.1$\pm$1.2  &         $-$\\
Mg\,{\sc xii}  & 7.0&8.42 &   245.9$\pm$13.6 &108.9 $\pm$16.1    &3.4 $\pm$0.9  &2.1$\pm$0.7  & 5.5$\pm$ 0.8  &  0.3$\pm$0.2 \\
Mg\,{\sc xi}   & 7.0&9.17 &     319.5$\pm$11.3 & 167.5$\pm$13.5   & 12.9$\pm$1.4 &9.6 $\pm$0.7 & 13.2$\pm$ 0.8& 1.5$\pm$0.2 \\
Si\,{\sc xiv}  & 7.2&6.18 &    130.6$\pm$ 24.7  & 63.0$\pm$11.9       &  $-$         & $-$          & $ -     $    &       $-$        \\
Si\,{\sc xiii} & 7.0&6.65 &   207.3$\pm$ 75.0& 104.5$\pm$14.0   &6.5$\pm$1.2  &  4.5$\pm$0.6  & 5.5$\pm$0.9  & 0.7$\pm$0.1 \\
S\,{\sc xvi}   & 7.4&4.72 &     52.9$\pm$8.1  &         $-$        &  $-$         & $-$          & $-$          &   $-$        \\
S\,{\sc xv}    & 7.2&5.04 &     105.1 $\pm$18.9  &  34.1$\pm$6.4       &  $-$         & $-$          & $-$          &        $-$        \\
\enddata
\tablenotetext{a}{For each line, the measured luminosity is given in $10^{26}$~erg~s$^{-1}$  
          using the APEC database. Note that the entries for the Fe lines contain blends
          of Fe around the given lines}
\tablenotetext{b}{Maximum line formation temperature}
\tablenotetext{c}{For $\beta$ Com none of the Ne lines could be reliably measured by fitting $\delta$ lines using 
          the baseline model from method 1 and the APEC database within XSPEC}
\end{deluxetable}

\begin{deluxetable}{llllllllll}
\tablecaption{Mean coronal temperatures\tablenotemark{a}~ and ranges\label{temp}}
\tabletypesize{\scriptsize}
\tablewidth{0pt}
\tablehead{
\colhead{} &\colhead{} &\colhead{} &\colhead{} &\colhead{SPEX} &\colhead{} &\colhead{}&\colhead{} &\colhead{APEC} &\colhead{} \\
\cline{4-6}  \cline{8-10}
\colhead{Star}    & \colhead{$\log L_X$\tablenotemark{b}} &\colhead{} & \colhead{$\log \bar{T}$}   & \colhead{$\log T_{\rm min}$} & \colhead{$\log T_{\rm max}$}&\colhead{}& \colhead{$\log \bar{T}$}   & \colhead{$\log T_{\rm min}$} & \colhead{$\log T_{\rm max}$}}
\startdata
47 Cas B   &   30.39  & & 7.03 & 6.51& 7.60 & & 7.02 & 6.52& 7.65\\
EK Dra      &  30.08  & & 6.96 & 6.46& 7.45 & & 6.99 & 6.47& 7.58 \\
$\pi^1$ UMa  &  29.06 & & 6.65 & 6.16& 6.98 & & 6.69 & 6.18& 7.17 \\
$\chi^1$ Ori &   28.95 & & 6.64 & 6.14& 7.03 & & 6.66 & 6.22& 6.98 \\
$\kappa^1$ Cet &   28.95 & & 6.66 & 6.18& 7.05 & & 6.65 & 6.13& 7.10 \\
$\beta$ Com  &  28.26 & & 6.59 & 6.15& 6.92 & & 6.55 & 6.07& 6.89 \\
\enddata
\tablenotetext{a}{$\log \bar{T}$ is the EM-weighted average of $\log T$, $T_{\rm min}$ and $T_{\rm max}$ are the minimum and maximum temperatures 
                  that contain 90\% of the EM above and below $\bar{T}$, respectively}
\tablenotetext{b}{Luminosity in erg~s$^{-1}$ determined with {\it XMM-Newton} in the 0.1--10~keV band}
\end{deluxetable}

\begin{deluxetable}{lllll}
\tablecaption{Reduced $\chi^2$ for the synthetic best-fit spectra\tablenotemark{a} \label{redchi}}
\tabletypesize{\scriptsize}
\tablewidth{0pt}
\tablehead{
\colhead{Star}    & \colhead{ $\chi^2_{\rm red}$\tablenotemark{b}}   & \colhead{ $\chi^2_{\rm red}$\tablenotemark{c}}  & \colhead{ $\chi^2_{\rm red}$\tablenotemark{d}} & \colhead{d.o.f \tablenotemark{e}}}
\startdata
47 Cas B      & 1.22 & 1.37  & 1.28 & 1091\\
              & 1.14 & 1.35  & $-$  & 1006\\
EK Dra        & 1.25 & 1.34  & 1.25 & 769\\
              & 1.21 & 1.29  & $-$  & 671\\
$\pi^1$ UMa   & 1.66 & 1.51  & $-$  & 415\\
              & 1.28 & 1.42  & $-$  & 327\\
$\chi^1$ Ori  & 1.58 & 1.71  & $-$  & 628\\
              & 1.28 & 1.39  & $-$  & 514\\
$\kappa^1$ Cet& 1.40 & 1.58  & $-$  & 635\\
              & 1.35 & 1.43  & $-$  & 538\\
$\beta$ Com   & 1.33 & 1.42  & $-$  & 405\\
              & 1.24 &  1.27 & $-$  & 313\\
\enddata
\tablenotetext{a}{The first line for a given star is based on MEKAL emissivities, the second line on APEC emissivities}
\tablenotetext{b}{spectrum obtained from method 1 EMD, using the regions listed in Table~\ref{region}}
\tablenotetext{c}{spectrum obtained from method 2 EMD, iterated to $\psi^2$=1.0 (based on MEKAL)}  
\tablenotetext{d}{spectrum obtained from method 2 EMD, iterated to $\psi^2$=0.5 (based on MEKAL)}  
\tablenotetext{e}{degrees of freedom}  
\end{deluxetable}

\begin{deluxetable}{llllllll}
\tabletypesize{\scriptsize}
\tablecaption{Abundances with respect to Fe, and absolute Fe abundances\tablenotemark{a}\label{abuntable}}
\tablewidth{0pt}
\tablehead{
\colhead{Abundance (ratio)} &\colhead{Method} & \colhead{47 Cas B} & \colhead{EK Dra} & \colhead{$\pi^1$ UMa} & \colhead{$\chi^1$ Ori} & \colhead{$\kappa^1$ Cet }& \colhead{$\beta$ Com }}
\startdata
C/Fe & M1. MEKAL    &0.95$_{-0.17}^{+0.18}$&0.54$_{-0.11}^{+0.14}$&0.20$_{-0.07}^{+0.09}$&0.29$_{-0.06}^{+0.07}$&0.34$_{-0.08}^{+0.08}$&0.29$_{-0.17}^{+0.29}$\\
         & M2. MEKAL    &0.80$_{-0.18}^{+0.18}$&0.63$_{-0.19}^{+0.19}$&0.19$_{-0.24}^{+0.24}$&0.34$_{-0.11}^{+0.11}$&0.43$_{-0.12}^{+0.12}$&0.66$_{-0.29}^{+0.29}$\\
         & M1. APEC   &0.76$_{-0.13}^{+0.13}$&0.60$_{-0.24}^{+0.20}$&0.23$_{-0.13}^{+0.13}$&0.19$_{-0.05}^{+0.11}$&0.22$_{-0.06}^{+0.08}$&0.10$_{-0.14}^{+0.10}$\\
         & M2. APEC   &0.65$_{-0.17}^{+0.17}$&0.44$_{-0.13}^{+0.13}$&0.14$_{-0.09}^{+0.09}$&0.24$_{-0.07}^{+0.07}$&0.24$_{-0.08}^{+0.08}$&0.39$_{-0.22}^{+0.22}$\\
\tableline
N/Fe & M1. MEKAL    &1.18$_{-0.21}^{+0.21}$&0.84$_{-0.17}^{+0.18}$&0.36$_{-0.12}^{+0.15}$&0.25$_{-0.06}^{+0.07}$&0.43$_{-0.09}^{+0.10}$&0.12$_{-0.12}^{+0.15}$\\
         & M2. MEKAL    &1.07$_{-0.28}^{+0.28}$&0.78$_{-0.25}^{+0.25}$&0.44$_{-0.28}^{+0.28}$&0.26$_{-0.09}^{+0.09}$&0.42$_{-0.12}^{+0.12}$&0.30$_{-0.29}^{+0.29}$\\
         & M1. APEC   &1.20$_{-0.20}^{+0.21}$&0.71$_{-0.21}^{+0.26}$&0.32$_{-0.17}^{+0.14}$&0.18$_{-0.04}^{+0.06}$&0.24$_{-0.06}^{+0.08}$&0.03$_{-0.05}^{+0.06}$\\
         & M2. APEC   &1.19$_{-0.30}^{+0.30}$&0.73$_{-0.21}^{+0.21}$&0.20$_{-0.14}^{+0.14}$&0.20$_{-0.08}^{+0.08}$&0.33$_{-0.11}^{+0.11}$&0.10$_{-0.15}^{+0.15}$\\
\tableline
O/Fe & M1. MEKAL    &0.79$_{-0.10}^{+0.09}$&0.60$_{-0.09}^{+0.09}$&0.27$_{-0.06}^{+0.07}$&0.35$_{-0.06}^{+0.05}$&0.36$_{-0.06}^{+0.07}$&0.28$_{-0.13}^{+0.13}$\\
         & M2. MEKAL    &0.70$_{-0.11}^{+0.11}$&0.51$_{-0.15}^{+0.15}$&0.32$_{-0.09}^{+0.09}$&0.33$_{-0.08}^{+0.08}$&0.39$_{-0.09}^{+0.09}$&0.41$_{-0.14}^{+0.14}$\\
         & M1. APEC   &0.64$_{-0.07}^{+0.08}$&0.41$_{-0.10}^{+0.10}$&0.17$_{-0.09}^{+0.06}$&0.17$_{-0.05}^{+0.03}$&0.18$_{-0.04}^{+0.04}$&0.06$_{-0.08}^{+0.06}$\\
         & M2. APEC   &0.64$_{-0.11}^{+0.11}$&0.40$_{-0.07}^{+0.07}$&0.25$_{-0.06}^{+0.06}$&0.28$_{-0.06}^{+0.06}$&0.27$_{-0.09}^{+0.09}$&0.26$_{-0.12}^{+0.12}$\\
\tableline
Ne/Fe & M1. MEKAL    &1.78$_{-0.18}^{+0.19}$&1.05$_{-0.15}^{+0.15}$&0.30$_{-0.10}^{+0.11}$&0.32$_{-0.07}^{+0.06}$&0.50$_{-0.09}^{+0.10}$&0.09$_{-0.10}^{+0.12}$\\
           & M2. MEKAL    &1.68$_{-0.29}^{+0.29}$&1.01$_{-0.21}^{+0.21}$&0.62$_{-0.21}^{+0.21}$&0.73$_{-0.16}^{+0.16}$&0.95$_{-0.20}^{+0.20}$&0.41$_{-0.35}^{+0.35}$\\
           & M1 APEC   &1.65$_{-0.18}^{+0.18}$&0.96$_{-0.25}^{+0.20}$&0.32$_{-0.15}^{+0.13}$&0.32$_{-0.07}^{+0.09}$&0.44$_{-0.10}^{+0.10}$&0.06$_{-0.10}^{+0.13}$\\
           & M2. APEC   &1.75$_{-0.45}^{+0.45}$&0.79$_{-0.17}^{+0.17}$&0.40$_{-0.15}^{+0.15}$&0.59$_{-0.12}^{+0.12}$&0.69$_{-0.19}^{+0.19}$&$-$                                \\
\tableline
Mg/Fe & M1. MEKAL    &1.71$_{-0.17}^{+0.17}$&1.31$_{-0.17}^{+0.17}$&1.22$_{-0.27}^{+0.29}$&0.98$_{-0.13}^{+0.13}$&1.49$_{-0.23}^{+0.23}$&1.00$_{-0.51}^{+0.55}$\\
           & M2. MEKAL    &2.21$_{-0.38}^{+0.38}$&1.54$_{-0.28}^{+0.28}$&1.24$_{-0.31}^{+0.31}$&1.12$_{-0.18}^{+0.18}$&1.94$_{-0.30}^{+0.30}$&1.20$_{-0.24}^{+0.24}$\\
           & M1 APEC   &1.50$_{-0.16}^{+0.16}$&1.21$_{-0.29}^{+0.26}$&1.03$_{-0.54}^{+0.53}$&0.81$_{-0.20}^{+0.21}$&1.17$_{-0.27}^{+0.27}$&0.90$_{-1.26}^{+1.42}$\\
           & M2. APEC   &1.97$_{-0.23}^{+0.23}$&1.41$_{-0.26}^{+0.26}$&0.73$_{-0.21}^{+0.21}$&0.73$_{-0.17}^{+0.17}$&1.19$_{-0.18}^{+0.18}$&0.81$_{-0.18}^{+0.18}$\\
\tableline
Si/Fe & M1. MEKAL    &0.89$_{-0.10}^{+0.11}$&0.89$_{-0.13}^{+0.12}$&0.69$_{-0.17}^{+0.19}$&0.74$_{-0.12}^{+0.12}$&0.87$_{-0.15}^{+0.15}$&1.13$_{-0.68}^{+0.64}$\\
           & M2. MEKAL    &1.07$_{-0.17}^{+0.17}$&0.95$_{-0.15}^{+0.15}$&0.68$_{-0.14}^{+0.14}$&0.82$_{-0.16}^{+0.16}$&0.95$_{-0.17}^{+0.17}$&0.95$_{-0.38}^{+0.38}$                               \\
           & M1 APEC   &0.79$_{-0.10}^{+0.10}$&0.85$_{-0.21}^{+0.18}$&0.60$_{-0.28}^{+0.27}$&0.66$_{-0.16}^{+0.19}$&0.73$_{-0.17}^{+0.17}$&0.77$_{-1.11}^{+0.63}$\\
           & M2. APEC   &1.06$_{-0.20}^{+0.20}$&0.78$_{-0.12}^{+0.12}$&0.50$_{-0.14}^{+0.14}$&0.57$_{-0.11}^{+0.11}$&0.63$_{-0.11}^{+0.11}$&0.69$_{-0.27}^{+0.27}$\\
\tableline
S/Fe & M1. MEKAL    &0.77$_{-0.14}^{+0.15}$&0.59$_{-0.15}^{+0.15}$&$-$                               &$-$                               &$-$                               &$-$                               \\
         & M2. MEKAL    &1.03$_{-0.32}^{+0.32}$&0.59$_{-0.18}^{+0.18}$&$-$                               &$-$                               &$-$                               &$-$                               \\
         & M1. APEC   &0.60$_{-0.11}^{+0.11}$&0.54$_{-0.17}^{+0.15}$&$-$                               &$-$                               &$-$                               &$-$                               \\
         & M2. APEC   &1.09$_{-0.26}^{+0.26}$&0.50$_{-0.15}^{+0.15}$&$-$                               &$-$                               &$-$                               &$-$                               \\
\tableline
Ar/Fe & M1. MEKAL    &1.58$_{-0.42}^{+0.43}$&0.44$_{-0.44}^{+0.44}$&$-$&$-$&$-$&$-$\\
     & M1. APEC   &1.29$_{-0.33}^{+0.33}$&0.55$_{-0.45}^{+0.47}$&$-$&$-$&$-$&$-$\\
\tableline
Fe & M1. MEKAL    &0.51$_{-0.04}^{+0.03}$&0.63$_{-0.05}^{+0.06}$&0.73$_{-0.09}^{+0.12}$&0.63$_{-0.04}^{+0.06}$&0.71$_{-0.06}^{+0.08}$&0.43$_{-0.11}^{+0.17}$\\
     & M2. MEKAL    &0.50$_{-0.05}^{+0.05}$&0.72$_{-0.08}^{+0.08}$&0.81$_{-0.17}^{+0.17}$&0.87$_{-0.14}^{+0.14}$&1.18$_{-0.25}^{+0.25}$&1.27$_{-1.53}^{+1.53}$\\
     & M1. APEC   &0.69$_{-0.05}^{+0.06}$&0.74$_{-0.05}^{+0.17}$&1.07$_{-0.32}^{+0.46}$&0.98$_{-0.18}^{+0.17}$&0.93$_{-0.17}^{+0.12}$&1.76$_{-0.41}^{+0.43}$\\
     & M2. APEC   &0.55$_{-0.05}^{+0.05}$&0.96$_{-0.11}^{+0.11}$&1.26$_{-0.31}^{+0.31}$&0.83$_{-0.13}^{+0.13}$&1.83$_{-0.48}^{+0.48}$&$-$                               \\
\enddata
\tablenotetext{a}{All abundance ratios and Fe abundances are with respect to the solar photospheric abundances
                  given by \citet{anders89} except for Fe, for which the photospheric value given
                 by \citet{grevesse99} has been adopted}
                 
\end{deluxetable}

\begin{deluxetable}{lllllll}
\tablecaption{Light curve modeling \label{tab:lc}}
\tabletypesize{\scriptsize}
\tablewidth{0pt}
\tablehead{
\colhead{}    & \colhead{47 Cas B }  & \colhead{EK Dra}  & \colhead{$\pi^1$ UMa} & \colhead{$\chi^1$ Ori}  & \colhead{$\kappa^1$ Cet}& \colhead{$\beta$ Com}}
\startdata
Slope for $T<T_m$\tablenotemark{a} &2.10-2.63 & 2.22-3.09 & 1.51-2.46 & 1.88-2.86 & 1.85-2.80 & 2.12-3.67 \\  
Slope for $T>T_m$\tablenotemark{b} &(-1.34)-(-1.48) & (-1.55)-(-1.62) & (-2.45)-(-3.63) & (-3.26)-(-3.68)& (-2.25)-(-2.60) &  (-2.95)-(-3.53)\\  
$\phi$ & -0.3 &-0.3 &-0.3 &-0.3 &-0.3 &-0.3 \\
$\zeta$&0.95-0.76 & 0.90-0.65 & 1.32-1.23 & 1.06-0.70 & 1.08-0.71 & 0.94-0.54 \\
$\alpha (\beta=0)$ & 2.21-2.28 & 2.25-2.31 & 2.43-2.78 & 2.60-2.80 & 2.39-2.54 & 2.54-2.76\\ 
$\alpha (\beta=0.25)$ & 2.16-2.20 & 2.19-2.23 & 2.33-2.59 & 2.45-2.60 & 2.30-2.41 & 2.40-2.57\\ 
$\log E_2/E_1 (\beta=0)$ & (-4.2)-(-5.3) & (-2.8)-(-3.4) & (-1.8)-(-2.2) & $-$ & $-$ & $-$ \\
$\log E_2/E_1 (\beta=0.25)$ & (-6.2)-(-6.8) & (-3.5)-(-4.1) & (-1.8)-(-2.5) & $-$ & $-$ & $-$ \\
$\log E_2$    & 30.20 & 30.00 & 29.0 & $-$ & $-$ & $-$ \\
mod. depth\tablenotemark{c} & 0.028 & 0.056 & 0.122 &$-$ & $-$ & $-$ \\
\enddata
\tablenotetext{a}{Determined in the logarithmic energy range between $\log T = 6.2$ and ($\log T_m$-0.1)}
\tablenotetext{b}{Determined in the logarithmic energy range between ($\log T_m$+0.1) and an upper 
limit that depends on the star ($\log T = 7.5$ for 47 Cas and EK Dra, $\log T = 7.3$ for $\pi^1$ UMa, and $\log T = 7.1$ for the other stars)}
\tablenotetext{c}{Modulation depth of the observed light curves}
\end{deluxetable}

\clearpage

\clearpage

\clearpage

\clearpage


\clearpage

\clearpage

\clearpage

\clearpage

\clearpage

\clearpage

\clearpage

\clearpage

\clearpage

\end{document}